\documentclass[11pt]{article}
\usepackage[title]{appendix}
\usepackage[margin={1.0in}]{geometry}
\usepackage{rotating, lscape}
\usepackage{float}
\usepackage{graphicx, color}

\usepackage{threeparttable, booktabs, longtable, threeparttablex, tabularx, multirow, pdflscape}

 \usepackage{natbib}
\usepackage[sc]{mathpazo}
\usepackage{courier}
\usepackage[utf8]{inputenc}
\usepackage[T1]{fontenc}
\usepackage{setspace} 

\usepackage{amssymb,amsmath,mathrsfs,amsthm, bbm}
\usepackage{accents}

\usepackage{microtype}
\usepackage[colorlinks=true, citecolor=black, urlcolor=blue]{hyperref}
\usepackage{afterpage}
\usepackage{caption}
\usepackage{amsmath}
\usepackage{subfigure}

\RequirePackage[font=small,format=plain,labelfont=bf,textfont=it]{caption}
\usepackage{chngcntr}
\usepackage{palatino}
\usepackage{threeparttable}
\usepackage{amsthm}
\usepackage{ragged2e}
\usepackage{subcaption}
\newtheorem{theorem}{Theorem}

\newtheorem{assumption}{Assumption}

\usepackage{textcomp}
\usepackage{tkz-graph}
\usetikzlibrary{shapes.geometric}

\usepackage{hyperref}
\usepackage{xcolor}
\hypersetup{
  colorlinks   = true, %Colours links instead of ugly boxes
  urlcolor     = blue, %Colour for external hyperlinks
  linkcolor    = blue, %Colour of internal links
  citecolor   = blue %Colour of citations
}

\newcolumntype{Y}{>{\centering\arraybackslash}X}

\newcolumntype{L}{>{\raggedleft\arraybackslash}X}

\newcolumntype{R}{>{\raggedright\arraybackslash}X}

\newenvironment{notes}%
{\begin{minipage}{\linewidth}%
\smallskip\footnotesize\emph{Notes:}}%
{\end{minipage}}

\usetikzlibrary{shapes,decorations,arrows,calc,arrows.meta,fit,positioning}
\tikzset{
    -Latex,auto,node distance =1 cm and 1 cm,semithick,
    state/.style ={ellipse, draw, minimum width = 0.7 cm},
    point/.style = {circle, draw, inner sep=0.04cm,fill,node contents={}},
    bidirected/.style={Latex-Latex,dashed},
    el/.style = {inner sep=2pt, align=left, sloped}
}

\usepackage{etoolbox}
\makeatletter
\def\@fnsymbol#1{\ifcase#1\or *\or **\or ***\else\@ctrerr\fi}
\makeatother

\title{A joint test of unconfoundedness and common trends\thanks{We grateful to Oliver G\"urtler, Jon Roth, Kaspar W\"uthrich, and Tom Zimmermann for helpful comments.}}
\author{{Martin Huber}\footnote{University of Fribourg; martin.huber@unifr.ch} \space and {Eva-Maria Oeß}\footnote{Universit\"{a}t zu K\"{o}ln; oess@wiso.uni-koeln.de}}
\date{\today}
\begin{document}

% Redefine the \thefootnote command to use Arabic numerals
\renewcommand{\thefootnote}{\arabic{footnote}}

\maketitle 
\thispagestyle{empty}

\begin{abstract}
\noindent 

This paper introduces an overidentification test of two alternative assumptions to identify the average treatment effect on the treated in a two-period panel data setting: unconfoundedness and common trends. Under the unconfoundedness assumption, treatment assignment and post-treatment outcomes are independent, conditional on control variables and pre-treatment outcomes, which motivates including pre-treatment outcomes in the set of controls. Conversely, under the common trends assumption, the trend and the treatment assignment are independent, conditional on control variables. This motivates employing a Difference-in-Differences (DiD) approach by comparing the differences between pre- and post-treatment outcomes of the treatment and control group. Given the non-nested nature of these assumptions and their often ambiguous plausibility in empirical settings, we propose a joint test using a doubly robust statistic that can be combined with machine learning to control for observed confounders in a data-driven manner. We discuss various causal models that imply the satisfaction of either common trends, unconfoundedness, or both assumptions jointly, and we investigate the finite sample properties of our test through a simulation study. Additionally, we apply the proposed method to five empirical examples using publicly available datasets and find the test to reject the null hypothesis in two cases.

\end{abstract}

	
	\vspace{0.3cm} \noindent 
\newpage
\setcounter{page}{1}

\section{Introduction}

For panel data where outcomes are observed both before and after treatment, there exist two popular strategies that exploit pre-treatment outcomes to identify the average treatment effect on the treated (ATET) on post-treatment outcomes. The first approach uses pre-treatment outcomes as control variables and relies on the unconfoundedness assumption, which invokes that, conditional on pre-treatment outcomes and possibly additional covariates, the treatment is independent of mean potential outcome under non-treatment. The second approach utilizes the difference between pre- and post-treatment outcomes as dependent variable when assessing the ATET, a so-called Difference-in-Differences (DiD) method that relies on the common trends assumption. This assumption asserts that treatment is independent of the time trend in mean potential outcome under non-treatment, potentially conditional on covariates. Unconfoundedness and common trends are non-nested assumptions, as discussed in \cite{ChabeFerret2017} and \cite{xu2023causal}.

For example, the unconfoundedness assumption allows pre-treatment outcomes to influence treatment but mandates that conditional on the pre-treatment outcome and possibly additional covariates, there are no unobserved confounders affecting treatment and post-treatment outcomes, not even time-constant ones. In contrast, the common trends assumption allows for time-constant unobserved confounders affecting treatment and outcome, as long as their effects are additively separable with regards to period fixed effects, thereby ruling out interaction effects between time-constant confounders and period fixed effects. However, the common trends assumption generally does not permit pre-treatment outcomes to influence treatment if the outcome is affected by time-varying (random) errors. The preference for one assumption over the other is typically not clearly motivated in applied work, as also acknowledged by \cite{NBERw31942} and in many empirical contexts, it appears far from clear  which of the two assumptions is more plausible, if any.

Given the frequent ambiguity surrounding the plausibility of common trends and unconfoundedness assumptions, this paper proposes a method to jointly test the common trends and unconfoundedness assumptions for ATET identification. The test is based on the difference in estimated average non-treated counterfactuals of treated units under common trends and unconfoundedness assumptions. Estimation relies on doubly robust (DR) statistics, as discussed in \citet{Robins+94} and \citet{RoRo95}, which is combined with machine learning to control for (possibly high-dimensional) covariates in a data-driven manner, following the double machine learning framework (DML) of \citet{Chetal2018}. As shown in the appendix, our DR statistic satisfies \cite{Neyman1959}-orthogonality, implying that the test statistic is asymptotically normal and $\sqrt{n}$-consistent under specific regularity conditions for machine learners applied to control for important covariates acting as confounders. We also discuss causal models in which either common trends or unconfoundedness hold, implying the failure of the null hypothesis tested by our procedure, or both assumptions hold jointly, such that the null hypothesis is satisfied. Furthermore, we provide a simulation study to investigate the finite sample properties. 

Moreover, we apply the test to four different publicly available datasets, where the treatment is not randomly assigned.
First, we consider the non-experimental \cite{LaLonde86} data, designed to estimate the effect of a job training program on real earnings, when comparing the original treatment group to a control group sourced from the Panel Study of Income Dynamics. Second, we investigate data from the Job Corps program, which supports disadvantaged youths through various vocational training opportunities. Youths enrolled in the program have the option to participate in educational programs. We apply the test to the effect of educational program participation on weeks employed and general health. Third, we analyze data from a natural experiment in the United States. \cite{CardKrueger1994} examined the effect of the minimum wage increase on employment, utilizing survey data from fast food restaurants in New Jersey, where the the policy was implemented, and Pennsylvania, where the minimum wage remained the same. Lastly, we apply the test to data from the United States' National Health and Nutrition Examination Survey to examine robustness of ATET regarding the effect of smoking cessation on weight using a textbook example from \cite{Hernan2020}. Selection into smoking cessation is likely confounded. Yet, the panel provides a larger number of covariates, to leverage selection on observables. In our empirical findings, the test indicates that the null hypothesis of joint unconfoundedness and common trends does not hold in the Job Corps data. However, we cannot reject the null hypothesis in the remaining examples.

% Literature
Our test follows the Durbin–Wu–Hausman framework, as outlined in \cite{Hausman1978}, designed to test two distinct model specifications. Such tests have been utilized for assessing nested model assumptions, where one causal model is more efficient while the other is more robust in the sense that it relies on weaker assumptions. For example, it has been employed to compare the random effects model, assuming exogeneity of treatment with respect to time-constant unobserved confounders, against the fixed effects model, which permits treatment correlation with time-constant confounders, as discussed in \cite{Wooldridge02book}. Additionally, the test has been used to compare OLS versus linear instrumental variable model assumptions, examining whether coefficients in one model significantly differ from those in the other. Our approach stands out by not imposing functional form assumptions such as linearity, thus accommodating heterogeneous treatment effects across observed or unobserved background characteristics. It also allows for controlling covariates and functional forms in a data-driven manner.

Moreover, our paper contributes to the literature on the bracketing relationship between estimation methods controlling for pre-treatment outcomes and DiD estimation. \cite{angrist_mostly_2008} demonstrate that if the true ATET is positive, falsely assuming common trends while unconfoundedness holds will overestimate the ATET, whereas falsely assuming unconfoundedness while common trends hold will underestimate the ATET. This relationship reverses if the true ATET is negative. \cite{Ding2019} extend this result to nonparametric and semiparametric settings, such as ours. However, our test does not identify which of the two assumptions is violated; it only indicates if one of them is not met.

Our method is most closely related to the Durbin-Wu-Hausman-type test suggested by \cite{DoHsLi2014}. The latter utilizes inverse probability weighting (IPW) to assess the unconfoundedness assumption in the presence of a valid instrumental variable (IV) that influences the treatment but is unrelated to the outcome except through the treatment, conditional on covariates. In scenarios featuring one-sided noncompliance, where the treatment is only received when the instrument is present, their test compares the IV-based estimate of the local average treatment effect on the treated (LATT) with the ATET estimate under unconfoundedness, as both causal parameters coincide under such circumstances.  While our approach shares similarities with this overidentification strategy, it also exhibits some distinctions. Firstly, we jointly test the unconfoundedness and common trend (rather than IV) assumptions and refrain from assuming that one is a priori more plausible or known to be satisfied. Secondly, instead of employing IPW, we utilize DML techniques to control for possibly high-dimensional covariates within a doubly robust framework tailored for panel data.

The remainder of this study is organized as follows. Section \ref{ass} discusses the common trend and unconfoundedness assumptions required for ATET identification based on DiD or controlling for pre-treatment outcomes, respectively, and introduces our doubly robust test statistic. Section \ref{est} discusses the empirical implementation of our test based on DML. Section \ref{sim} provides a simulation  study. Section \ref{app} presents empirical applications, and Section \ref{conclusion} concludes. 

% \cite{imbens2024lalonde} einfuegen?

\section{Identifying assumptions and test statistic}\label{ass}

We aim at identifying the average treatment effect on the treated (ATET), i.e., the average causal impact of a binary treatment, denoted by $D$, on an outcome variable in a post-treatment period among those receiving the treatment. We will make use of capital letters for denominating random variables ($D$) and lowercase letters to express specific values of these variables ($d=0,1$ for not receiving or receiving the treatment, respectively). To distinguish observed  outcomes in terms of pre- and post-treatment periods, we make use of time index $T$. The latter is equal to zero in the pre-treatment period, when no subject receives the treatment (yet), and one in the post-treatment period, after introducing the treatment in the treated group with $D=1$, but not in the nontreated group with $D=0$. We denote by $Y_t$ the observed outcome in period $T=t$, with $t=0,1$ for outcomes observed in the pre-treatment and post-treatment periods, respectively. We assume to have panel data available in which we observe both pre- and post-treatment outcomes of all subjects in those data. 

We also assume to observe a set of covariates, denoted as $X$, which are allowed to jointly affect the treatment and outcome, but must not be affected by the treatment. In many treatment evaluations, $X$ includes variables measured prior to treatment assignment to rule out effect of $D$ on $X$. However, in numerous DiD applications, the covariates are measured in the same period as the (pre- or post-treatment) outcome, implying that for $T=1$, $X$ is measured after the introduction of the treatment. In this case, $X$ must not be influenced by $D$, otherwise treatment evaluation is generally biased if $X$ also affects $Y$ and/or is associated with unobservables affecting $Y$. See also the discussion in \cite*{caetano2022difference} on alternative evaluation strategies for time-varying covariates depending on whether the latter are or are not influenced by the treatment. In this paper, we will consider conditioning on pre-treatment covariates only. This implies that for panel data including subject-specfic pre- and post-treatment outcomes and covariates, $X$ only includes pre-treatment covariates measured in period $t=0$.

For discussing the identification of the ATET, we make use of the potential outcomes framework, see for instance \cite{Neyman23} and \cite{Rubin74}. Notably, $Y_t(d)$ denotes the potential outcome that would be hypothetically realized under treatment assignment $D=d$ in period $t=T$, with $d,t$ $\in$ $\{0,1\}$. This permits defining the ATET in post-treatment period $t=1$, which we denote by $\Delta_{D=1}$:
\begin{align}\label{ATET}
\Delta_{D=1}=E[Y_1(1)-Y_1(0)|D=1]
\end{align}
Throughout the discussion, we assume that the `stable unit treatment value assumption' (SUTVA) holds,  see \cite{Cox58} and \cite{Rubin80}, imposing that any subject's potential outcomes are not affected by the treatment status of other subjects. Furthermore, we rule out any anticipation effects of the treatment on the pre-treatment outcomes, implying that subjects do not anticipate their treatment in a way that affects their pre-treatment outcomes.

We henceforth discuss two types of assumptions that both yield the identification of $\Delta_{D=1}$. We first consider the unconfoundedness assumption, a popular condition in the treatment evaluation literature as for instance discussed in \cite {Im04} and \cite{ImWo08}. It implies treatment selection based on observables in the sense that the treatment is conditionally mean independent of the potential outcome under non-treatment, $Y_1(0)$, given the covariates $X$ and pre-treatment outcome $Y_0$:
\begin{assumption}[Unconfoundedness]\label{A1}
\begin{align*}
  & E[Y_1(0)|D=1,X,Y_0]=E[Y_1(0)|D=0,X,Y_0],\\
  & \Pr(D=1|X,Y_0)<1.
\end{align*}
\end{assumption}
The first line of Assumption \ref{A1} formalizes the unconfoundedness assumption,  stating that given $X$ and $Y_0$, no unobservables jointly affect  the treatment and the (post-treatment) mean potential outcome under non-treatment. The second line imposes a common support condition, requiring that for any value of $X$ and $Y_0$ among the treated, non-treated subjects with comparable values in $X$ and $Y_0$ must exist. 

Under Assumption \ref{A1},  $\Delta_{D=1}$ is identified by 
\begin{align}\label{identunconf}
\Delta_{Unconf} = \underbrace{E[Y_1|D=1]}_{E[Y(1)|D=1]}-\underbrace{E[E[Y_1|X,Y_0,D=0]|D=1]}_{E[Y(0)|D=1]}, 
\end{align}
because
\begin{align}
E[Y_1|X,Y_0,D=0]=E[Y_1(0)|X,Y_0,D=0]=E[Y_1(0)|X,Y_0,D=1],
\end{align}
where the first equality follows from the observational rule ($Y_1(0)=Y_1$ given $D=0$) and the second equality follows from unconfoundedness. Subsequently, we denote the ATET identified by Assumption \ref{A1} as $\Delta_{Unconf}$.

The second type of assumptions imposes conditional common trends, as discussed in \cite{Abadie2005} and \cite{Lechner2010}. It requires that that the treatment is conditionally mean independent of trends in the potential outcome under non-treatment, $Y_1(0)-Y_0(0)$, given the covariates $X$:
\begin{assumption}[Conditional common trends]\label{A2}
\begin{align*}
  & E[Y_1(0)-Y_0(0)|D=1,X]=E[Y_1(0)-Y_0(0)|D=0,X],\\
  & \Pr(D=1|X)<1.
\end{align*}
\end{assumption}
The first line in Assumption \ref{A2} formalizes the conditional common trend assumption,  stating that given $X$, no unobservables jointly affect the treatment and the trend of mean potential outcomes under non-treatment. This is a selection-on-observables assumption on $D$, however, w.r.t.\ the changes in mean potential outcomes under non-treatment over time, rather than the post-treatment levels of potential outcomes under non-treatment as under the previously discussed unconfoundedness assumption. Subsequently we use the terms conditional common trends and common trends interchangeably, when referring to \ref{A2}. The two types of  selection-on-observables assumptions are not nested in the sense that neither implies the other. Furthermore, they cannot be fruitfully combined to develop a more general treatment effect model, as shown in \cite{ChabeFerret2017}. The second line in Assumption \ref{A2} imposes common support, requiring that for any value of $X$ among the treated, non-treated subjects with comparable values in $X$ must exist.

Under Assumption \ref{A2},  $\Delta_{D=1}$ is identified by the following difference-in-differences (DiD) approach in panel data: 
\begin{align}\label{identdid}
\Delta_{DiD} = \underbrace{E[Y_1|D=1]}_{E[Y(1)|D=1]}-\underbrace{\{E[Y_0|D=1]+E[E[Y_1-Y_0|X,D=0]|D=1]\}}_{E[Y(0)|D=1]}. 
\end{align}
This is because by the observational rule and the absence of anticipation effects, 
\begin{align}
E[Y_1|D=1]-E[Y_0|D=1]=E[Y_1(1)|D=1]-E[Y_0(0)|D=1],
\end{align}
while 
\begin{equation}
\begin{split}
E[Y_1-Y_0|X,D=0] & = E[Y_1(0)|X,D=0] - E[Y_0(0)|X,D=0] \\
& = E[Y_1(0)|X,D=1] - E[Y_0(0)|X,D=1],
\end{split}
\end{equation}
where the first equality follows from the observational rule and the second equality follows from common trends. We label the ATET identified by Assumption \ref{A2} as $\Delta_{DiD}$.

\begin{figure}[!htp]
	\centering \caption{\label{dag1}  Causal model violating unconfoundedness and common trends \bigskip}
\centering 
\begin{subfigure}
\centering
\begin{tikzpicture}
  % x node set with absolute coordinates
  \node[state] (Y0) at (0,0) {$Y_0$};
  % y node set relative to x.
  % Locations can be:
  % right,left,above,below,
  % above left,below right, etc
  \node[state] (D) [right =of Y0] {$D$};
  \node[state] (Y1) [right =of D] {$Y_1$};
  \node[state, dashed] (V0) [above =of Y0] {$V_0$};
  \node[state, dashed] (V1) [above =of Y1] {$V_1$};
  \node[state, dashed] (U)  [below =of D] {$U$};
  \node[state, dashed] (Q)  [above =of D, yshift=-0.45cm] {$Q$};
  %\node[dashed] (Q)  [above =of D] {$Q$};
   % Directed edge
   \path (D) edge (Y1);
   \path (U) edge (D);
   \path (U) edge (Y0);
   \path (U) edge (Y1);
   \path (V0) edge (Y0);
   \path (Y0) edge (D);
   \path (V1) edge (Y1);
   \path (Q) edge (D);
   \path (V0) edge (V1);
\end{tikzpicture} 
\end{subfigure}
\quad
\begin{subfigure}
\centering
\begin{tikzpicture}
  % x node set with absolute coordinates
  \node[state] (Y0) at (0,0) {$Y_0$};
  % y node set relative to x.
  % Locations can be:
  % right,left,above,below,
  % above left,below right, etc
  \node[state] (D) [right =of Y0] {$D$};
  \node[state] (Y1) [right =of D] {$Y_1-Y_0$};
  \node[state, dashed] (V0) [above =of Y0] {$V_0$};
  \node[state, dashed] (V1) [above =of Y1] {$V_1-V_0$};
  \node[state, dashed] (U)  [below =of D] {$U$};
    \node[state, dashed] (Q)  [above =of D,yshift=-0.45cm] {$Q$};
  %\node[dashed] (Q)  [above =of D] {$Q$};
   % Directed edge
   \path (D) edge (Y1);
   \path (U) edge (D);
   \path (U) edge (Y0);
   \path (U) edge (Y1);
   \path (V0) edge (Y0);
   \path (V1) edge (Y1);
   \path (V0) edge (V1);
   \path (Y0) edge (D);
    \path (Q) edge (D);
\end{tikzpicture} 
\end{subfigure}
\end{figure}
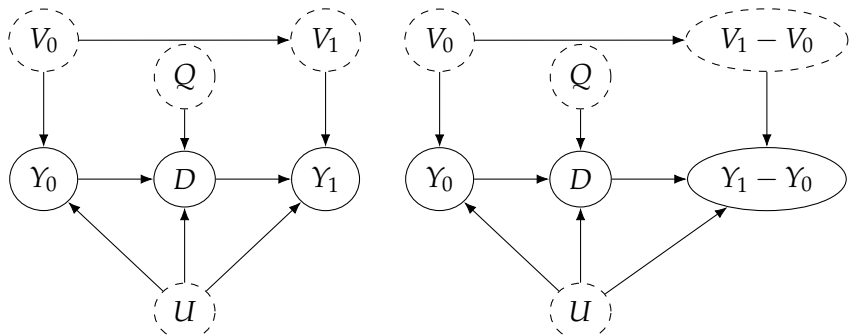

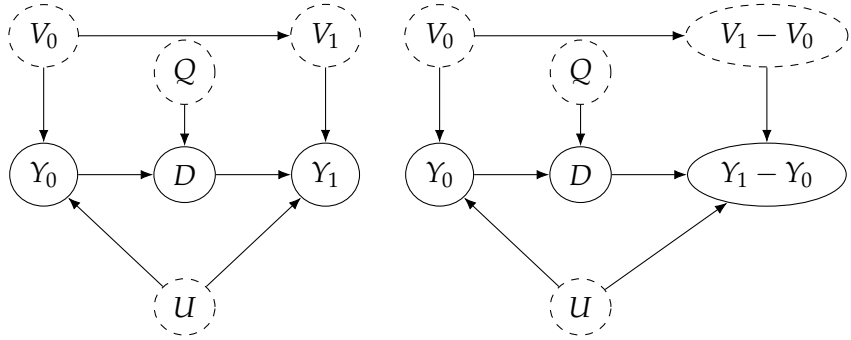
\begin{figure}[!htp]
	\centering \caption{\label{dag2}  Causal model satisfying unconfoundedness and violating common trends \bigskip}
\centering 
\begin{subfigure}
\centering
\begin{tikzpicture}
  % x node set with absolute coordinates
  \node[state] (Y0) at (0,0) {$Y_0$};
  % y node set relative to x.
  % Locations can be:
  % right,left,above,below,
  % above left,below right, etc
  \node[state] (D) [right =of Y0] {$D$};
  \node[state] (Y1) [right =of D] {$Y_1$};
  \node[state, dashed] (V0) [above =of Y0] {$V_0$};
  \node[state, dashed] (V1) [above =of Y1] {$V_1$};
  \node[state, dashed] (U)  [below =of D] {$U$};
  \node[state, dashed] (Q)  [above =of D, yshift=-0.45cm] {$Q$};
  %\node[dashed] (Q)  [above =of D] {$Q$};
   % Directed edge
   \path (D) edge (Y1);
   \path (U) edge (Y0);
   \path (U) edge (Y1);
   \path (V0) edge (Y0);
   \path (Y0) edge (D);
   \path (V1) edge (Y1);
   \path (Q) edge (D);
    \path (V0) edge (V1);
\end{tikzpicture} 
\end{subfigure}
\quad
\begin{subfigure}
\centering
\begin{tikzpicture}
  % x node set with absolute coordinates
  \node[state] (Y0) at (0,0) {$Y_0$};
  % y node set relative to x.
  % Locations can be:
  % right,left,above,below,
  % above left,below right, etc
  \node[state] (D) [right =of Y0] {$D$};
  \node[state] (Y1) [right =of D] {$Y_1-Y_0$};
  \node[state, dashed] (V0) [above =of Y0] {$V_0$};
  \node[state, dashed] (V1) [above =of Y1] {$V_1-V_0$};
  \node[state, dashed] (U)  [below =of D] {$U$};
    \node[state, dashed] (Q)  [above =of D,yshift=-0.45cm] {$Q$};
  %\node[dashed] (Q)  [above =of D] {$Q$};
   % Directed edge
   \path (D) edge (Y1);
   \path (U) edge (Y0);
   \path (U) edge (Y1);
   \path (V0) edge (Y0);
   \path (V1) edge (Y1);
   \path (V0) edge (V1);
   \path (Y0) edge (D);
    \path (Q) edge (D);
\end{tikzpicture} 
\end{subfigure}
\end{figure}

Figures \ref{dag1} to \ref{dag4} provide a causal models that help understanding the implications of the unconfoundedness and common trend assumptions based on directed acyclic graphs (DAG), see for instance \cite{Pearl00}. Solid nodes represent observed variables, dashed nodes unobserved variables, and arrows average causal effects between variables. The subfigures on the left represent causal models in which the post-treatment outcome $Y_1$ is considered as outcome variable, and subfigures on the right correspond to causal models in which the difference $Y_1-Y_0$ is considered as outcome, like in DiD approaches. Considering the left graph of Figure \ref{dag1}, we see that treatment $D$ affects the $Y_1$, which is the causal effect of interest, and is itself affected by the pre-treatment outcome $Y_0$. In addition, we have two time varying unobservables $V_0$ and $V_1$ that affect the outcomes $Y_0$ and $Y_1$, respectively, and the pre-treatment unobservable $V_0$ may also affect the post-treatment unobservable $V_1$. Furthermore, we assume an independent unobservable $Q$ that affects the treatment $D$. Finally, the unobserved time constant variable $U$ jointly affects the pre-treatment outcome $Y_0$, the treatment $D$, and the post-treatment outcome $Y_1$. In the right graph, the outcome change over time, $Y_1-Y_0$, is affected by the change in time varying unobservables, $V_1-V_0$, rather than $V_1$. $V_1-V_0$ is affected by $V_0$, as the difference between two variables in general may depend on both variables, which motivates the causal arrow from $V_0$ to $V_1-V_0$. Only under the specific serial dependence $E[V_1|V_0]=V_0$ would $V_0$ be uncorrelated with $V_1-V_0$, because $E[V_1-V_0|V_0]=E[V_1|V_0]-V_0=V_0-V_0=0$ would hold in this special case of a random walk. We note that we could additionally include observed covariates $X$ in the causal graphs that might jointly affect the treatment and the outcomes, which is omitted for the sake of simplicity. When one controls for such covariates $X$, the graphs can be interpreted as causal associations conditional on $X$.

In the causal framework given in Figure \ref{dag1}, neither unconfoundedness, nor common trends hold. Unconfoundedness fails because conditional on  $Y_0$ and possibly $X$, $U$ affects both $Y_1$ and $D$ in the left graph. Common trends fail for two reasons. First, the time constant unobservable $U$ affects both $D$  and the outcome change $Y_1-Y_0$ (implying heterogenous time trends in $U$). Second, the time varying unobservable $V_0$ affects both $D$ (via $Y_0$) and $Y_1-Y_0$ (via $V_1-V_0$), unless $E[V_1|V_0]=V_0$ holds, as also discussed in  \cite{ghanem2022selection}. 
Both assumptions impose different, non-nested conditions, see also the discussion in \cite{ChabeFerret2017}. Unconfoundedness invokes that conditional on $Y_0$ (and $X$), no unobserved time constant characteristics like $U$ or time varying characteristics like $V_0$ jointly affect $D$ and $Y_1$. This is satisfied in the causal model presented in Figure \ref{dag2}, where $U$ only affects the outcomes $Y_0$ and $Y_1$, but not directly $D$, such that $D$ is not affected by $U$ conditional on $Y_0$. Likewise, $V_0$, which affects $Y_1$ via $V_1$, does not affect $D$ conditional on $Y_0$. However, the common trend assumption fails because of $U$ jointly affecting $D$ (via $Y_0$) and $Y_1-Y_0$, as well as $V_0$ jointly affecting $D$ (via $Y_0$) and $Y_1-Y_0$ (via $V_1-V_0$). In the causal model in Figure \ref{dag3}, it is the common trend assumption that holds. Here, taking the outcome difference $Y_1-Y_0$ 
removes the impact of $U$ if the average effect of $U$ on the outcome does not interact with (or is homogeneous in) the time trend, which holds if $U$ is additively separable in the outcome model. Furthermore, the pre-treatment outcome $Y_0$ does not affect $D$ such that $V_0$ does not jointly affect $Y_1-Y_0$ and $D$. In contrast, unconfoundedness fails as $U$ directly affects $D$ and $Y_1$.  Finally, Figure \ref{dag4} provides a causal model in which both the unconfoundedness and common trends assumptions hold, such that identification based on Assumption \ref{A1} and \ref{A2} both yield the true ATET. Unconfoundedness holds because $U$, which influences $Y_1$, and $V_0$, which also influences $Y_1$ via $V_1$ both affect $D$ only via $Y_0$. Common trends hold because $U$ does not affect the outcome change  $Y_1-Y_0$, while  $V_0$ is not associated with $Y_1-Y_0$ via $V_1-V_0$, implying a specific serial correlation of $V_1$ and $V_0$,  $E[V_1|V_0]=V_0$.

\begin{figure}[!htp]
	\centering \caption{\label{dag3}  Causal model violating unconfoundedness and satisfying common trends \bigskip}
\centering 
\begin{subfigure}
\centering
\begin{tikzpicture}
  % x node set with absolute coordinates
  \node[state] (Y0) at (0,0) {$Y_0$};
  % y node set relative to x.
  % Locations can be:
  % right,left,above,below,
  % above left,below right, etc
  \node[state] (D) [right =of Y0] {$D$};
  \node[state] (Y1) [right =of D] {$Y_1$};
  \node[state, dashed] (V0) [above =of Y0] {$V_0$};
  \node[state, dashed] (V1) [above =of Y1] {$V_1$};
  \node[state, dashed] (U)  [below =of D] {$U$};
  \node[state, dashed] (Q)  [above =of D, yshift=-0.45cm] {$Q$};
  %\node[dashed] (Q)  [above =of D] {$Q$};
   % Directed edge
   \path (D) edge (Y1);
   \path (U) edge (D);
   \path (U) edge (Y0);
   \path (U) edge (Y1);
   \path (V0) edge (Y0);
   \path (V1) edge (Y1);
   \path (Q) edge (D);
  \path (V0) edge (V1);
\end{tikzpicture} 
\end{subfigure}
\quad
\begin{subfigure}
\centering
\begin{tikzpicture}
  % x node set with absolute coordinates
  \node[state] (Y0) at (0,0) {$Y_0$};
  % y node set relative to x.
  % Locations can be:
  % right,left,above,below,
  % above left,below right, etc
  \node[state] (D) [right =of Y0] {$D$};
  \node[state] (Y1) [right =of D] {$Y_1-Y_0$};
  \node[state, dashed] (V0) [above =of Y0] {$V_0$};
  \node[state, dashed] (V1) [above =of Y1] {$V_1-V_0$};
  \node[state, dashed] (U)  [below =of D] {$U$};
  \node[state, dashed] (Q)  [above =of D,yshift=-0.45cm] {$Q$};
  %\node[dashed] (Q)  [above =of D] {$Q$};
   % Directed edge
   \path (D) edge (Y1);
   \path (U) edge (D);
   \path (U) edge (Y0);
   \path (V0) edge (Y0);
   \path (V1) edge (Y1);
   \path (V0) edge (V1);
    \path (Q) edge (D);
\end{tikzpicture} 
\end{subfigure}

\end{figure}
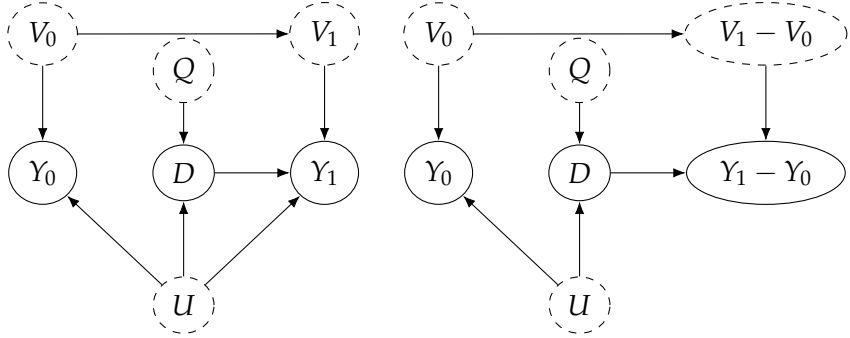
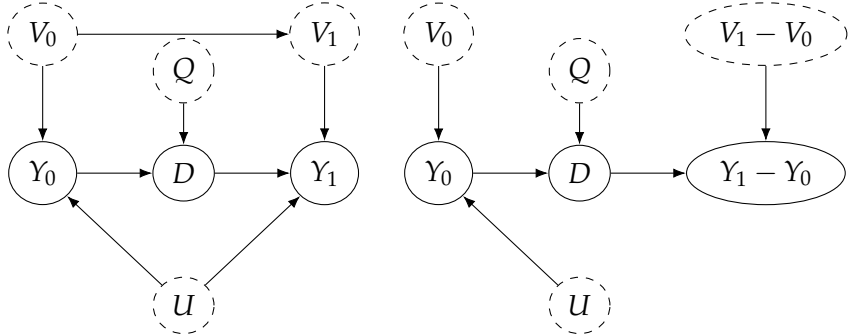
\begin{figure}[!htp]
	\centering \caption{\label{dag4}  Causal model satisfying unconfoundedness and common trends \bigskip}
\centering 
\begin{subfigure}
\centering
\begin{tikzpicture}
  % x node set with absolute coordinates
  \node[state] (Y0) at (0,0) {$Y_0$};
  % y node set relative to x.
  % Locations can be:
  % right,left,above,below,
  % above left,below right, etc
  \node[state] (D) [right =of Y0] {$D$};
  \node[state] (Y1) [right =of D] {$Y_1$};
  \node[state, dashed] (V0) [above =of Y0] {$V_0$};
  \node[state, dashed] (V1) [above =of Y1] {$V_1$};
  \node[state, dashed] (U)  [below =of D] {$U$};
   \node[state, dashed] (Q)  [above =of D, yshift=-0.45cm] {$Q$};
  %\node[dashed] (Q)  [above =of D] {$Q$};
   % Directed edge
   \path (D) edge (Y1);
    \path (U) edge (Y0);
   \path (U) edge (Y1);
   \path (V0) edge (Y0);
   \path (V1) edge (Y1);
   \path (Y0) edge (D);
   \path (Q) edge (D);
   \path (V0) edge (V1);
   \end{tikzpicture} 
\end{subfigure}
\quad
\begin{subfigure}
\centering
\begin{tikzpicture}
  % x node set with absolute coordinates
  \node[state] (Y0) at (0,0) {$Y_0$};
  % y node set relative to x.
  % Locations can be:
  % right,left,above,below,
  % above left,below right, etc
  \node[state] (D) [right =of Y0] {$D$};
  \node[state] (Y1) [right =of D] {$Y_1-Y_0$};
  \node[state, dashed] (V0) [above =of Y0] {$V_0$};
  \node[state, dashed] (V1) [above =of Y1] {$V_1-V_0$};
  \node[state, dashed] (U)  [below =of D] {$U$};
    \node[state, dashed] (Q)  [above =of D, yshift=-0.45cm] {$Q$};
  %\node[dashed] (Q)  [above =of D] {$Q$};
   % Directed edge
   \path (D) edge (Y1);
   \path (U) edge (Y0);
   \path (V0) edge (Y0);
   \path (V1) edge (Y1);
   \path (Y0) edge (D);
      \path (Q) edge (D);
\end{tikzpicture} 
\end{subfigure}
\end{figure}

Comparing expressions  \eqref{identdid} and \eqref{identunconf} demonstrates that the identification results under common trends and unconfoundedness differ in terms of how the mean potential outcome under non-treatment, $E[Y(0)|D=1]$, is computed. We define the parameter $\theta$ as the difference between expressions  \eqref{identdid} and \eqref{identunconf}, which is the target parameter of our testing approach: 
\begin{align}\label{thetameans}
\theta&=E[E[Y_1|X,Y_0,D=0]|D=1]-\{E[Y_0|D=1]+E[E[Y_1-Y_0|X,D=0]|D=1]\}\notag\\
&= E\left[ \frac{D}{\Pr(D=1)} \cdot \{E[Y_1|X,Y_0,D=0]- Y_0 - E[Y_1-Y_0|X,D=0]\}\right].
\end{align}
Expression \eqref{thetameans} makes use of conditional mean outcomes for defining $\theta$. Alternatively, a numerically equivalent expression for $\theta$ is the following, so-called doubly robust (DR) statistic, which is based on both, conditional mean outcomes and conditional treatment probabilities:  
\begin{align}\label{thetaDR}
\theta&=E\left[ \frac{D}{\Pr(D=1)} \cdot [\mu(X,Y_0)-Y_0 - m(X)]\right] \notag\\
&+E\left[\frac{1-D}{\Pr(D=1)} \cdot  \left\{ \frac{ p(X,Y_0) \cdot (Y_1-\mu(X,Y_0))}{1-p(X,Y_0)} - \frac{ \pi(X) \cdot [Y_1-Y_0-m(X)]}{1-\pi(X)}\right\}\right],
\end{align} 
with $\mu(x,y_0)=E[Y|X=x,Y_0=y_0,D=0]$, $m(x)=E[Y_1-Y_0|X=x,D=0]$ denoting the conditional mean outcomes or differences in outcomes under non-treatment, respectively, and $p(x,y_0)=\Pr(D=1|X=x, Y_0=y_0)$, $\pi(x)=\Pr(D=1|X=x)$ denoting the conditional treatment probabilities, also known as propensity scores. \footnote{By the law of iterated expectations and basic probability theory,
\begin{align*}
&E\left[ \frac{1-D}{\Pr(D=1)} \cdot \frac{p(X,Y_0)\cdot (Y_1-\mu(X,Y_0))}{1-p(X,Y_0)}\right]=E\left[ \frac{p(X,Y_0)}{\Pr(D=1)} \cdot E\left[ \frac{(1-D)\cdot (Y_1-\mu(X,Y_0))}{1-p(X,Y_0)}\Big| X,Y_0\right]  \right]\notag\\
&=E[E [ Y_1-\mu(X,Y_0) | X,Y_0, D=0 ] | D=1 ]= E [ \mu(X,Y_0)-\mu(X,Y_0))| D=1 ]=0.
\end{align*} In an analogous manner,  
\begin{align*}
&E\left[ \frac{1-D}{\Pr(D=1)} \cdot \frac{\pi(X)\cdot [Y_1-Y_0-m(X)]}{1-\pi(X)} \right]=E [ m(X)-m(X) | D=1 ]=0.
\end{align*} 
For this reason, expressions \eqref{thetameans} and \eqref{thetaDR} are equivalent.} Expression \eqref{thetaDR} coincides with the difference in DR identification of the ATET based on DiD in panel data, see e.g.\ equation (2.6) of \cite{SantAnnaZhao2018}, and DR identification of the ATET based on unconfoundedness when controlling for covariates $X$ and pre-treatment outcome $Y_0$, see e.g. equation  (4.45) in \cite{huber2023causal}. This test-statistic is our main result. 

DR approaches bear the attractive property that estimation is consistent (under standard regularity conditions) if either the models for the conditional mean outcomes or the propensity scores, henceforth denoted as nuisance parameters are correctly specified, see e.g.\ \cite{RobinsMarkNewey1992},  \cite{Robins+94}, \cite{RoRoZa95}, and \cite{BaRo05}. This implies that estimation of $\theta$ based on expression \eqref{thetaDR} is consistent if either  $\mu(X,Y_0)$ or $p(X,Y_0)$ when imposing unconfoundedness, and either $m_T(X,0)$ or $\pi(X)$ when imposing common trends is correctly specified. Furthermore, DR estimators are quite robust to moderate misspecifications in all of the nuisance parameters, a property known as  \cite{Neyman1959}-orthogonality. The latter implies that the estimation of $\theta$ based on expression \eqref{thetaDR} is first order insensitive to approximation errors in the estimation of the propensity scores and conditional mean outcomes. This appears attractive in big data contexts with high-dimensional covariates $X$, as it permits estimating the nuisance parameters by machine learning, which generally introduces regularization bias, but nevertheless permits $\sqrt{n}$-consistent estimation of $\theta$ under specific regularity conditions outlined in the next section. 

\section{Estimation}\label{est}

This section proposes an estimator of $\theta$ as given in the DR expression \eqref{thetaDR}, based on double machine learning (DML) as suggested in \cite{Chetal2018}. Let $\mathcal{W} = \{W_i|1\leq i \leq n\}$ with $W_i = (Y_{1i}, Y_{0i}, X_{i}, D_{i})$ for all $ i $ denote the set of observations in an i.i.d.\ sample of size $n$. We define $\eta=\{p(X, Y_0), \pi(X), \mu (X,Y_0,D), m(X,D))\}$ as the vector of nuisance parameter functions, i.e.\ the propensity scores and the conditional means of the outcomes or outcome differences, respectively. Their respective estimates are denoted as $\hat{\eta} =\{\hat p(X, Y_0), \hat\pi(X), \hat\mu (X,Y_0,D), \hat m(X,D))\}$ and the respective true functions by $\eta_0=\{p_0(X, Y_0), \pi_0(X), \mu_0 (X,Y_0,D), m_0(X,D))\}$.

We estimate $\theta$ by the following algorithm that combines the estimation of Neyman-orthogonal scores with sample splitting or cross-fitting and is $\sqrt{n}$-consistent under conditions outlined further below. \vspace{8pt}\newline
\textbf{Algorithm 1: Estimation of $\theta$ based on expression \eqref{thetaDR} }\label{algo1}
\begin{enumerate}
	\item Split $\mathcal{W}$ in $ K $ subsamples. For each subsample $ k $, let $n_k$ denote its size, $\mathcal{W}_k$ the set of observations in the sample and $\mathcal{W}_k^{C}$ the complement set of all observations not in $k$.
	\item For each $k$, use  $\mathcal{W}_k^{C}$ to estimate the model parameters of $\eta$ in order to predict the nuisance parameters for each observation $j$ in $\mathcal{W}_k$, where the predictions are denoted by $\hat{\eta}_j^k =\{\hat p^k(X_j, Y_{0j}), \hat\pi^k(X_j), \hat\mu^k(X_j,Y_{0j},D_j), \hat m^k(X_j,D_j))\}$.
	\item Stack the estimated nuisance parameters $\hat{\eta}_j^k$ across all $j$ and $k$ and denote by $\hat{\eta}_i$ the nuisance parameter estimate of observation $i$ in the total sample (with $1\leq i \leq n$) after stacking. 
 \item Estimate expression \eqref{thetaDR} based on its normalized sample analogue:
\begin{align*}
\hat \theta &= \sum_{i=1}^n\left[\frac{ D_i}{\sum_{i=1}^n D_i} \cdot [\hat \mu(X_i,Y_{0i})-Y_{0i} - \hat m(X_i)] + \frac{1-D_i}{ \sum_{i=1}^n \frac{(1-D_i)\cdot \hat p(X_i,Y_{0i})}{1-\hat p(X_i,Y_{0i})}} \cdot  \left\{ \frac{ \hat p(X_i,Y_{0i}) \cdot [Y_{1i}-\hat\mu(X_i,Y_{0i})]}{1-\hat p(X_i,Y_{0i})}\right\}\right.\\
&-\left. \frac{1-D_i}{ \sum_{i=1}^n \frac{(1-D_i)\cdot \hat \pi(X_i)}{1-\hat \pi(X_i)}} \cdot \left\{\frac{ \hat \pi(X_i) \cdot [Y_{1i}-Y_{0i}-\hat m(X_i)]}{1-\hat\pi(X_i)}\right\}\right].
\end{align*}
\vspace{5pt}\newline
\end{enumerate}
To see that the normalized estimator in step 4 corresponds in expectation to expression \eqref{thetaDR} under consistent estimation of the nuisance parameters, we note that by the law of iterated expectations,
$E\left[\sum_{i=1}^n D_i\right]=E\left[\sum_{i=1}^n \frac{(1-D_i)\cdot \hat p(X_i,Y_{0i})}{1-\hat p(X_i,Y_{0i})}\right]=E\left[\sum_{i=1}^n \frac{(1-D_i)\cdot \hat \pi(X_i)}{1-\hat \pi(X_i)}\right]=n \cdot \Pr(D=1)$.

In order to obtain root-n consistency for the estimation of $\theta$, we impose the following assumption consisting of regularity conditions w.r.t.\ the prediction quality of machine learning for estimating the nuisance parameters. Following  \cite{Chetal2018}, we introduce some further notation: let $(\delta_n)_{n=1}^{\infty}$ and $(\Delta_n)_{n=1}^{\infty}$ denote sequences of positive constants with $\lim_{n\rightarrow \infty} \delta_n = 0 $ and $\lim_{n\rightarrow \infty} \Delta_n = 0.$ Furthermore, let $c, \epsilon, C$ and $q$ be positive constants such that $q>2,$ and let $K \geq 2$ be a fixed integer. Also, for any random vector $R = (R_1,...,R_l)$, let $\left\| R \right\|_{q} = \max_{1\leq j \leq l}\left\| R_l \right\|_{q},$ where
$\left \| R_l \right\|_{q}  =  \left( E\left[ \left| R_l \right|^q \right] \right)^{\frac{1}{q}}$.
In order to ease notation, we assume that $n/K$ is an integer. For the sake of brevity we omit the dependence of probability $\Pr_P,$ expectation $E_P(\cdot),$ and norm $\left\| \cdot  \right\|_{P,q}$ on the probability measure $P.$
\begin{assumption}[Regularity conditions]\label{A3}
For all probability laws $P \in \mathcal{P},$ where $\mathcal{P}$ is the set of all possible probability laws the following conditions hold for the random vector $(Y_1,Y_0,D,X)$ for positive constants $C$, $c$, and $\epsilon$:
\begin{enumerate}
\item[(a)] $  \left\| Y_1 \right\|_{q} \leq C$, $\left\| Y_1-Y_0 \right\|_{q} \leq C$,
$\left\|E[Y_1^2| X, D=0 ] \right\|_{\infty} \leq C^2$, $\left\|E[(Y_1-Y_0)^2| X, D=0 ] \right\|_{\infty} \leq C^2$,
\item[(b)] $\Pr(\epsilon \leq p_0(X, Y_0) \leq 1-\epsilon) = 1$, $\Pr(\epsilon \leq \pi_0(X) \leq 1-\epsilon) = 1$,
\item[(c)] $\left\| Y_1-\mu_0(X,Y_0)  \right\|_{2} = E_{ } \Big[\left(Y_1-\mu_0(X,Y_0) \right)^2 \Big]^{\frac{1}{2}} \geq c$,\\$\left\| Y_1-Y_0-m_0(X) \right\|_{2} = E_{ } \Big[\left(Y_1-Y_0-m_0(X) \right)^2 \Big]^{\frac{1}{2}} \geq c$,
\item[(d)] Given a random subset $I$ of $[n]$ of size $n_k=n/K,$ the nuisance parameter estimator $\hat \eta = \hat \eta((W_i)_{i \in I^C})$ satisfies the following conditions. With $P$-probability no less than $1-\Delta_n:$
\begin{eqnarray}
\left\|  \hat \eta - \eta_0 \right\|_{q} &\leq& C,  \notag \\
\left\|  \hat \eta - \eta_0 \right\|_{2} &\leq& \delta_n, \notag \\
\left\|  \hat  p(X, Y_0)-1/2\right\|_{\infty}  &\leq& 1/2-\epsilon, \notag\\
\left\|  \hat  \pi(X)-1/2)\right\|_{\infty} &\leq & 1/2-\epsilon, \notag \\
\left\|  \hat  \mu(X,Y_0,D)-\mu_0(X,Y_0,D)\right\|_{2} \times \left\| \hat   p(X, Y_0)-p_0(X, Y_0)\right\|_{2}  &\leq & \delta^{}_n n^{-1/2}, \notag \\
\left\|  \hat  m(X,D)-m_0(X,D)\right\|_{2} \times \left\| \hat   \pi(X)-\pi_{0}(X)\right\|_{2} &\leq & \delta^{}_n n^{-1/2}.\notag
\end{eqnarray}
\end{enumerate}
\end{assumption}
Condition (a) in Assumption \ref{A3} states that the distribution of outcomes or differences in outcomes does not have unbounded moments. Condition (b) refines the common support condition such that the treatment propensity scores are bounded away from $0$ and $1$. Condition (c) requires that the covariates $X$ do not perfectly predict the conditional means of the outcomes or differences in outcomes, respectively. Condition (d) is the only non-primitive condition and puts restrictions on the quality of the nuisance parameter estimators. We note that for the satisfaction of the last two restrictions of condition (d), $\left\|  \hat  \mu(X,Y_0,D)-\mu_0(X,Y_0,D)\right\|_{2} \times \left\| \hat   p(X, Y_0)-p_0(X, Y_0)\right\|_{2}  \leq  \delta^{}_n n^{-1/2}$ and
$\left\|  \hat  m(X,D)-m_0(X,D)\right\|_{2} \times \left\| \hat   \pi(X)-\pi_{0}(X)\right\|_{2} \leq  \delta^{}_n n^{-1/2}$, it suffices that the  nuisance estimators converge with rate $ o(n^{-1/4})$, which is attainable by many commonly used machine learners under specific conditions (like approximate sparsity), such as lasso, random forests, boosting and neural nets, see for instance \cite{Bellonietal2014}, \cite{huber2023testing}, \cite{WagerAthey2018}, and \cite{FarrellLiangMisra2018}. 
% kueck2023estimation --> huber2023testing und  farrell2021deep -->FarrellLiangMisra2018

Theorem \ref{theorem1} formally states the asymptotic  normality and  root-n consistency of the estimator $\hat \theta$. The proof is based on demonstrating that our estimation approach satisfies the conditions of  DML estimation given in \cite{Chetal2018}, namely their Assumption 3.1 on the linearity and Neyman-orthogonality of the DR expression underlying the identification of $\theta$ and Assumption 3.2 on regularity conditions and the quality of nuisance parameter estimators, related to Assumption \ref{A3}.  
\begin{theorem}[Asymptotic normality]\label{theorem1}
Under  Assumptions \ref{A1}, \ref{A2}, and \ref{A3}, it holds for estimating $\theta$ based on Algorithm 1:  \\
$\sqrt{n} \Big(\hat \theta -  \theta \Big) \rightarrow N(0,\sigma^2$),\quad where\\$\sigma^2= E\left[\left(\frac{D}{\Pr(D=1)} \cdot [\mu(X,Y_0)-Y_0 - m(X)] +\frac{1-D}{\Pr(D=1)} \cdot  \left\{ \frac{ p(X,Y_0) \cdot (Y_1-\mu(X,Y_0))}{1-p(X,Y_0)} - \frac{ \pi(X) \cdot [Y_1-Y_0-m(X)]}{1-\pi(X)}\right\}-\theta\right)^2\right]$. 	
\end{theorem}
\noindent
As DR expression \eqref{thetaDR} corresponds to the difference in DR expressions for the ATET under Assumptions \ref{A2} (common trends) and \ref{A1} (unconfoundedness), respectively, Theorem \ref{theorem1} follows from combining Theorem 3.1 in \cite{Chang2020} and Theorem 5.1 in \cite{Chetal2018} on ATET estimation by cross-fitting under common trends and unconfoundedness, respectively. Appendix \ref{Neyman1} provides a proof of the linearity and Neyman orthogonality of our DR expression, i.e., Assumption 3.1 in \cite{Chetal2018}, and refers to \cite{Chang2020} and \cite{Chetal2018} for proofs of Assumption 3.2 in \cite{Chetal2018}.

\section{Simulation}\label{sim}
This section presents a simulation study designed to examine the finite sample properties of our proposed testing approach as outlined in Theorem \ref{theorem1}. The underlying data generating process (DGP) models pre-treatment and post-treatment outcomes $Y_0$ and $Y_1$, along with the treatment assignment $D$, as follows:

	\begin{eqnarray*}\label{GDPsim}
		Y_0 &=& U + V_0,\quad D = I\{X'\beta_{X}+\gamma  U + \delta Y_0 + Q >0\},\\
		Y_1 &=&  D + X'\beta_{X}  +   U  + V_1, \\
		X &\sim& N(0,\sigma^2_X), V_0\sim N(0,1),  V_1\sim N(0,1), U\sim N(0,1),  Q\sim N(0,1), \\ 
		&&\textrm{ with }X,V_0,V_1,U,Q\textrm{ being independent of each other}.
	\end{eqnarray*}

In this setup, the pre-treatment outcome $Y_0$ is a linear function of the fixed effect $U$ and the period-specific unobservable $V_0$. The binary treatment $D$ is a function of covariates $X$ for the coefficient vector $\beta_X \neq 0$, the fixed effect $U$ if $\gamma\neq 0$, the pre-treatment outcome $Y_0$ if $\delta\neq 0$, and further unobservables $Q$. The post-treatment outcome $Y_1$ is a linear function of treatment $D$, whose effect is 1, of the fixed effect $U$, covariates $X$ if $\beta_X \neq 0$, and the period-specific unobservable $V_1$. We note that for $\gamma \neq 0$, the unconfoundedness assumption is violated, due to the confounding of $D$ and $Y_1$ by $U$. For $\delta \neq 0$, the common trend assumption is violated. To see this, note that if $Y_0$, which is a function of the period-specific error $V_0$, affects $D$, then $D$ also becomes a function of $V_0$. Furthermore, the difference in outcomes over time, $Y_1 - Y_0$, is a function of the difference in the period-specific errors, $V_1 - V_0$, and thus, of $V_0$. For this reason, $V_0$ confounds $D$ and $Y_1 - Y_0$ if $Y_0$ affects $D$, such that common trends fail in our framework with independently drawn errors $V_1$ and $V_0$. \footnote{Only in the special case of $E(V_1 - V_0) = 0$ would common trends hold, but this would imply a strong serial correlation between $V_1$ and $V_0$, characterized by a random walk.}

The covariate vector $X$ follows a multivariate normal distribution with a zero mean and a covariance matrix $\sigma^2_X$, where the covariance between the $i$th and $j$th covariate is set to $0.5^{|i-j|}$. The variables $V_0$, $V_1$, $U$, and $Q$ each follow standard normal distributions and are independent of each other, as well as of $X$. The vector $\beta_X$ gauges the effects of the covariates on the treatment assignment $D$, on the post-treatment outcome $Y_1$, and the outcome trend $Y_1-Y_0$ alike, which is true, because $X$ does not influence $Y_0$. Thus, $\beta_X$ precisely captures confounding due to observables, yet it does not capture potential unobserved effects introduced by $U$ and $Q$. The $i$th element in the coefficient vector $\beta_X$ is set to $0.6/i$ for $i=1, \dots, p$. This implies a linear decay of covariate importance in terms of confounding.

We define three distinct scenarios to evaluate the performance of our approach, which mirror the causal models depicted in Figures \ref{dag2}-\ref{dag4}: 
\begin{itemize}
    \item Scenario 1: $\gamma = 0, \; \delta = 0$ – unconfoundedness \ref{A1} and common trends \ref{A2} hold.
    \item Scenario 2: $\gamma = 0.25, \; \delta = 0$ – unconfoundedness \ref{A1} is violated and common trends \ref{A2} hold.
    \item Scenario 3: $\gamma = 0, \; \delta = 0.25$ – unconfoundedness \ref{A1} holds and common trends \ref{A2} are violated.
\end{itemize}

To assess convergence rates, we vary the sample sizes, using either $n = 1000$ or $n = 4000$ and keep the remaining parameters constant. We maintain 1000 iterations and $p = 100$ covariates. The test statistic $\hat{\theta}$ is computed using Algorithm \ref{algo1}. Under the null hypothesis both selection on observables assumptions hold. Thus, a significant deviation of $\hat{\theta}$ from zero indicates a violation of at least one assumption. We reject the null if the p-value is smaller than $0.05$.

% 2*Referenz einfuegen!
We employ various models for training the nuisance parameters. Our parametric approach utilizes ordinary least squares (OLS) for estimating outcomes $\hat{\mu}$ and $\hat{m}$, and logistic regression for computing the propensity scores $\hat{\pi}$ and $\hat{p}$. Additionally, we explore four ML models: Lasso, Random Forest (RF), Support Vector Machine (SVM), and the Super Learner, which determines the optimal weighted average across an ensemble of Lasso, RF, and SVM. This approach facilitates the selection of the best model for the nuisance parameters without the need for manual performance comparisons and selection of the best model \citep{VanderLaanHubbard2007}. To prevent extreme weights, we trim on the propensity score, and exclude observations where $\hat{\pi}(X_i)$ and $\hat{p}(X_i, Y_{0i})$ are greater than or equal to 0.99. The nuisance parameters are estimated through cross-fitting to ensure honest estimates, following the methodology proposed by \cite{AtheyImbens2016}.

\begin{figure}
    \caption{Convergence of simulation results}
    
    \begin{minipage}[t]{0.32\textwidth}
        \centering
        \begin{tikzpicture}
            \node[anchor=south west,inner sep=0] (image) at (0,0) {\includegraphics[width=\textwidth]{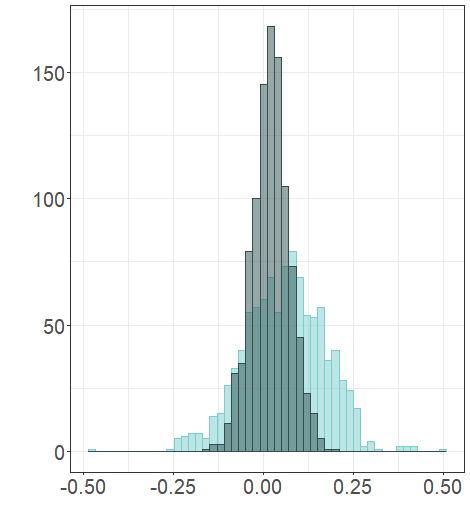}};
            \begin{scope}[x={(image.south east)},y={(image.north west)}]
                % Add annotation
                \node[black, font=\small] at (0.575,0) {$\hat{\theta}$};
                % Add "Count" annotation
                \node[black, font=\small, rotate=90] at (0,0.5) {Count};
            \end{scope}
        \end{tikzpicture}
        \caption*{(a) Scenario 1: $\gamma = 0$ and $\delta = 0$}
        \label{subfig1}
    \end{minipage}
    \hfill
    \begin{minipage}[t]{0.32\textwidth}
        \centering
        \begin{tikzpicture}
            \node[anchor=south west,inner sep=0] (image) at (0,0) {\includegraphics[width=\textwidth]{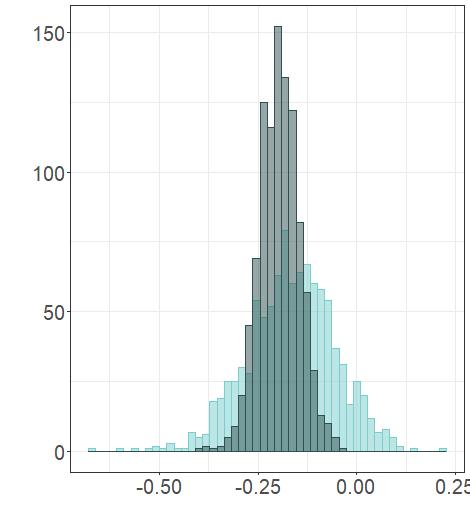}};
            \begin{scope}[x={(image.south east)},y={(image.north west)}]
                % Add annotation
                \node[black, font=\small] at (0.575,0) {$\hat{\theta}$};
                % Add "Count" annotation
                \node[black, font=\small, rotate=90] at (0,0.5) {Count};
            \end{scope}
        \end{tikzpicture}
        \caption*{(b) Scenario 2: $\gamma = 0.25$ and $\delta = 0$}
        \label{subfig2}
    \end{minipage}
    \hfill
    \begin{minipage}[t]{0.32\textwidth}
        \centering
        \begin{tikzpicture}
            \node[anchor=south west,inner sep=0] (image) at (0,0) {\includegraphics[width=\textwidth]{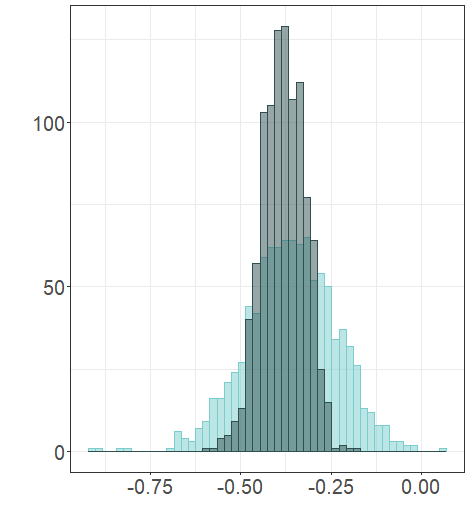}};
            \begin{scope}[x={(image.south east)},y={(image.north west)}]
                % Add annotation
                \node[black, font=\small] at (0.575,0) {$\hat{\theta}$};
                % Add "Count" annotation
                \node[black, font=\small, rotate=90] at (0,0.5) {Count};
            \end{scope}
        \end{tikzpicture}
        \caption*{(c) Scenario 3: $\gamma = 0$ and $\delta = 0.25$}
        \label{subfig3}
    \end{minipage}

    \label{fig:simplot}
    \begin{notes}
        The Figure displays the estimated test-statistic $\hat{\theta}$ over 1000 iterations using either $n = 1000$ observations, displayed in blue, or $n = 4000$ observations, displayed in gray. In sub-figure (a) the unconfoundedness and conditional common trends assumptions hold, in sub-figure (b) the unconfoundedness assumption is violated, and in sub-figure (c) the conditional common trends assumption is violated.
    \end{notes}
\end{figure}

% Describe Figure -> finite sample distribution.
We first analyze the finite sample distribution of $\hat{\theta}$. Figure \ref{fig:simplot} presents a histogram with the distribution of $\hat{\theta}$ for the three distinct simulation scenarios determined by the parameters $\delta$ and $\gamma$. These scenarios are visualized with $n = 1000$ observations in blue and $n = 4000$ observations in gray. Nuisance parameters for this analysis are estimated by the ensemble learner. The Figure effectively demonstrates the asymptotic properties of $\hat{\theta}$, showing that the estimator converges to a normal distribution, with decreasing varaince as N increases. Specifically, Panel (a) indicates that $\hat{\theta}$ aligns with an expectation of zero when both unconfoundedness and conditional common trends are met. Conversely, Panels (b) and (c) reveal that the expectation of $\hat{\theta}$ deviates from zero, when either assumption is compromised.

Next, we conduct an in-depth analysis of the test's performance. Table \ref{tabsim} provides a comprehensive summary of the simulation results across the three specifications. It details the most important statistics of the test, under `Test results', and the two doubly robust approaches to estimate ATET, `Unconf' and `DiD'. The Table reports the test-statistic $\hat{\theta}$ along with the standard deviation, standard errors, and p-values, which equal the mean values across iterations within the same specification, and the corresponding rejection rates. Instead of reporting the point estimates for $\Delta_{Unconf}$ and $\Delta_{DiD}$, we present the bias, which can be interpreted as percentage deviation from $\Delta_{D=1} = 1$. Similar to the test results, we report the bias, RMSE, and standard errors as mean values across all iterations within the same parameter specification. Additionally, we include the empirical standard deviation as a robustness measure for the standard errors, which is expected to stabilize as the number of iterations increases.

In the first scenario, unconfoundedness and common trends are satisfied. The results are reported in the upper panel of Table \ref{tabsim}. The average of the DR-based difference $\hat{\theta}$, the test-statistic, is close to zero for either sample sizes. We find that the estimator seems to converge at $\sqrt{n}$-rate, as the standard deviation roughly reduces by half when quadrupling the sample size from 1000 to 4000. Moreover, the standard deviation is similar to the standard error, at least for the larger sample size. The average p-values are all larger than $0.3$ and the test rejects the null in 5-15$\%$ of the cases if $n = 1000$ and and 6-12$\%$ if $n = 4000$, respectively. This implies a decent empirical size
for $n = 4000$. The bias of $\Delta_{Unconf}$ and $\Delta_{\text{DiD}}$ vanishes with increasing sample size and the empirical standard deviation resembles the mean standard errors.

In the second scenario, we set $\gamma = 0.25$ such that treatment assignment $D$ and post-treatment outcome $Y_1$ are confounded by the fixed effect $U$. The results are reported in the middle panel of Table \ref{tabsim}. Here, $\hat{\theta}$ is moderately different from zero for both sample sizes. The average estimate for $\hat{\theta}$ now amounts to roughly $0.2$ to $0.3$, across both sample sizes. In the small sample case, the rejection rate varies from 33-65$\%$. However, it increases to 92-100$\%$ in the large n case. Thus, in this specification the precision gain with larger sample sizes is more pronounced, due to the smaller standard errors. For this reason, also the average p-values are close to zero, if $n=4000$. The bias in $\Delta_{Unconf}$ remains persistent, while the bias in $\Delta_{DiD}$ is smaller with increasing sample size. 

% Table with Simulationresults
\begin{landscape}
\begin{table}[htbp]
  \centering
  \caption{Simulation results}
  \label{tabsim}

   \footnotesize
  %\small
  % Table with simulation results
  % latex table generated in R 4.3.1 by xtable 1.8-4 package
% Wed May 29 12:24:34 2024
\begin{tabularx}{\linewidth}{cccl*{13}{Y}}
   \toprule
  \toprule
 \multicolumn{3}{c}{\textit{Sim.-par.}} & & \multicolumn{5}{c}{\textit{Test results}} & \multicolumn{4}{c}{\textit{Unconf}} & \multicolumn{4}{c}{\textit{DiD}} \\
 \cmidrule(lr){1-3} \cmidrule(lr){5-9} \cmidrule(lr){10-13} \cmidrule(lr){14-17}
  $\gamma$ & $\delta$ & $N$ & Learner & $\hat{\theta}$ & Std. & SE & p-val & Reject & Bias & RMSE & SE & E.std  & Bias & RMSE & SE. & E.std \\ \midrule
 \multirow{10}{*}{0.00} &  \multirow{10}{*}{0.00} &  \multirow{5}{*}{1,000} & Ensemble & 0.043 & 0.105 & 0.086 & 0.392 & 0.148 & 0.166 & 0.221 & 0.135 & 0.146 & 0.209 & 0.269 & 0.154 & 0.170 \\ 
   &  &  & Lasso & 0.021 & 0.099 & 0.083 & 0.425 & 0.107 & 0.132 & 0.193 & 0.134 & 0.140 & 0.153 & 0.224 & 0.152 & 0.163 \\ 
   &  &  & Parametric & 0.002 & 0.178 & 0.156 & 0.451 & 0.060 & -0.006 & 0.295 & 0.264 & 0.295 & -0.004 & 0.340 & 0.300 & 0.340 \\ 
   &  &  & RF & -0.023 & 0.109 & 0.095 & 0.452 & 0.090 & 0.694 & 0.706 & 0.135 & 0.128 & 0.672 & 0.685 & 0.140 & 0.134 \\ 
   &  &  & SVM & -0.005 & 0.158 & 0.144 & 0.473 & 0.052 & 0.337 & 0.391 & 0.199 & 0.199 & 0.332 & 0.393 & 0.211 & 0.211 \\ 
   \cmidrule(lr){3-3} \cmidrule(lr){4-9} \cmidrule(lr){10-13} \cmidrule(lr){14-17} &  &  \multirow{5}{*}{4,000} & Ensemble & 0.020 & 0.052 & 0.039 & 0.390 & 0.157 & 0.044 & 0.089 & 0.067 & 0.078 & 0.064 & 0.112 & 0.076 & 0.091 \\ 
   &  &  & Lasso & 0.007 & 0.044 & 0.040 & 0.480 & 0.075 & 0.032 & 0.078 & 0.069 & 0.071 & 0.038 & 0.091 & 0.079 & 0.083 \\ 
   &  &  & Parametric & -0.001 & 0.052 & 0.050 & 0.486 & 0.064 & -0.005 & 0.090 & 0.087 & 0.090 & -0.006 & 0.105 & 0.100 & 0.104 \\ 
   &  &  & RF & -0.021 & 0.052 & 0.046 & 0.434 & 0.120 & 0.533 & 0.537 & 0.067 & 0.063 & 0.511 & 0.516 & 0.070 & 0.068 \\ 
   &  &  & SVM & 0.004 & 0.063 & 0.060 & 0.488 & 0.061 & 0.151 & 0.175 & 0.088 & 0.089 & 0.155 & 0.184 & 0.098 & 0.099 \\ 
   \cmidrule(lr){1-3} \cmidrule(lr){4-9} \cmidrule(lr){10-13} \cmidrule(lr){14-17} \multirow{10}{*}{0.25} &  \multirow{10}{*}{0.00} &  \multirow{5}{*}{1,000} & Ensemble & -0.182 & 0.106 & 0.092 & 0.154 & 0.528 & 0.387 & 0.413 & 0.136 & 0.145 & 0.205 & 0.263 & 0.150 & 0.164 \\ 
   &  &  & Lasso & -0.203 & 0.098 & 0.088 & 0.104 & 0.654 & 0.352 & 0.378 & 0.135 & 0.138 & 0.149 & 0.217 & 0.149 & 0.158 \\ 
   &  &  & Parametric & -0.249 & 0.198 & 0.177 & 0.260 & 0.325 & 0.213 & 0.362 & 0.267 & 0.293 & -0.036 & 0.337 & 0.295 & 0.336 \\ 
   &  &  & RF & -0.307 & 0.108 & 0.097 & 0.033 & 0.877 & 0.966 & 0.974 & 0.137 & 0.122 & 0.659 & 0.672 & 0.139 & 0.134 \\ 
   &  &  & SVM & -0.247 & 0.158 & 0.154 & 0.220 & 0.389 & 0.583 & 0.614 & 0.200 & 0.195 & 0.336 & 0.395 & 0.203 & 0.207 \\ 
   \cmidrule(lr){3-3} \cmidrule(lr){4-9} \cmidrule(lr){10-13} \cmidrule(lr){14-17} &  &  \multirow{5}{*}{4,000} & Ensemble & -0.196 & 0.051 & 0.042 & 0.003 & 0.986 & 0.262 & 0.273 & 0.067 & 0.078 & 0.066 & 0.111 & 0.075 & 0.089 \\ 
   &  &  & Lasso & -0.213 & 0.045 & 0.043 & 0.001 & 0.992 & 0.253 & 0.262 & 0.069 & 0.069 & 0.040 & 0.089 & 0.077 & 0.079 \\ 
   &  &  & Parametric & -0.220 & 0.056 & 0.053 & 0.005 & 0.975 & 0.218 & 0.235 & 0.086 & 0.087 & -0.002 & 0.099 & 0.096 & 0.099 \\ 
   &  &  & RF & -0.304 & 0.053 & 0.047 & 0.000 & 1.000 & 0.803 & 0.805 & 0.068 & 0.062 & 0.499 & 0.503 & 0.069 & 0.066 \\ 
   &  &  & SVM & -0.228 & 0.064 & 0.065 & 0.017 & 0.926 & 0.384 & 0.394 & 0.088 & 0.087 & 0.156 & 0.182 & 0.094 & 0.095 \\ 
   \cmidrule(lr){1-3} \cmidrule(lr){4-9} \cmidrule(lr){10-13} \cmidrule(lr){14-17} \multirow{10}{*}{0.00} &  \multirow{10}{*}{0.25} &  \multirow{5}{*}{1,000} & Ensemble & -0.376 & 0.117 & 0.104 & 0.020 & 0.927 & 0.185 & 0.240 & 0.142 & 0.152 & -0.190 & 0.251 & 0.145 & 0.164 \\ 
   &  &  & Lasso & -0.405 & 0.106 & 0.100 & 0.012 & 0.950 & 0.149 & 0.210 & 0.139 & 0.147 & -0.256 & 0.301 & 0.144 & 0.159 \\ 
   &  &  & Parametric & -0.445 & 0.246 & 0.221 & 0.145 & 0.559 & 0.003 & 0.285 & 0.270 & 0.285 & -0.441 & 0.542 & 0.289 & 0.315 \\ 
   &  &  & RF & -0.592 & 0.104 & 0.099 & 0.000 & 1.000 & 0.899 & 0.907 & 0.138 & 0.122 & 0.307 & 0.336 & 0.137 & 0.136 \\ 
   &  &  & SVM & -0.473 & 0.168 & 0.174 & 0.058 & 0.793 & 0.435 & 0.481 & 0.217 & 0.204 & -0.037 & 0.209 & 0.203 & 0.205 \\ 
   \cmidrule(lr){3-3} \cmidrule(lr){4-9} \cmidrule(lr){10-13} \cmidrule(lr){14-17} &  &  \multirow{5}{*}{4,000} & Ensemble & -0.381 & 0.059 & 0.049 & 0.000 & 1.000 & 0.045 & 0.095 & 0.070 & 0.083 & -0.336 & 0.347 & 0.072 & 0.087 \\ 
   &  &  & Lasso & -0.408 & 0.052 & 0.051 & 0.000 & 1.000 & 0.038 & 0.083 & 0.072 & 0.074 & -0.370 & 0.379 & 0.075 & 0.081 \\ 
   &  &  & Parametric & -0.409 & 0.066 & 0.063 & 0.000 & 1.000 & -0.001 & 0.091 & 0.089 & 0.091 & -0.410 & 0.422 & 0.092 & 0.097 \\ 
   &  &  & RF & -0.574 & 0.054 & 0.049 & 0.000 & 1.000 & 0.708 & 0.711 & 0.068 & 0.064 & 0.134 & 0.151 & 0.068 & 0.069 \\ 
   &  &  & SVM & -0.433 & 0.072 & 0.075 & 0.001 & 0.995 & 0.195 & 0.217 & 0.096 & 0.093 & -0.237 & 0.256 & 0.091 & 0.095 \\ 
   \bottomrule
\end{tabularx}

  % Add description to the table
  \begin{notes}
  This table summarizes the simulation results. If $\gamma$ is different from zero, the unconfoundedness assumption is violated; if $\delta$ is different from zero the conditional common trends assumption is violated. We generate data using $n = 1000$ or $n = 4000$ observations, and $k = 100$ covariates across 1000 iterations. The learners indicate the models to estimate the nuisance parameters. The test statistics, standard errors, p-values, bias, and the RMSE, are averaged across simulations within the same specification. The rejection rate equals the proportion of iterations in which the test rejects. The bias measures the deviation from $\Delta_{D = 1} = 1$.
  \end{notes}  

\end{table}
\end{landscape}

In the third scenario, where $\delta = 0.25$, the treatment assignment $D$ becomes a function of the period specific error term $V_0$, mediated by the pre-treatment outcome $Y_0$. The test robustly detects violations of the common trends assumption, even in smaller samples. For the small sample case $\hat{\theta}$ varies between $0.38$ to $0.60$ and does not change much if n increases. Thus, the precision gain from the smaller standard errors is less pronounced. The test correctly rejects the null hypothesis in 55-100$\%$ of the cases using the smaller sample, and almost 100$\%$ of the cases in the larger sample. Here, the bias in $\Delta_{Unconf}$ diminishes, while the bias in $\Delta_{DiD}$ remains present with increasing sample size.

% Martin fragen ob das so stimmt.
It is crucial to recognize that the accuracy of nuisance parameters significantly influences the statistical power of the test. Specifically, either $\hat{p}(X, Y_0)$ or $\hat{\mu}(X, Y_0)$ and $\hat{\pi}(X)$ or $\hat{m}(X)$ must be accurately specified to yield a precise estimate of $\hat{\theta}$. While the test statistic is doubly robust with respect to the outcome and propensity score estimates, achieving true robustness requires that at least one set of nuisance parameters be approximately correctly specified for both $Unconf$ and $DiD$ scenarios. Applying the test to simulated data corroborates the previously derived asymptotic sample properties and underscores the critical importance of accurately estimated nuisance parameters for achieving a robust performance of the test. This foundation is crucial when extending the application of the test to real-world data.

\section{Application}\label{app}
This section presents an empirical examination of the test using publicly accessible datasets previously employed to estimate the average treatment effect on the treated (ATET). The test rejects the null hypothesis in two of the five cases. Emphasis is placed on scenarios with non-random assignment mechanisms into treatment and control group to highlight the practical applicability of the test.

\subsection{LaLonde (1986) data}\label{sec:lalonde}
In our first example, we apply the test to one of the most prominently used data from the field of labor economics, the \cite{LaLonde86} data. Within the framework of the National Supported Work Demonstration (NSW), LaLonde conducted a field experiment to evaluate the impact of a job training program on the real earnings of participants during the mid-1970s in the United States. 

In the experiment, applicants that were randomly assigned to the treatment group were guaranteed employment during the program. In contrast, applicants that were randomly assigned to the control group did not receive any form of support from the NSW. The job training initiative started in January 1978, with male participants completing the program after nine and female participants after twelve months. Initial interviews with participants were conducted in 1974, and their real earnings were documented for the years 1974, 1975, and 1978. The treatment is participation in the job program and the outcome of interest are participant's real earnings in the year 1978.

To assess the accuracy with which results from experimental data could be replicated using observational data LaLonde replaced the original control group with observations from the Panel Study of Income Dynamics (PSID). The comparison between estimates based on the experimental LaLonde data and the nonexperimental control group from the PSID revealed significant discrepancies, leading to the formulation of what is now widely acknowledged as the LaLonde critique.

Using the experimental LaLonde data as a benchmark and PSID-data to assess the performance of causal estimates, have been used in various settings, including propensity score matching (\cite{DehejiaWahba99} and \cite{DeWa02}), DiD-matching (\cite{HeIcTo98} and \cite{SmithTodd00}), and finally, the doubly robust score function for $\Delta_{DiD}$ which enters the test-statistic described in \cite{SantAnnaZhao2018}. Most recently, \cite{imbens2024lalonde} use the LaLonde data to reassess the LaLonde critique focusing on the performance of doubly robust methods. They stress the importance of validation exercises, such as the evaluation of common support, to obtain credible estimates.

We perform the test using the experimental treatment group and the non-experimental PSID control group, which only covers male participants. Under the null hypothesis unconfoundedness and conditional common trends jointly hold. Our analysis considers real earnings in 1975 as the pre-treatment outcome.  Included control variables are employment status in 1974 and 1975, alongside demographic factors such as as age, education, marital status, and race. We include real earnings in 1974 in the set of controls for estimating $\hat{p}(X, Y_0)$ and $\hat{\mu}(X, Y_0)$, where the previous earnings are likely an important predictor to account for dynamic selection bias. However, we exclude real earnings in 1974 for the estimation of $\hat{\pi}(X)$ and $\hat{m}(X)$ to avoid M-bias \cite{ChabeFerret2017}\footnote{The \textit{DiD} estimator assumes that selection bias remains constant over time. By including pre-treatment outcomes, we attempt to account for time-varying selection bias within the trend itself. However, this approach can lead to spurious correlation between the pre-treatment outcome and the treatment effect, as they are influenced by the same confounding factors that affect the treatment effect and result in bias. This correlation would not exist had we not opened the backdoor pathway in the first place.}. To estimate the nuisance parameters we use four different learners: Lasso, RF, and the parametric specification in combination with three-fold cross-validation. The sample consists of \(n = 2650\) observations out of which 185 are treated\footnote{We use the \texttt{lalonde.psid} dataset accessible via the \texttt{causalsens} R package \url{https://CRAN.R-project.org/package=causalsens}.}. We employ different learners for the nuisance parameters, which are reported in the table and the plots\footnote{Across all empirical examples we use Lasso, Random Forest, Support Vector Machine, an ensemble of the latter, and a parametric specification. If SVM does not converge we exclude it from the ensemble learner and as a single learner to obtain the nuisance parameter estimates.}. We find satisfactory overlap in the distribution of the propensity scores. The appendix \ref{appcs} provides plots for all propensity score estimates of our empirical examples.

\begin{figure}[h!]
    \centering
    \caption{LaLonde (1986)-PSID data: test results and comparison of ATET}
    \label{figlalonde}
    \includegraphics[width=0.8\textwidth]{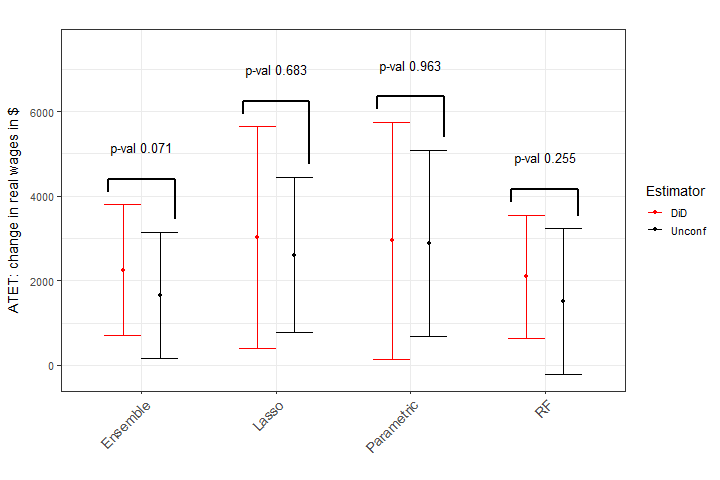}
    \vspace{10pt}
    
\begin{notes}
The plot shows ATET point estimates and their $95\%$ confidence intervals for the impact of the job training program on 1978 real earnings (USD). Results are color-coded: black for estimates under the unconfoundedness and red for estimates under common trends. P-values reflect the significance of the test statistic. Estimates are categorized by the learners used to obtain the nuisance parameters.
\end{notes}
\end{figure}

Table \ref{tablalonde} reports the test results, the point estimates for $\hat{\theta}$, the standard errors, and the p-values; and the point estimates and standard errors for $\Delta_{Unconf}$ and $\Delta_{DiD}$ using the LaLonde-PSID sample. Across the chosen specifications, the test statistic ranges from 65 to 600 and the standard errors from 330 to 1400. As a consequence, we cannot reject the null hypothesis that unconfoundedness and conditional common trends jointly hold at the 5$\%$ significance level. However, the ensemble learner yields a comparably small p-value of $0.072$. The point estimates suggest that participation in the job training program leads to an approximate increase by $1,500\$ $ to $3,000\$ $ in annual real wages, one year after participation. Although, there is variation across the different learners, the estimates for $\Delta_{Unconf}$ and $\Delta_{DiD}$ lie in a similar range, and exhibit largely overlapping confidence intervals. A visual representation can be found in Figure \ref{figlalonde}, where the point estimator and corresponding 95-$\%$ confidence bands for $\Delta_{Unconf}$ are color-coded in black, and the point estimator and corresponding 95-$\%$ confidence bands for $\Delta_{DiD}$ are color-coded in red. The p-values denote the significance-level that $\hat{\theta}$ is different from zero. To sum up, the distinct learners yield estimates for $\hat{\Delta}_{D = 1}$ in different ranges. Yet, the joint test does not reject in this example.

% Plots with estimated results
% Table with test results: ggf Nachkommastellen anpassen. 
\begin{table}[h!]
 \small
  \centering
    \caption{LaLonde (1986) data: test results and ATET estimates}
    \label{tablalonde}
    % latex table generated in R 4.3.1 by xtable 1.8-4 package
% Sat May 25 15:34:04 2024
\begin{tabularx}{\textwidth}{l*{7}{Y}}
   \toprule
  \toprule 
 & \multicolumn{3}{c}{\textit{Test results}} & \multicolumn{2}{c}{$Unconf$} & \multicolumn{2}{c}{$DiD$} \\
 \cmidrule(lr){2-4} \cmidrule(lr){5-6} \cmidrule(lr){7-8} 
Learner & $\hat{\theta}$ & SE & p-val & $\hat{\Delta}_{Unconf}$ & SE & $\hat{\Delta}_{DiD}$ & SE \\
 \cmidrule(lr){1-4} \cmidrule(lr){5-6} \cmidrule(lr){7-8}   Ensemble & 601.341 &  333.931 & 0.072 & 1646.837 &  761.483 & 2248.178 &  789.668 \\ 
  Lasso & 412.576 & 1012.622 & 0.684 & 2607.053 &  938.962 & 3019.628 & 1337.208 \\ 
  Parametric &  64.647 & 1428.279 & 0.964 & 2882.355 & 1120.957 & 2947.002 & 1429.548 \\ 
  RF & 593.185 &  521.413 & 0.255 & 1499.356 &  879.784 & 2092.542 &  745.496 \\ 
   \bottomrule
\end{tabularx}

    \vspace{10pt}   

\begin{notes} % Detailliertere Notes?
The table summarises the test results and ATET of the job training program on real earnings in 1978. The pre-treatment outcome is real earnings in 1975, and we control for real earnings in 1974, age, education, marital status, and race. We estimate the nuisance parameters using the learners displayed in the table. 
\end{notes}  

\end{table}

\subsection{Job Corps data}\label{sec:JC}
In our second example we use data from the United States largest education and training program for disadvantaged youths, the Job Corps program (JC). This program provides vocational training of different forms, such as academic education, health education, training in independent living, and assistance to find employment after the training was completed. JC was initiated in the 1960s and receives the largest share of funds from the US Department of Labor that targets training of young adults. To evaluate the impact of the program, the National Job Corps Study conducted the first nationally representative experiment in the beginning of the 1990s.

% Hier noch etwas einfügen
During the enrollment phase between November 1994 to December 1995, all youths that applied to JC were randomly assigned to treatment or control group. Youths in the treatment group were able to enroll in the training program. More precisely, the final group of participants consist of compliers within the randomly chosen treatment group. \cite{ScBuGl01} provide a detailed description of the experiment. Applications of the data can be found in \cite{FlFl09} who investigate the effect of the program on earnings, a discussion about the programs success \citep{FlFl10}, a decomposition of the ATE on general health using mediation analysis via inverse probability weighting \citep{Huber2012}, a discussion about the long-run effect of the program \citep{ScBuMc2008}.  

We are interested in the effect of the program on the employment rates, measured as proportion of weeks employed, after one year. The sample consists of \(n = 9240\) observations out of which 6574 are enrolled in an education program in the first year of JC\footnote{We use the \texttt{JC} dataset accessible via the \texttt{causalweight} R package \url{https://CRAN.R-project.org/package=causalweight}.}. 
The pre-treatment outcome is proportion of weeks employed before the program started, and we control for gender, age, ethnicity, education, GED degree, highschool degree, mother's and father's education, English as mother tongue, marital or cohabiting status, if participant has at least one child, and ever worked at the time of assignment, average weekly gross earnings, household size, welfare receipt during childhood, general health, extent of smoking and alcohol consumption.

\begin{figure}[h!]
    \centering
    \caption{Job Corps data (proportion of weeks employed): test results and ATET estimates}
    \label{figjc}
    \includegraphics[width=0.8\textwidth]{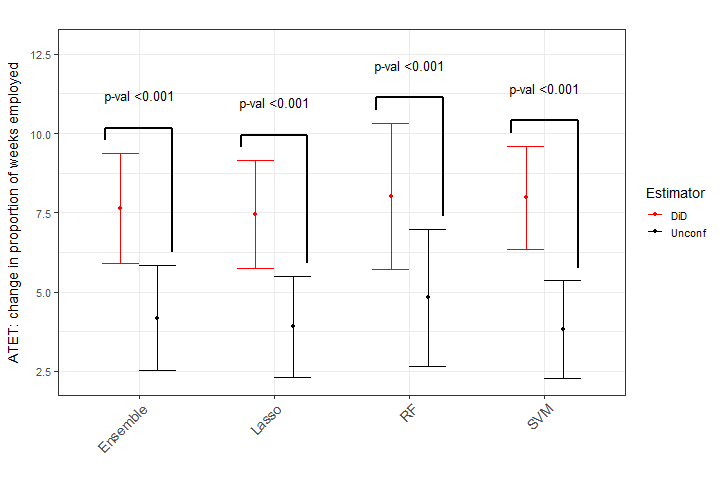}
    \vspace{10pt}
    
\begin{notes}
The plot shows ATET point estimates and their $95\%$ confidence intervals for the impact of the Job Corps program participation on proportion of weeks employed. Results are color-coded: black for estimates under the unconfoundedness and red for estimates under common trends. P-values reflect the significance of the test statistic. Estimates are categorized by the learners used to obtain the nuisance parameters.
\end{notes}
\end{figure}

The results are summarized in Table \ref{tabjc}. The test statistic $\hat{\theta}$ ranges between $3.2$ to $4.1$ with corresponding standard errors below $0.78$. Consequently, the test rejects across all implemented specifications at a lower than $0.001$ level. The variation in the point estimates for $\hat{\Delta}_{Unconf}$ and $\hat{\Delta}_{DiD}$ is demonstrated explicitly in Figure \ref{figjc}. Thus, the unconfoundedness and conditional common trends assumption do not jointly hold in this application.

% Plots with estimated results
% Table with test results: ggf Nachkommastellen anpassen. 
\begin{table}[h!]
  \centering
    \caption{Job Corps data (proportion of weeks employed): test results and ATET estimates}
    \label{tabjc}
    % latex table generated in R 4.3.1 by xtable 1.8-4 package
% Mon May 27 10:35:51 2024
\begin{tabularx}{\textwidth}{l*{7}{Y}}
   \toprule
  \toprule 
 & \multicolumn{3}{c}{\textit{Test results}} & \multicolumn{2}{c}{$Unconf$} & \multicolumn{2}{c}{$DiD$} \\
 \cmidrule(lr){2-4} \cmidrule(lr){5-6} \cmidrule(lr){7-8} 
Learner & $\hat{\theta}$ & SE & p-val & $\hat{\Delta}_{Unconf}$ & SE & $\hat{\Delta}_{DiD}$ & SE \\
 \cmidrule(lr){1-4} \cmidrule(lr){5-6} \cmidrule(lr){7-8}   Ensemble & 3.454 & 0.443 & 0.000 & 4.177 & 0.847 & 7.631 & 0.883 \\ 
  Lasso & 3.533 & 0.396 & 0.000 & 3.908 & 0.813 & 7.440 & 0.869 \\ 
  RF & 3.203 & 0.785 & 0.000 & 4.819 & 1.105 & 8.022 & 1.174 \\ 
  SVM & 4.166 & 0.405 & 0.000 & 3.810 & 0.789 & 7.975 & 0.826 \\ 
   \bottomrule
\end{tabularx}

    \vspace{10pt}   

\begin{notes} % Detailliertere Notes?
The table summarises the test results and ATET of the Job Corps program on proportion of weeks employed, one year after enrollment. The pre-treatment outcome is proportion of weeks employed before the program started, and we control for gender, age, ethnicity, education, GED degree, highschool degree, mother's and father's education, English as mother tongue, marital or cohabiting status, if participant has at least one child, and ever worked at the time of assignment, average weekly gross earnings, household size, welfare receipt during childhood, general health, extent of smoking and alcohol consumption. We estimate the nuisance parameters using the learners displayed in the table. 
\end{notes}  

\end{table}

\begin{figure}[h!]
    \centering
    \caption{Job Corps data (general health status): test results and ATET estimates}
    \label{figJChealth}
    \includegraphics[width=0.8\textwidth]{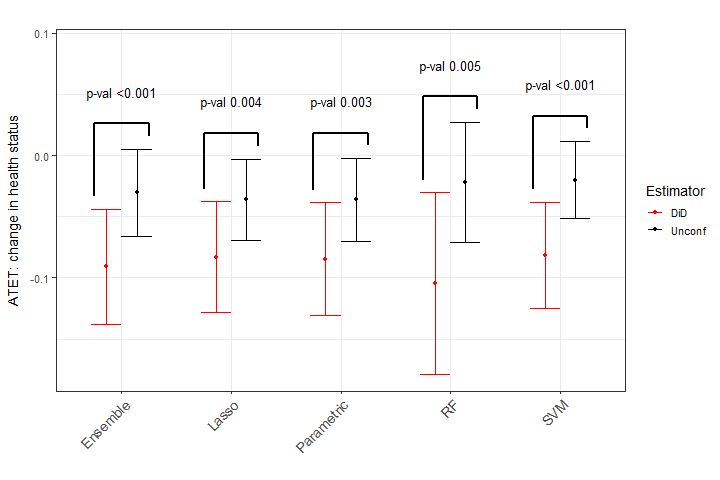}
    \vspace{10pt}
    
\begin{notes}
The plot shows ATET point estimates and their $95\%$ confidence intervals for the impact of the Job Corps program participation on general health status. Results are color-coded: black for estimates under the unconfoundedness and red for estimates under common trends. P-values reflect the significance of the test statistic. Estimates are categorized by the learners used to obtain the nuisance parameters.
\end{notes}
\end{figure}

As the JC program had multiple objectives, we employ a second variable, health outcomes, as the dependent variable employing the same vector of control variables. Similar to proportion of weeks employed, the test rejects the null hypothesis, as $\hat{\theta}$ is significantly different from zero. The detailed results are summarized in Table \ref{tabjchealth} and Figure \ref{figJChealth}.

% Plots with estimated results
% Table with test results: ggf Nachkommastellen anpassen. 
\begin{table}[h!]
  \centering
    \caption{Job Corps data (general health status): test results and ATET estimates}
    \label{tabjchealth}
    % latex table generated in R 4.3.1 by xtable 1.8-4 package
% Mon May 27 11:22:20 2024
\begin{tabularx}{\textwidth}{l*{7}{Y}}
   \toprule
  \toprule 
 & \multicolumn{3}{c}{\textit{Test results}} & \multicolumn{2}{c}{$Unconf$} & \multicolumn{2}{c}{$DiD$} \\
 \cmidrule(lr){2-4} \cmidrule(lr){5-6} \cmidrule(lr){7-8} 
Learner & $\hat{\theta}$ & SE & p-val & $\hat{\Delta}_{Unconf}$ & SE & $\hat{\Delta}_{DiD}$ & SE \\
 \cmidrule(lr){1-4} \cmidrule(lr){5-6} \cmidrule(lr){7-8}   Ensemble & -0.060 & 0.017 & 0.001 & -0.030 & 0.018 & -0.091 & 0.024 \\ 
  Lasso & -0.047 & 0.016 & 0.004 & -0.036 & 0.017 & -0.083 & 0.023 \\ 
  Parametric & -0.049 & 0.017 & 0.003 & -0.036 & 0.017 & -0.085 & 0.023 \\ 
  RF & -0.083 & 0.030 & 0.006 & -0.022 & 0.025 & -0.105 & 0.038 \\ 
  SVM & -0.062 & 0.017 & 0.000 & -0.020 & 0.016 & -0.082 & 0.022 \\ 
   \bottomrule
\end{tabularx}

    \vspace{10pt}   

\begin{notes} % Detailliertere Notes?
The table summarises the test results and ATET of the Job Corps program on proportion of weeks employed, one year after enrollment. The pre-treatment outcome is proportion of weeks employed before the program started, and we control for gender, age, ethnicity, education, GED degree, highschool degree, mother's and father's education, English as mother tongue, marital or cohabiting status, if participant has at least one child, and ever worked at the time of assignment, average weekly gross earnings, household size, welfare receipt during childhood, general health, extent of smoking and alcohol consumption. We estimate the nuisance parameters using the learners displayed in the table. 
\end{notes}  

\end{table}

\newpage
\subsection{Card and Krueger (1994) data}\label{sec:njmin}

This section investigates a natural experiment that examines the impact of minimum wage increases on employment. In April 1992, the state of New Jersey raised its minimum wage from $\$4.25$ to $\$5.05$. \cite{CardKrueger1994} used this policy change to investigate the hypothesis that an increase in the minimum wage leads to decreased employment. Their seminal study focused on the fast food industry, which is the largest employer of low-wage workers. Utilizing fast food restaurants from the neighboring state of Pennsylvania, the authors established control observations that were not subject to the policy shift. In total, they conducted two survey waves involving 410 establishments: the initial one in early spring 1992, just before the wage increase took effect, and the second in late autumn of the same year.

We analyze a subset of the data comprising $n = 334$ observations with complete records for the employment of full-time and part-time workers, the number of managers, the percentage of employees affected by the wage increase, and the initial wage upon starting the new job\footnote{The data and replication code from the original paper are available under \url{https://davidcard.berkeley.edu/data_sets.html}.}. Following the methodology of \cite{CardKrueger1994}, we assign full-time employment as dependent variable. The treated units are precisely identified as those fast-food restaurants in New Jersey that were mandated to increase wages due to the policy change. We incorporate a total of 27 control variables, accounting for potential confounding factors such as the composition of workers employed, product pricing, operating hours, co-ownership, and the type of fast-food chain. 

\begin{figure}[h!]
    \centering
    \caption{Card and Krueger (1994) data: test results and ATET estimates}
    \label{fignjmin}
    \includegraphics[width=0.8\textwidth]{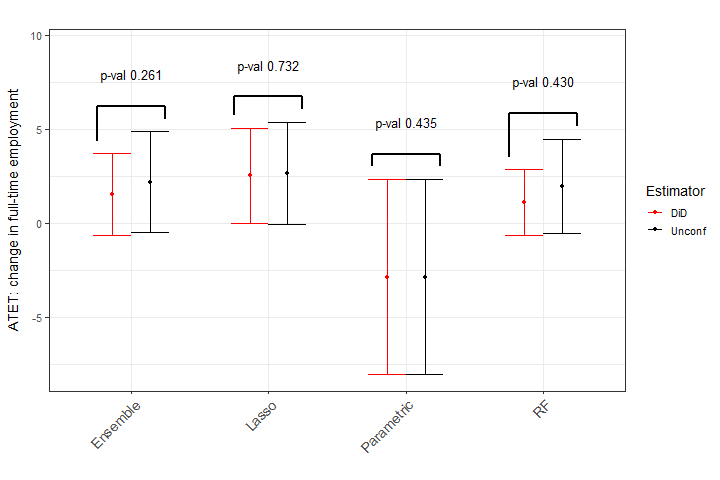}
    \vspace{10pt}
    
\begin{notes}
The plot shows ATET point estimates and their $95\%$ confidence intervals for the impact of the minimum wage increase on employment. Results are color-coded: black for estimates under the unconfoundedness and red for estimates under common trends. P-values reflect the significance of the test statistic. Estimates are categorized by the learners used to obtain the nuisance parameters.
\end{notes}
\end{figure}

The test results are displayed in Table \ref{tabnjmin} and indicate that we cannot reject the null hypothesis as $\hat{\theta}$ is close to zero across all specifications. Consistent with the findings of the original study, the two doubly robust point estimates for $\Delta_{D=1}$ suggest that the increase in the minimum wage does not significantly affect full-time employment. The point estimates along with their confidence intervals, and the p-values of the test are displayed in Figure \ref{fignjmin}.

% Plots with estimated results
% Table with test results: ggf Nachkommastellen anpassen. 
\begin{table}[h!]
  \centering
    \caption{Card and Krueger (1994) data: test results and ATET estimates}
    \label{tabnjmin}
    % latex table generated in R 4.3.1 by xtable 1.8-4 package
% Mon May 27 12:07:13 2024
\begin{tabularx}{\textwidth}{l*{7}{Y}}
   \toprule
  \toprule 
 & \multicolumn{3}{c}{\textit{Test results}} & \multicolumn{2}{c}{$Unconf$} & \multicolumn{2}{c}{$DiD$} \\
 \cmidrule(lr){2-4} \cmidrule(lr){5-6} \cmidrule(lr){7-8} 
Learner & $\hat{\theta}$ & SE & p-val & $\hat{\Delta}_{Unconf}$ & SE & $\hat{\Delta}_{DiD}$ & SE \\
 \cmidrule(lr){1-4} \cmidrule(lr){5-6} \cmidrule(lr){7-8}   Ensemble & -0.640 & 0.570 & 0.261 &  2.170 & 1.373 &  1.530 & 1.105 \\ 
  Lasso & -0.124 & 0.363 & 0.732 &  2.648 & 1.394 &  2.523 & 1.283 \\ 
  Parametric &  0.000 & 0.000 & 0.435 & -2.870 & 2.657 & -2.870 & 2.657 \\ 
  RF & -0.853 & 1.082 & 0.431 &  1.960 & 1.288 &  1.107 & 0.889 \\ 
   \bottomrule
\end{tabularx}

    \vspace{10pt}   

\begin{notes}
The table summarizes the test results and the ATET of the minimum wage increase on full-time employment.The pre-treatment outcome is defined as the number of full-time workers employed before the implementation of the policy. We control for co-ownership, the number of calls in the first survey wave, the numbers of full-time and part-time employees and managers, total employment, initial wage upon employment, duration until the first wage increase, bonuses, and subsidized meals. Additionally, we include prices for soda, fries, and main course, as well as the number of cash registers operational at 11:00 AM and the total number of cash registers. Nuisance parameters were estimated using the learners detailed in the table. 
\end{notes}  

\end{table}

\subsection{National Health and Nutrition Examination Survey Data}\label{sec:nhefs}

Next, we assess whether parallel trends and unconfoundedness hold when estimating the effect of smoking cessation on weight, following the analysis from a textbook example by \cite{Hernan2020}. The authors used data from the National Health and Nutrition Examination Survey I Epidemiologic Follow-up Study (NHEFS), initiated by the National Center for Health Statistics, National Institute on Aging, and United States Public Health Service. This survey collected various health measures over time, such as pulse rate, weight, and blood pressure, and documented hospital and nursing home stays, as well as alcohol and cigarette consumption, resulting in a rich panel dataset.

We use a subset of the NHEFS data to compare the ATET of smoking cessation on weight, measured in kilograms. Our final sample includes \(n = 1566\) participants who were smokers in the first wave of the survey in the early 1970s, with complete individual-level characteristics. The treatment is defined as smoking cessation between the initial survey and the follow-up visit ten years later. The sample includes 403 quitters. We control for 47 variables reflecting socioeconomic characteristics, health measures, and cigarette expenses. The data is publicly available\footnote{We use a subsample from the \texttt{causaldata} R package at \url{https://CRAN.R-project.org/package=causaldata}. The full sample can be accessed at \url{https://wwwn.cdc.gov/nchs/nhanes/nhefs/}.}.

\begin{figure}[h!]
\centering
\caption{NHEFS data: test results and ATET estimates}
\label{figNHEFS}
\includegraphics[width=0.8\textwidth]{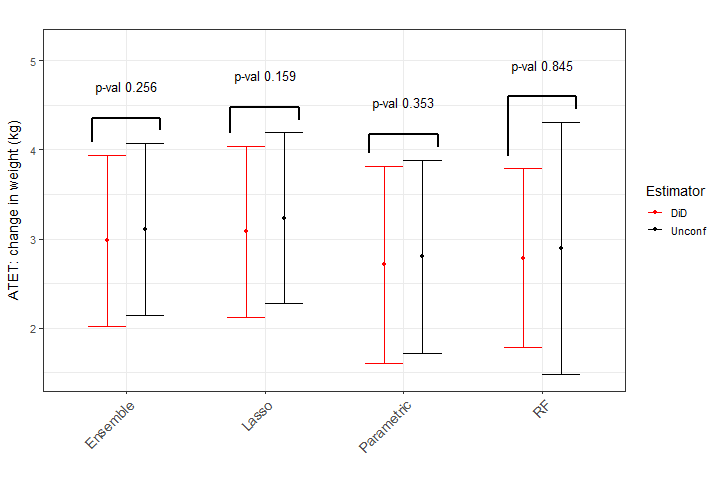}
\vspace{10pt}

\begin{notes}
The plot shows ATET point estimates and their $95\%$ confidence intervals for the impact of smoking cessation on weight. Results are color-coded: black for estimates under the unconfoundedness and red for estimates under common trends. P-values reflect the significance of the test statistic. Estimates are categorized by the learners used to obtain the nuisance parameters.
\end{notes}
\end{figure}

The results are summarized Table \ref{tabNHEFS} and indicate that we cannot reject the null hypothesis at any sensible level, as $\hat{\theta}$ is close to zero. A visual representation can be found in Figure \ref{figNHEFS}. This observational study provides an example, where selection bias is likely present, as quitters probably self-select into the treatment group. However, the high-dimensional controls seem to account for confounding effects. 

\begin{table}[h!]
\centering
\caption{NHEFS data: test results and ATET estimates}
\label{tabNHEFS}
% latex table generated in R 4.3.1 by xtable 1.8-4 package
% Sun May 26 16:32:21 2024
\begin{tabularx}{\textwidth}{l*{7}{Y}}
   \toprule
  \toprule 
 & \multicolumn{3}{c}{\textit{Test results}} & \multicolumn{2}{c}{$Unconf$} & \multicolumn{2}{c}{$DiD$} \\
 \cmidrule(lr){2-4} \cmidrule(lr){5-6} \cmidrule(lr){7-8} 
Learner & $\hat{\theta}$ & SE & p-val & $\hat{\Delta}_{Unconf}$ & SE & $\hat{\Delta}_{DiD}$ & SE \\
 \cmidrule(lr){1-4} \cmidrule(lr){5-6} \cmidrule(lr){7-8}   Ensemble & -0.126 & 0.111 & 0.256 & 3.106 & 0.491 & 2.980 & 0.488 \\ 
  Lasso & -0.152 & 0.108 & 0.159 & 3.234 & 0.488 & 3.081 & 0.489 \\ 
  Parametric & -0.090 & 0.097 & 0.353 & 2.803 & 0.552 & 2.713 & 0.564 \\ 
  RF & -0.107 & 0.550 & 0.846 & 2.893 & 0.723 & 2.787 & 0.511 \\ 
   \bottomrule
\end{tabularx}

\vspace{10pt}

\begin{notes}
The table summarizes the test results and ATET of smoking cessation on weight. We estimate the nuisance parameters using the learners displayed in the table. We control for 47 variables, including gender, age, marital status, ethnicity, income, education, height, smoking intensity, alcohol consumption, and various diseases.
\end{notes}
\end{table}

The applications in this section demonstrate that the test can serve as a robustness check to causal estimates using panel data, when using an ex-post matched control group, the treatment group consists of compliers, the data derives from a natural experiment, or has no degree of randomness to it.

% \subsection{Oregon Health Plan}\label{sec:OHP}
% In early 2008, the state of Oregon used public funds to finance participation in Mediacid for selected low-income adults as part of the Oregon Health Plan (OHP). As the demand for health insurance acceeded the available public funds could cover, the state opened a wating list from which the final applicants to OHP were chosen by a lottery. Those adults who won the lottery were allowed to apply for the health insurance program. Thus, the final treatment group consists of compliers who won the lottery and were eligible to participate. Importantly, application to the list required only limited information, and final approval to enter the program took place after the draws.

% The Oregon health insurance program was used to study the effect of health insurance on several outcomes, such as cost efficiency, health outcomes, and ... . We focus on the effect on health outcomes and costs as discussed in \cite{Finkelstein2012}. The authors argue that lowering the price of health care has incraeses the utilization of health care. On the one hand, overall health of users should improve, on the other hand, the increased use must be financed and higher utilization might increase the overall costs.The authors used the lottery draws as an instrument to analyze these effects.

\section{Conclusion}\label{conclusion}

We introduce a Durbin–Wu–Hausman type test leveraging selection on observables to assess the joint validity of unconfoundedness and conditional common trends in panel data. Our test statistic is derived from the difference between counterfactual estimation when applying doubly robust estimation to control for pre-treatment outcomes and covariates, and a doubly robust DiD approach controlling for covariates. The former assumes unconfoundedness, while the latter assumes conditional common trends.  A significant deviation of the test statistic from zero indicates a violation of at least one assumption.

We establish Neyman-orthogonality of our test statistic, which implies asymptotic normality and $\sqrt{n}$-consistency of estimation under specific regularity conditions. The latter permit controlling for possibly high dimensional covariates in a data-adaptive manner based on a double machine learning (DML) approach, as we suggest in the paper. We also investigate the finite sample performance of the DML-based testing procedure in a simulation study and find it to perform satisfactorily across the considered designs. Additionally, we apply our test to five empirical examples in which the treatment is not randomly assigned and find the test to reject the null hypothesis in two examples.

\clearpage

\newpage

% --- biblatex
% \printbibliography
 \bibliography{literature.bib}

\begin{thebibliography}{48}
\newcommand{\enquote}[1]{``#1''}
\expandafter\ifx\csname natexlab\endcsname\relax\def\natexlab#1{#1}\fi

\bibitem[\protect\citeauthoryear{Abadie}{Abadie}{2005}]{Abadie2005}
\textsc{Abadie, A.} (2005): \enquote{Semiparametric Difference-in-Differences
  Estimators,} \emph{Review of Economic Studies}, 72, 1--19.

\bibitem[\protect\citeauthoryear{Angrist and Pischke}{Angrist and
  Pischke}{2008}]{angrist_mostly_2008}
\textsc{Angrist, J.~D. and J.-S. Pischke} (2008): \emph{Mostly Harmless
  Econometrics: An Empiricist's Companion}, Princeton University Press.

\bibitem[\protect\citeauthoryear{Arkhangelsky and Imbens}{Arkhangelsky and
  Imbens}{2023}]{NBERw31942}
\textsc{Arkhangelsky, D. and G.~Imbens} (2023): \enquote{Causal Models for
  Longitudinal and Panel Data: A Survey,} \emph{NBER Working Paper 31942}.

\bibitem[\protect\citeauthoryear{Athey and Imbens}{Athey and
  Imbens}{2016}]{AtheyImbens2016}
\textsc{Athey, S. and G.~Imbens} (2016): \enquote{Recursive partitioning for
  heterogeneous causal effects,} \emph{Proceedings of the National Academy of
  Sciences}, 113, 7353--7360.

\bibitem[\protect\citeauthoryear{Bang and Robins}{Bang and
  Robins}{2005}]{BaRo05}
\textsc{Bang, H. and J.~Robins} (2005): \enquote{Doubly Robust Estimation in
  Missing Data and Causal Inference Models,} \emph{Biometrics}, 61, 962--972.

\bibitem[\protect\citeauthoryear{Belloni, Chernozhukov, and Hansen}{Belloni
  et~al.}{2014}]{Bellonietal2014}
\textsc{Belloni, A., V.~Chernozhukov, and C.~Hansen} (2014): \enquote{Inference
  on Treatment Effects after Selection among High-Dimensional Controls,}
  \emph{The Review of Economic Studies}, 81, 608--650.

\bibitem[\protect\citeauthoryear{Caetano, Callaway, Payne, and
  Rodrigues}{Caetano et~al.}{2022}]{caetano2022difference}
\textsc{Caetano, C., B.~Callaway, S.~Payne, and H.~S. Rodrigues} (2022):
  \enquote{Difference in Differences with Time-Varying Covariates,} \emph{arXiv
  preprint 2202.02903}.

\bibitem[\protect\citeauthoryear{Card and Krueger}{Card and
  Krueger}{1994}]{CardKrueger1994}
\textsc{Card, D. and A.~B. Krueger} (1994): \enquote{Minimum wages and
  employment: A case study of the fast-food industry in New Jersey and
  Pennsylvania,} \emph{The American Economic Review}, 84, 772--793.

\bibitem[\protect\citeauthoryear{Chab\'{e}-Ferret}{Chab\'{e}-Ferret}{2017}]{ChabeFerret2017}
\textsc{Chab\'{e}-Ferret, S.} (2017): \enquote{Should We Combine Difference In
  Differences with Conditioning on Pre-Treatment Outcomes,} \emph{working
  paper, Toulouse School of Economics}.

\bibitem[\protect\citeauthoryear{Chang}{Chang}{2020}]{Chang2020}
\textsc{Chang, N.-C.} (2020): \enquote{{Double/debiased machine learning for
  difference-in-differences models},} \emph{The Econometrics Journal}, 23,
  177--191.

\bibitem[\protect\citeauthoryear{Chernozhukov, Chetverikov, Demirer, Duflo,
  Hansen, Newey, and Robins}{Chernozhukov et~al.}{2018}]{Chetal2018}
\textsc{Chernozhukov, V., D.~Chetverikov, M.~Demirer, E.~Duflo, C.~Hansen,
  W.~Newey, and J.~Robins} (2018): \enquote{Double/debiased machine learning
  for treatment and structural parameters,} \emph{The Econometrics Journal},
  21, C1--C68.

\bibitem[\protect\citeauthoryear{Cox}{Cox}{1958}]{Cox58}
\textsc{Cox, D.} (1958): \emph{Planning of Experiments}, New York: Wiley.

\bibitem[\protect\citeauthoryear{Dehejia and Wahba}{Dehejia and
  Wahba}{1999}]{DehejiaWahba99}
\textsc{Dehejia, R.~H. and S.~Wahba} (1999): \enquote{Causal Effects in
  Non-experimental Studies: Reevaluating the Evaluation of Training
  Programmes,} \emph{Journal of American Statistical Association}, 94,
  1053--1062.

\bibitem[\protect\citeauthoryear{Dehejia and Wahba}{Dehejia and
  Wahba}{2002}]{DeWa02}
---\hspace{-.1pt}---\hspace{-.1pt}--- (2002):
  \enquote{Propensity-score-matching methods for nonexperimental causal
  studies,} \emph{The Review of Economics and Statistics}, 84, 151--161.

\bibitem[\protect\citeauthoryear{Ding and Li}{Ding and Li}{2019}]{Ding2019}
\textsc{Ding, P. and F.~Li} (2019): \enquote{A bracketing relationship between
  difference-in-differences and lagged-dependent-variable adjustment,} .

\bibitem[\protect\citeauthoryear{Donald, Hsu, and Lieli}{Donald
  et~al.}{2014}]{DoHsLi2014}
\textsc{Donald, S.~G., Y.-C. Hsu, and R.~P. Lieli} (2014): \enquote{Testing the
  Unconfoundedness Assumption via Inverse Probability Weighted Estimators of
  {(L)ATT},} \emph{Journal of Business \& Economic Statistics}, 32, 395--415.

\bibitem[\protect\citeauthoryear{Farrell, Liang, and Misra}{Farrell
  et~al.}{2021}]{FarrellLiangMisra2018}
\textsc{Farrell, M.~H., T.~Liang, and S.~Misra} (2021): \enquote{Deep Neural
  Networks for Estimation and Inference,} \emph{Econometrica}, 89, 181--213.

\bibitem[\protect\citeauthoryear{Flores and Flores-Lagunes}{Flores and
  Flores-Lagunes}{2009}]{FlFl09}
\textsc{Flores, C.~A. and A.~Flores-Lagunes} (2009): \enquote{Identification
  and Estimation of Causal Mechanisms and Net Effects of a Treatment under
  Unconfoundedness,} \emph{IZA DP No. 4237}.

\bibitem[\protect\citeauthoryear{Flores and Flores-Lagunes}{Flores and
  Flores-Lagunes}{2010}]{FlFl10}
---\hspace{-.1pt}---\hspace{-.1pt}--- (2010): \enquote{Nonparametric Partial
  Identification of Causal Net and Mechanism Average Treatment Effects,}
  \emph{mimeo, University of Florida}.

\bibitem[\protect\citeauthoryear{Ghanem, Sant'Anna, and W{\"u}thrich}{Ghanem
  et~al.}{2022}]{ghanem2022selection}
\textsc{Ghanem, D., P.~H. Sant'Anna, and K.~W{\"u}thrich} (2022):
  \enquote{Selection and parallel trends,} \emph{arXiv preprint 2203.09001}.

\bibitem[\protect\citeauthoryear{Hausman}{Hausman}{1978}]{Hausman1978}
\textsc{Hausman, J.~A.} (1978): \enquote{Specification Tests in Econometrics,}
  \emph{Econometrica}, 46, 1251--1271.

\bibitem[\protect\citeauthoryear{Heckman, Ichimura, and Todd}{Heckman
  et~al.}{1998}]{HeIcTo98}
\textsc{Heckman, J.~J., H.~Ichimura, and P.~Todd} (1998): \enquote{Matching as
  an econometric evaluation estimator,} \emph{Review of Economic Studies}, 65,
  261--294.

\bibitem[\protect\citeauthoryear{Hern\'{a}n and Robins}{Hern\'{a}n and
  Robins}{2020}]{Hernan2020}
\textsc{Hern\'{a}n, M. and J.~Robins} (2020): \emph{{Causal Inference: What
  If}}, Chapman \& Hall/CRC.

\bibitem[\protect\citeauthoryear{Huber}{Huber}{2014}]{Huber2012}
\textsc{Huber, M.} (2014): \enquote{Identifying causal mechanisms (primarily)
  based on inverse probability weighting,} \emph{Journal of Applied
  Econometrics}, 29, 920--943.

\bibitem[\protect\citeauthoryear{Huber}{Huber}{2023}]{huber2023causal}
---\hspace{-.1pt}---\hspace{-.1pt}--- (2023): \emph{Causal analysis: Impact
  evaluation and Causal Machine Learning with applications in R}, MIT Press.

\bibitem[\protect\citeauthoryear{Huber and Kueck}{Huber and
  Kueck}{2023}]{huber2023testing}
\textsc{Huber, M. and J.~Kueck} (2023): \enquote{Testing the identification of
  causal effects in observational data,} \emph{arXiv preprint 2203.15890}.

\bibitem[\protect\citeauthoryear{Imbens and Xu}{Imbens and
  Xu}{2024}]{imbens2024lalonde}
\textsc{Imbens, G. and Y.~Xu} (2024): \enquote{LaLonde (1986) after Nearly Four
  Decades: Lessons Learned,} .

\bibitem[\protect\citeauthoryear{Imbens}{Imbens}{2004}]{Im04}
\textsc{Imbens, G.~W.} (2004): \enquote{Nonparametric estimation of average
  treatment effects under exogeneity: a review,} \emph{The Review of Economics
  and Statistics}, 86, 4--29.

\bibitem[\protect\citeauthoryear{Imbens and Wooldridge}{Imbens and
  Wooldridge}{2009}]{ImWo08}
\textsc{Imbens, G.~W. and J.~M. Wooldridge} (2009): \enquote{Recent
  Developments in the Econometrics of Program Evaluation,} \emph{Journal of
  Economic Literature}, 47, 5--86.

\bibitem[\protect\citeauthoryear{LaLonde}{LaLonde}{1986}]{LaLonde86}
\textsc{LaLonde, R.} (1986): \enquote{Evaluating the econometric evaluations of
  training programs with experimental data,} \emph{American Economic Review},
  76, 604--620.

\bibitem[\protect\citeauthoryear{Lechner}{Lechner}{2011}]{Lechner2010}
\textsc{Lechner, M.} (2011): \enquote{The Estimation of Causal Effects by
  Difference-in-Difference Methods,} \emph{Foundations and Trends in
  Econometrics}, 4, 165--224.

\bibitem[\protect\citeauthoryear{Neyman}{Neyman}{1923}]{Neyman23}
\textsc{Neyman, J.} (1923): \enquote{On the Application of Probability Theory
  to Agricultural Experiments. Essay on Principles.} \emph{Statistical
  Science}, Reprint, 5, 463--480.

\bibitem[\protect\citeauthoryear{Neyman}{Neyman}{1959}]{Neyman1959}
---\hspace{-.1pt}---\hspace{-.1pt}--- (1959): \emph{Optimal asymptotic tests of
  composite statistical hypotheses}, Wiley, 416--444.

\bibitem[\protect\citeauthoryear{Pearl}{Pearl}{2000}]{Pearl00}
\textsc{Pearl, J.} (2000): \emph{Causality: Models, Reasoning, and Inference},
  Cambridge: Cambridge University Press.

\bibitem[\protect\citeauthoryear{Robins, Mark, and Newey}{Robins
  et~al.}{1992}]{RobinsMarkNewey1992}
\textsc{Robins, J.~M., S.~D. Mark, and W.~K. Newey} (1992): \enquote{Estimating
  exposure effects by modelling the expectation of exposure conditional on
  confounders,} \emph{Biometrics}, 48, 479--495.

\bibitem[\protect\citeauthoryear{Robins and Rotnitzky}{Robins and
  Rotnitzky}{1995}]{RoRo95}
\textsc{Robins, J.~M. and A.~Rotnitzky} (1995): \enquote{Semiparametric
  Efficiency in Multivariate Regression Models with Missing Data,}
  \emph{Journal of the American Statistical Association}, 90, 122--129.

\bibitem[\protect\citeauthoryear{Robins, Rotnitzky, and Zhao}{Robins
  et~al.}{1994}]{Robins+94}
\textsc{Robins, J.~M., A.~Rotnitzky, and L.~Zhao} (1994): \enquote{Estimation
  of Regression Coefficients When Some Regressors Are not Always Observed,}
  \emph{Journal of the American Statistical Association}, 90, 846--866.

\bibitem[\protect\citeauthoryear{Robins, Rotnitzky, and Zhao}{Robins
  et~al.}{1995}]{RoRoZa95}
---\hspace{-.1pt}---\hspace{-.1pt}--- (1995): \enquote{Analysis of
  Semiparametric Regression Models for Repeated Outcomes in the Presence of
  Missing Data,} \emph{Journal of the American Statistical Association}, 90,
  106--121.

\bibitem[\protect\citeauthoryear{Rubin}{Rubin}{1980}]{Rubin80}
\textsc{Rubin, D.} (1980): \enquote{Comment on 'Randomization Analysis of
  Experimental Data: The Fisher Randomization Test' by D. Basu,} \emph{Journal
  of American Statistical Association}, 75, 591--593.

\bibitem[\protect\citeauthoryear{Rubin}{Rubin}{1974}]{Rubin74}
\textsc{Rubin, D.~B.} (1974): \enquote{Estimating Causal Effects of Treatments
  in Randomized and Nonrandomized Studies,} \emph{Journal of Educational
  Psychology}, 66, 688--701.

\bibitem[\protect\citeauthoryear{Sant'Anna and Zhao}{Sant'Anna and
  Zhao}{2020}]{SantAnnaZhao2018}
\textsc{Sant'Anna, P. H.~C. and J.~B. Zhao} (2020): \enquote{Doubly Robust
  Difference-in-Differences Estimators,} \emph{Journal of Econometrics}, 219,
  101–122.

\bibitem[\protect\citeauthoryear{Schochet, Burghardt, and Glazerman}{Schochet
  et~al.}{2001}]{ScBuGl01}
\textsc{Schochet, P.~Z., J.~Burghardt, and S.~Glazerman} (2001):
  \enquote{National Job Corps Study: The Impacts of Job Corps on Participants'
  Employment and Related Outcomes,} \emph{Report (Washington, DC: Mathematica
  Policy Research, Inc.)}.

\bibitem[\protect\citeauthoryear{Schochet, Burghardt, and McConnell}{Schochet
  et~al.}{2008}]{ScBuMc2008}
\textsc{Schochet, P.~Z., J.~Burghardt, and S.~McConnell} (2008): \enquote{Does
  Job Corps Work? Impact Findings from the National Job Corps Study,} \emph{The
  American Economic Review}, 98, 1864--1886.

\bibitem[\protect\citeauthoryear{Smith and Todd}{Smith and
  Todd}{2005}]{SmithTodd00}
\textsc{Smith, J. and P.~Todd} (2005): \enquote{Does matching overcome
  LaLonde's critique of nonexperimental estimators?} \emph{Journal of
  Econometrics}, 125, 305--353.

\bibitem[\protect\citeauthoryear{van~der Laan, Polley, and Hubbard}{van~der
  Laan et~al.}{2007}]{VanderLaanHubbard2007}
\textsc{van~der Laan, M., E.~Polley, and A.~Hubbard} (2007): \enquote{Super
  Learner,} \emph{Statistical Applications in Genetics and Molecular Biology},
  6, 1--21.

\bibitem[\protect\citeauthoryear{Wager and Athey}{Wager and
  Athey}{2018}]{WagerAthey2018}
\textsc{Wager, S. and S.~Athey} (2018): \enquote{Estimation and Inference of
  Heterogeneous Treatment Effects using Random Forests,} \emph{Journal of the
  American Statistical Association}, 113, 1228--1242.

\bibitem[\protect\citeauthoryear{Wooldridge}{Wooldridge}{2002}]{Wooldridge02book}
\textsc{Wooldridge, J.} (2002): \emph{Econometric Analysis of Cross Section and
  Panel Data}, Cambridge: MIT Press.

\bibitem[\protect\citeauthoryear{Xu}{Xu}{2023}]{xu2023causal}
\textsc{Xu, Y.} (2023): \enquote{Causal inference with time-series
  cross-sectional data: a reflection,} \emph{SSRN 3979613}.

\end{thebibliography}

{\large \renewcommand{\theequation}{A-\arabic{equation}}
\setcounter{equation}{0} \appendix }
\appendix \numberwithin{equation}{section}

\begin{appendix}

{\small

\section{Proof of Theorem 1}\label{Neyman1}

To prove Theorem \ref{theorem1}, it is sufficient to verify the conditions of Assumptions 3.1 (on the linearity of the DR expression that permits identifying $\theta$ as well as its Neyman orthogonality) and 3.2 (on regularity conditions and quality of nuisance parameter estimators) of Theorems 3.1 and 3.2 as well as of Corollary 3.2 in \cite{Chetal2018}. All bounds hold uniformly over $P \in \mathcal{P},$ where $\mathcal{P}$ is the set of all possible probability laws, and we omit $P$ for brevity.

The nuisance parameters are given by $\eta=\{p(X, Y_0), \pi(X), \mu (X,Y_0), m(X)\}$,
with $p(X, Y_0)=\Pr(D=1 |X, Y_0)$, $\pi(X)=\Pr(D=1|X)$,  $\mu (X,Y_0)=E[Y |X,Y_0,D=0]$, and $m(X)=E[Y |X,D=0]$. The Neyman-orthogonal score function for $\theta=E[E[Y_1|X,Y_0,D=0]|D=1]-\{E[Y_0|D=1]+E[E[Y_1-Y_0|X,D=0]|D=1]\}$ is given by the following expression, with $W  = (Y_1,Y_0,D, X)$:
\begin{eqnarray}\label{scorefunction}
\psi(W,  \eta,  \theta) &=&  \frac{D}{\Pr(D=1)} \cdot \left\{\mu(X,Y_0)-Y_0 - m(X) \right\} \\
&+&  \frac{1-D}{\Pr(D=1)} \cdot  \left\{ \frac{ p(X,Y_0) \cdot [Y-\mu(X,Y_0)]}{1-p(X,Y_0)} - \frac{ \pi(X) \cdot [Y_1-Y_0-m(X)]}{1-\pi(X)}\right\}-\theta.\notag
\end{eqnarray}

We denote by $\eta_0$ the true functions of the nuisance parameters in $\eta$.  Let $\mathcal{T}_n$ be the set of all $\eta=(p, \pi, \mu, $m$)$ consisting of $P$-square integrable functions $p$, $\pi$, $\mu$, $m$ such that
\begin{eqnarray}\label{Tn}
\left\|  \eta - \eta_0 \right\|_{q} &\leq& C,  \\
\left\|  \eta - \eta_0 \right\|_{2} &\leq& \delta_n, \notag \\
\left\|   p(X, Y_0)-1/2\right\|_{\infty}  &\leq& 1/2-\epsilon, \notag\\
\left\|   \pi(X)-1/2)\right\|_{\infty} &\leq & 1/2-\epsilon, \notag \\
\left\|   \mu(X,Y_0,D)-\mu_0(X,Y_0,D)\right\|_{2} \times \left\|   p(X, Y_0)-p_0(X, Y_0)\right\|_{2}  &\leq & \delta^{}_n n^{-1/2}, \notag \\
\left\|   m(X,D)-m_0(X,D)\right\|_{2} \times \left\|   \pi(X)-\pi_{0}(X)\right\|_{2} &\leq & \delta^{}_n n^{-1/2}.\notag
\end{eqnarray}
We furthermore replace the sequence $(\delta_n)_{n \geq 1}$ by $(\delta_n')_{n \geq 1},$ where $\delta_n' = C_{\epsilon} \max(\delta_n,n^{-1/2}),$ where $C_{\epsilon}$ is sufficiently large constant that only depends on $C$ and $\epsilon.$

\textbf{Assumption 3.1:  Linear score function and Neyman orthogonality.}
\vspace{5pt}\newline
\textbf{Assumption 3.1(a) - moment condition:} The moment condition $E\Big[\psi(W, \eta_0,  \theta)\Big] = 0$ holds:
\begin{align}
E\Big[\psi(W,  \eta_0,  \theta)\Big] &= E\left[  \frac{D}{\Pr(D=1)} \cdot [\mu(X,Y_0)-Y_0 - m(X) ] + \frac{p(X,Y_0)}{\Pr(D=1)} \cdot E\left[ \frac{(1-D)\cdot (Y_1-\mu(X,Y_0))}{1-p(X,Y_0)}\Big| X,Y_0\right]\right.\notag\\
& - \left.\frac{\pi(X)}{\Pr(D=1)} \cdot E\left[ \frac{(1-D)  \cdot [Y_1-Y_0-m(X)]}{1-\pi(X)}\Big| X\right] -\theta\right] \\
&= E\left[   \frac{D}{\Pr(D=1)} \cdot [\mu(X,Y_0)-Y_0 - m(X)]+ \frac{p(X,Y_0)}{\Pr(D=1)} \cdot[\mu(X,Y_0)-\mu(X,Y_0)]\right.\notag\\
&- \left.\frac{\pi(X)}{\Pr(D=1)} \cdot [m(X)-m(X)]\right]- \theta\notag\\
&= E[  \mu(X,Y_0)-Y_0 - m(X)  + \mu(X,Y_0)-\mu(X,Y_0)-m(X)+m(X)|D=1]- \theta  \notag\\
&= E[  \mu(X,Y_0)-Y_0 - m(X) |D=1   ]- \theta =  0, \notag
\end{align}
where the first equality follows from the law of iterated expectations and the second and third from basic probability theory.

\textbf{Assumption 3.1(b)- linearity:} The score $ \psi(W,  \eta_0,  \theta) $ is linear in $\theta$:
$\psi(W, \eta_0, \theta) = \psi^a(W, \eta_0) \cdot\theta + \psi^b(W, \eta_0) $
with $\psi^a(W, \eta_0) = -1$ and
\begin{eqnarray}
\psi^b(W, \eta_0) &=&\frac{D}{\Pr(D=1)} \cdot [\mu(X,Y_0)-Y_0 - m(X) ] \\
&+&  \frac{1-D}{\Pr(D=1)} \cdot  \left\{ \frac{ p(X,Y_0) \cdot (Y_1-\mu(X,Y_0))}{1-p(X,Y_0)} - \frac{ \pi(X) \cdot [Y_1-Y_0-m(X)]}{1-\pi(X)}\right\}. \notag
\end{eqnarray}

\textbf{Assumption 3.1(c) - continuity:}
The expression for the second Gateaux derivative of a map $\eta \mapsto E[\psi(W,  \eta,  \theta)]$ is continuous.

\textbf{Assumption 3.1(d) - Neyman Orthogonality}: For any $\eta \in \mathcal{T}_n,$ the Gateaux derivative in the direction $ \eta - \eta_0 =
p(X, Y_0)-p_0(X, Y_0), (\pi(X)-\pi_0 (X),   \mu(X,Y_0)-\mu_0 (X,Y_0)$, $m(X)-m_0(X)$ is given by:
\begin{align}
&\partial E \big[\psi_{d}(W, \eta, \theta)\big] \big[\eta - \eta_0 \big]  = \notag\\
& \underbrace{E\left[ \frac{D}{\Pr(D=1)} \cdot [\mu(X,Y_0)-\mu_0(X,Y_0)] \right]  -  E\left[\frac{1-D}{\Pr(D=1)} \cdot  \left\{ \frac{ p_0(X,Y_0) \cdot [\mu(X,Y_0)-\mu_0(X,Y_0)]}{1-p_0(X,Y_0)}\right\} \right]}_{E[\mu(X,Y_0)-\mu_0(X,Y_0)-(\mu(X,Y_0)-\mu_0(X,Y_0))|D=1]=0}\notag\\
+&\underbrace{E\left[ \frac{p(X,Y_0)-p_0(X,Y_0)}{\Pr(D=1)} \cdot \frac{ (1-D)\cdot [Y_1-\mu_0(X,Y_0)]}{1-p_0(X,Y_0)} \right]}_{E\left[ \frac{p(X,Y_0)-p_0(X,Y_0)}{\Pr(D=1)} \cdot [\mu_0(X,Y_0)-\mu_0(X,Y_0)]\right]=0} -\underbrace{E\left[ \frac{p_0(X,Y_0)-p(X,Y_0)}{1-p_0(X,Y_0)} \cdot \frac{p_0(X,Y_0)}{\Pr(D=1)} \cdot\frac{ (1-D)  \cdot [Y_1-\mu_0(X,Y_0)]}{  1-p_0(X,Y_0)} \right]}_{E\left[ \frac{p_0(X,Y_0)-p(X,Y_0)}{1-p_0(X,Y_0)} \cdot \frac{p_0(X,Y_0)}{\Pr(D=1)} \cdot [\mu_0(X,Y_0)-\mu_0(X,Y_0)]\right]=0}\notag\\
& \underbrace{-E\left[ \frac{D}{\Pr(D=1)} \cdot [m(X)-m_0(X)] \right]  +  E\left[\frac{1-D}{\Pr(D=1)} \cdot  \left\{ \frac{ \pi_0(X) \cdot [m(X)-m_0(X)]}{1-\pi_0(X)}\right\} \right]}_{E[m(X)-m_0(X)-(m(X)-m_0(X))|D=1]=0}\notag\\
-&\underbrace{E\left[ \frac{\pi(X)-\pi_0(X)}{\Pr(D=1)} \cdot \frac{ (1-D)\cdot [Y_1-Y_0-m_0(X)]}{1-\pi_0(X)} \right]}_{E\left[ \frac{\pi(X)-\pi_0(X)}{\Pr(D=1)} \cdot [m_0(X)-m_0(X)]\right]=0} +\underbrace{E\left[ \frac{\pi_0(X)-\pi(X)}{1-\pi_0(X)} \cdot \frac{\pi_0(X)}{\Pr(D=1)} \cdot\frac{ (1-D)  \cdot [Y_1-Y_0-m_0(X)]}{  1-\pi_0(X)} \right]}_{E\left[ \frac{\pi_0(X)-\pi(X)}{1-\pi_0(X)} \cdot \frac{\pi_0(X)}{\Pr(D=1)} \cdot [m_0(X)-m_0(X)]\right]=0}\notag\\
&=0.
\end{align}

\textbf{Assumption 3.2:  Score regularity and quality of nuisance parameter estimators.}
\vspace{5pt}\newline
Reconsidering the Neyman orthogonal score function \eqref{scorefunction}, we note that it is equivalent to the following DR expression, which gives the differences in ATETs when relying on Assumption \ref{A2} (common trends) and Assumption \ref{A1} (unconfoundedness):
\begin{align}\label{scorefunction2}
\psi(W,  \eta,  \theta) &= \overbrace{\frac{D \cdot [Y_1-Y_0 - m(X) ]}{\Pr(D=1)} - \frac{(1-D)\cdot \pi(X) \cdot [Y_1-Y_0-m(X)]}{\Pr(D=1) \cdot (1-\pi(X))}}^{ATET\textrm{ }based\textrm{ }on\textrm{ }common\textrm{ }trends\textrm{ }(part\textrm{ }A)} \\
&- \underbrace{\left\{\frac{D \cdot [Y_1-\mu(X,Y_0)]}{\Pr(D=1)}- \frac{(1-D)\cdot p(X,Y_0) \cdot [Y-\mu(X,Y_0)]}{\Pr(D=1) \cdot (1-p(X,Y_0))}\right\}}_{ATET\textrm{ }based\textrm{ }on\textrm{ }unconfoundedness\textrm{ }(part\textrm{ }B)} -\theta.\notag
\end{align}
Part B corresponds to the score function of the ATET in equation (5.4) of \cite{Chetal2018} when utilizing $X,Y_0$ as conditioning set. By Theorem 5.1 in  \cite{Chetal2018}, estimation of part B based on cross-fitting satisfies Assumption 3.2 of \cite{Chetal2018} under the regularity conditions imposed on $Y_1$, $\mu_0(X,Y_0), \hat\mu(X,Y_0), p_0(X,Y_0), \hat p(X,Y_0)$ in Assumption \ref{A3}. Part A in equation \eqref{scorefunction2} is equivalent to the score function of the ATET provided in equation (3.1) \cite{Chang2020} for panel data (referred to as repeated outcomes). By Theorem 3.1 in \cite{Chang2020}, estimation of part A based on cross-fitting  satisfies Assumption 3.2 of \cite{Chetal2018} under the regularity conditions imposed on $Y_1-Y_0$, $m_0(X), \hat m(X), \pi_0(X), \hat \pi(X)$ in Assumption \ref{A3}. As the score function \eqref{scorefunction} (or \eqref{scorefunction2}) is a linear combination of the expressions in \cite{Chetal2018} and \cite{Chang2020}, it directly follows that our cross-fitted estimator $\hat \theta$ suggested in Algorithm 1 satisfies Assumption 3.2 of \cite{Chetal2018}, too. This concludes the proof of Theorem \ref{theorem1}

\section{An alternative DR test statistic}

We subsequently briefly present an alternative testing approach that does not directly compare the DiD- and unconfoundedness-based mean potential outcomes under non-treatment in the treated population but is based on a modified statistic that is equivalent to such a comparison. To derive this statistic, we reconsider the difference in the mean potential outcome \(E[Y_1(0)|D=1]\) under unconfoundedness (Assumption \ref{A1}) and common trends (Assumption \ref{A2}) in the first line of equation \eqref{thetameans}:  
\begin{align}\label{thetameans2}
&\underbrace{E[E[Y_1|X,Y_0,D=0]|D=1]}_{E[Y_1(0)|D=1]\textrm{ under Assumption \ref{A1}} }-\underbrace{\{E[Y_0|D=1]+E[E[Y_1-Y_0|X,D=0]|D=1]\}}_{E[Y_1(0)|D=1]\textrm{ under Assumption \ref{A2}} }\notag\\
&= E[E[Y_1-Y_0|X,Y_0,D=0]|D=1]-E[E[Y_1-Y_0|X,D=0]|D=1].
\end{align} 
The second line in expression \eqref{thetameans2} follows from $E[E[Y_1-Y_0|X,Y_0,D=0]|D=1]=E[E[Y_1|X,Y_0,D=0] - Y_0|D=1]$, because $E[Y_0|X,Y_0,D=0]=Y_0$, and is numerically equivalent to the following DR statistic $\theta'$:
\begin{eqnarray} \label{DRmodified}
\theta'&=&E\left[\frac{ D}{\Pr(D=1)}\cdot[q(Y_0,X)-m(X)]\right]\\
&+&E\left[\frac{1-D}{\Pr(D=1)}\cdot \left\{
\frac{ p(Y_0,X)\cdot(Y_1-Y_0-q(Y_0,X))}{1-p(Y_0,X)} - \frac{\pi(X) \cdot  (Y_1-Y_0-m(X))}{1-\pi(X)}\right\}\right],\notag
	\end{eqnarray}
	with $q(x,Y_0)=E[Y_1-Y_0|X=x, Y_0=y_0,D=0]$.

We may construct a DML estimator based on equation \eqref{DRmodified} which is normally distributed under the same regularity conditions as considered for the DML procedure outlined in Section \ref{est} of the main paper, which is based on the DR expression \eqref{thetaDR}. Asymptotically, such DML estimators based on \eqref{DRmodified} and \eqref{thetaDR} are equivalent under the satisfaction of appropriate regularity conditions for the machine learning-based estimation of the propensity scores and conditional means entering the respective DR functions.
}

\newpage
\section{Assessing common support}\label{appcs}
Subsequently, we provide histograms of the propensity scores $\hat{p}(X, Y_0)$ and $\hat{\pi}(X)$ by empirical application and for each implemented learner to demonstrate common support across the treatment and control groups.

% ------------------------------------------------------------------------------ %
% LaLonde
% ------------------------------------------------------------------------------ %

\subsection{LaLonde (1986) data}

\begin{figure}[H]
    \caption{LaLonde (1986) data: Common support of $\hat{p}(X, Y_0)$}
    
    \begin{minipage}[t]{0.43\textwidth}
        \centering
        \begin{tikzpicture}
            \node[anchor=south west,inner sep=0] (image) at (0,0) {\includegraphics[width=\textwidth]{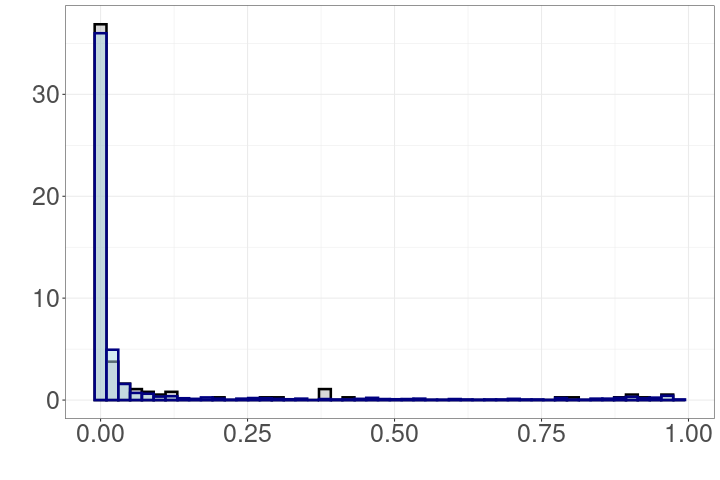}};
            \begin{scope}[x={(image.south east)},y={(image.north west)}]
                % Add annotation
                \node[black, font=\small] at (0.575,0) {$\hat{p}(X, Y_0)$};
                % Add "Count" annotation
                \node[black, font=\small, rotate=90] at (0,0.5) {Count};
            \end{scope}
        \end{tikzpicture}
        \caption*{(a) Ensemble learner}
        %\label{subfig1}
    \end{minipage}
    \hfill
    \begin{minipage}[t]{0.43\textwidth}
        \centering
        \begin{tikzpicture}
            \node[anchor=south west,inner sep=0] (image) at (0,0) {\includegraphics[width=\textwidth]{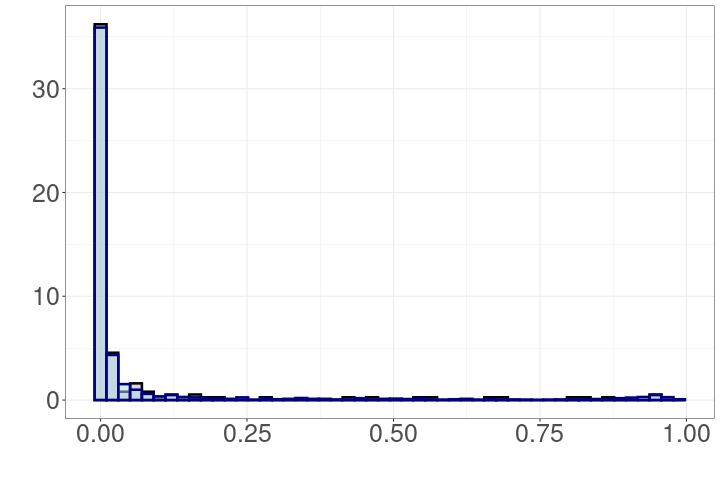}};
            \begin{scope}[x={(image.south east)},y={(image.north west)}]
                % Add annotation
                \node[black, font=\small] at (0.575,0) {$\hat{p}(X, Y_0)$};
                % Add "Count" annotation
                \node[black, font=\small, rotate=90] at (0,0.5) {Count};
            \end{scope}
        \end{tikzpicture}
        \caption*{(b) Lasso}
        %\label{subfig2}
    \end{minipage}
    \hfill
    \begin{minipage}[t]{0.43\textwidth}
        \centering
        \begin{tikzpicture}
            \node[anchor=south west,inner sep=0] (image) at (0,0) {\includegraphics[width=\textwidth]{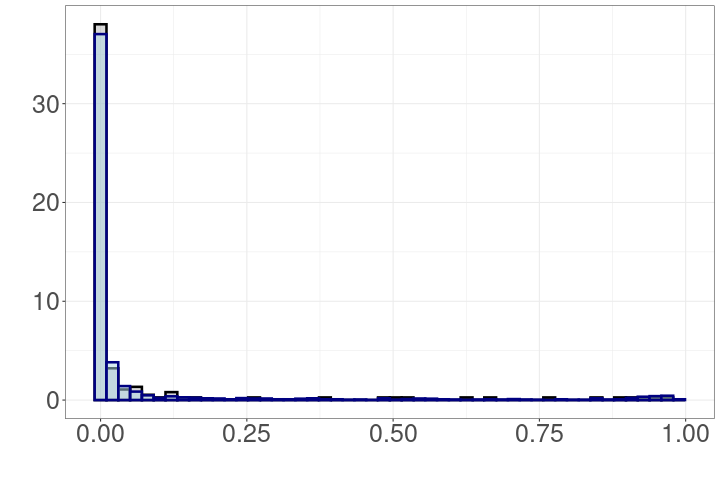}};
            \begin{scope}[x={(image.south east)},y={(image.north west)}]
                % Add annotation
                \node[black, font=\small] at (0.575,0) {$\hat{p}(X, Y_0)$};
                % Add "Count" annotation
                \node[black, font=\small, rotate=90] at (0,0.5) {Count};
            \end{scope}
        \end{tikzpicture}
        \caption*{(c) Logistic regression (Parametric)}
        %\label{subfig3}
    \end{minipage}
    \hfill
    \begin{minipage}[t]{0.43\textwidth}
        \centering
        \begin{tikzpicture}
            \node[anchor=south west,inner sep=0] (image) at (0,0) {\includegraphics[width=\textwidth]{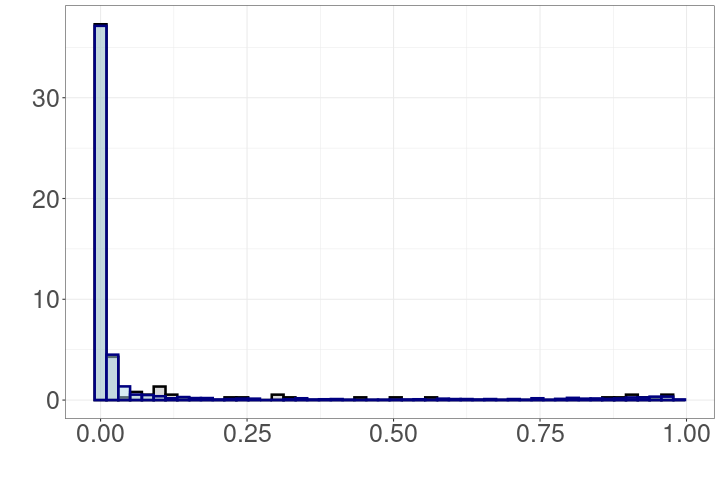}};
            \begin{scope}[x={(image.south east)},y={(image.north west)}]
                % Add annotation
                \node[black, font=\small] at (0.575,0) {$\hat{p}(X, Y_0)$};
                % Add "Count" annotation
                \node[black, font=\small, rotate=90] at (0,0.5) {Count};
            \end{scope}
        \end{tikzpicture}
        \caption*{(d) Random forest}
        %\label{subfig3}
    \end{minipage}

    \label{fig:psLalondeunconf}
    \begin{notes}
    The figure displays the common support for the propensity scores $\hat{p}(X, Y_0)$. Treated observations are color-coded in gray, and control observations are in blue. The sample size is $n = 2675$, with $185$ treated units.
    \end{notes}

\end{figure}

% --------------------------------- DiD ---------------------------------------------- %

\begin{figure}[H]
    \caption{LaLonde (1986) data: Common support of $\hat{\pi}(X)$}
    
    \begin{minipage}[t]{0.43\textwidth}
        \centering
        \begin{tikzpicture}
            \node[anchor=south west,inner sep=0] (image) at (0,0) {\includegraphics[width=\textwidth]{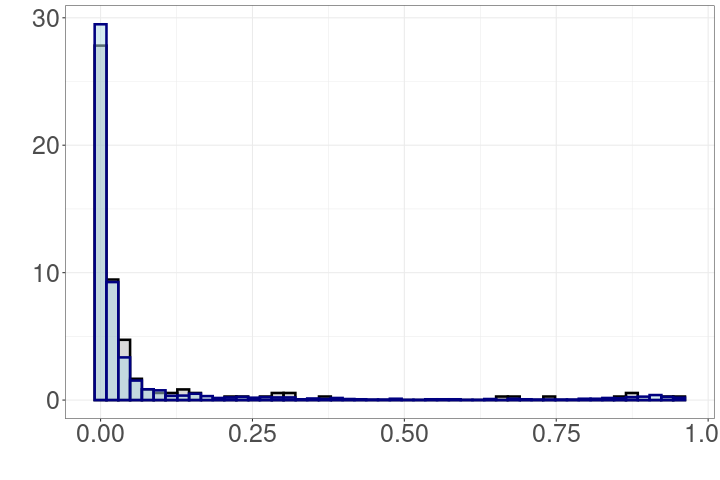}};
            \begin{scope}[x={(image.south east)},y={(image.north west)}]
                % Add annotation
                \node[black, font=\small] at (0.575,0) {$\hat{\pi}(X)$};
                % Add "Count" annotation
                \node[black, font=\small, rotate=90] at (0,0.5) {Count};
            \end{scope}
        \end{tikzpicture}
        \caption*{(a) Ensemble learner}
        %\label{subfig1}
    \end{minipage}
    \hfill
    \begin{minipage}[t]{0.43\textwidth}
        \centering
        \begin{tikzpicture}
            \node[anchor=south west,inner sep=0] (image) at (0,0) {\includegraphics[width=\textwidth]{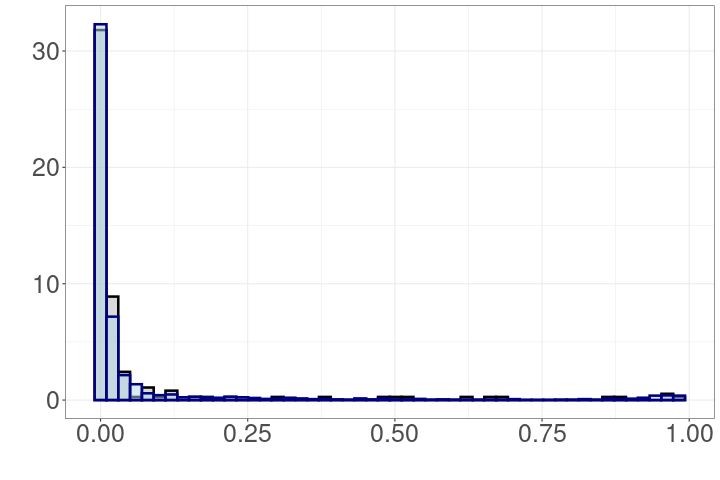}};
            \begin{scope}[x={(image.south east)},y={(image.north west)}]
                % Add annotation
                \node[black, font=\small] at (0.575,0) {$\hat{\pi}(X)$};
                % Add "Count" annotation
                \node[black, font=\small, rotate=90] at (0,0.5) {Count};
            \end{scope}
        \end{tikzpicture}
        \caption*{(b) Lasso}
        %\label{subfig2}
    \end{minipage}
    \hfill
    \begin{minipage}[t]{0.43\textwidth}
        \centering
        \begin{tikzpicture}
            \node[anchor=south west,inner sep=0] (image) at (0,0) {\includegraphics[width=\textwidth]{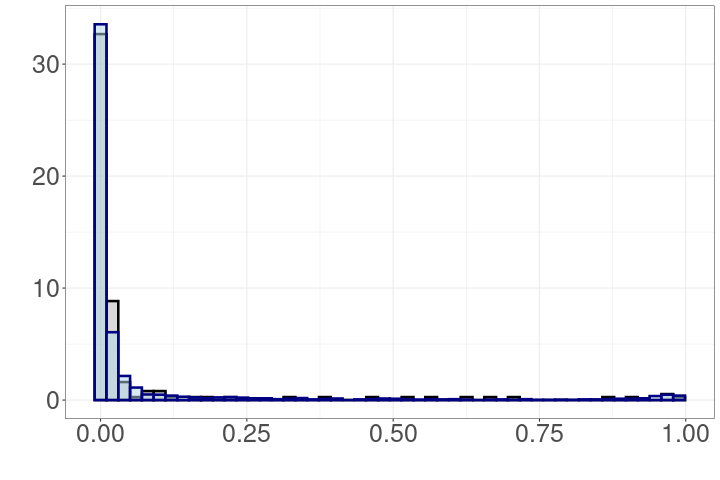}};
            \begin{scope}[x={(image.south east)},y={(image.north west)}]
                % Add annotation
                \node[black, font=\small] at (0.575,0) {$\hat{\pi}(X)$};
                % Add "Count" annotation
                \node[black, font=\small, rotate=90] at (0,0.5) {Count};
            \end{scope}
        \end{tikzpicture}
        \caption*{(c) Logistic regression (Parametric)}
        %\label{subfig3}
    \end{minipage}
    \hfill
    \begin{minipage}[t]{0.43\textwidth}
        \centering
        \begin{tikzpicture}
            \node[anchor=south west,inner sep=0] (image) at (0,0) {\includegraphics[width=\textwidth]{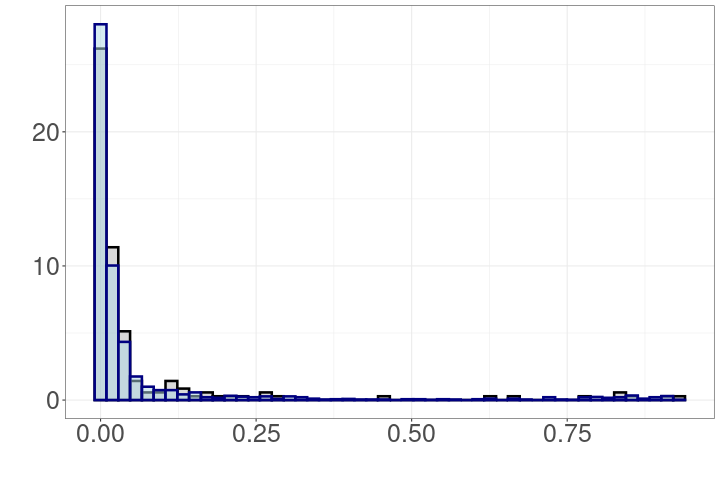}};
            \begin{scope}[x={(image.south east)},y={(image.north west)}]
                % Add annotation
                \node[black, font=\small] at (0.575,0) {$\hat{\pi}(X)$};
                % Add "Count" annotation
                \node[black, font=\small, rotate=90] at (0,0.5) {Count};
            \end{scope}
        \end{tikzpicture}
        \caption*{(d) Random forest}
        %\label{subfig3}
    \end{minipage}

    \label{fig:psLalondeDiD}
    \begin{notes}
    The figure displays the common support for the propensity scores $\hat{\pi}(X)$. Treated observations are color-coded in gray, and control observations are in blue. The sample size is $n = 2675$, with $185$ treated units.
    \end{notes}

\end{figure}

% ------------------------------------------------------------------------------ %
% Job Corps
% ------------------------------------------------------------------------------ %

\subsection{Job Corps data}
\begin{figure}[H]
    \caption{Job Corps data (proportion of weeks employed): Common support of $\hat{p}(X, Y_0)$}
    
    \begin{minipage}[t]{0.43\textwidth}
        \centering
        \begin{tikzpicture}
            \node[anchor=south west,inner sep=0] (image) at (0,0) {\includegraphics[width=\textwidth]{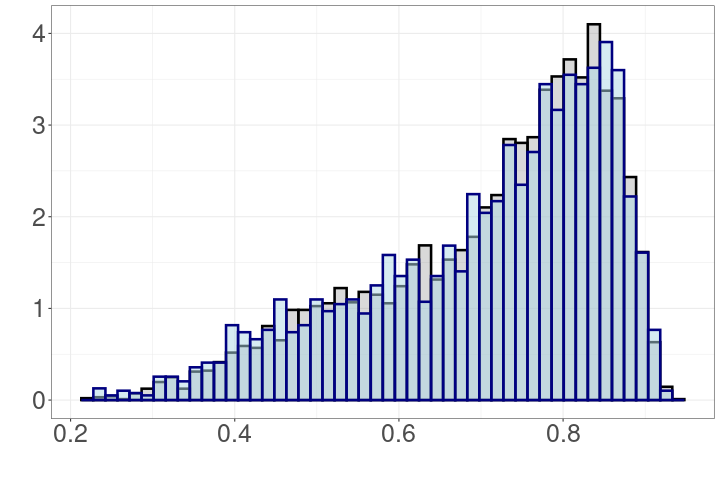}};
            \begin{scope}[x={(image.south east)},y={(image.north west)}]
                % Add annotation
                \node[black, font=\small] at (0.575,0) {$\hat{p}(X, Y_0)$};
                % Add "Count" annotation
                \node[black, font=\small, rotate=90] at (0,0.5) {Count};
            \end{scope}
        \end{tikzpicture}
        \caption*{(a) Ensemble learner}
        %\label{subfig1}
    \end{minipage}
    \hfill
    \begin{minipage}[t]{0.43\textwidth}
        \centering
        \begin{tikzpicture}
            \node[anchor=south west,inner sep=0] (image) at (0,0) {\includegraphics[width=\textwidth]{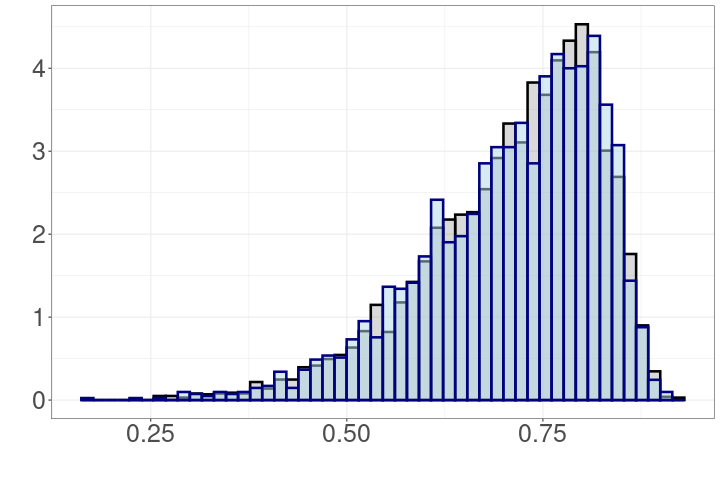}};
            \begin{scope}[x={(image.south east)},y={(image.north west)}]
                % Add annotation
                \node[black, font=\small] at (0.575,0) {$\hat{p}(X, Y_0)$};
                % Add "Count" annotation
                \node[black, font=\small, rotate=90] at (0,0.5) {Count};
            \end{scope}
        \end{tikzpicture}
        \caption*{(b) Lasso}
        %\label{subfig2}
    \end{minipage}
    \hfill
    \begin{minipage}[t]{0.43\textwidth}
        \centering
        \begin{tikzpicture}
            \node[anchor=south west,inner sep=0] (image) at (0,0) {\includegraphics[width=\textwidth]{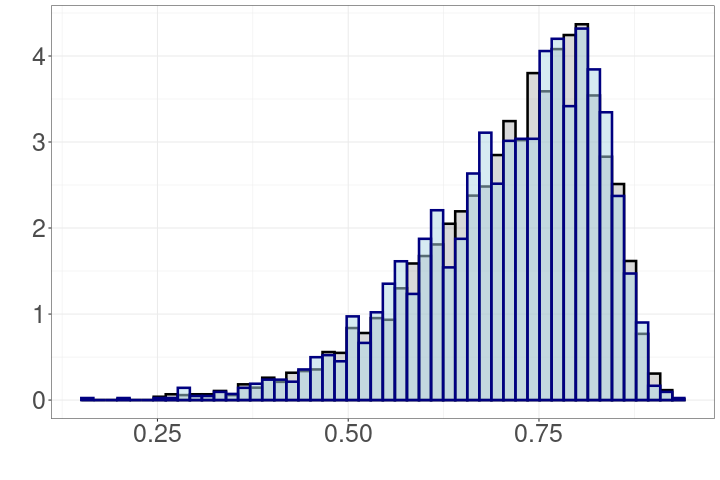}};
            \begin{scope}[x={(image.south east)},y={(image.north west)}]
                % Add annotation
                \node[black, font=\small] at (0.575,0) {$\hat{p}(X, Y_0)$};
                % Add "Count" annotation
                \node[black, font=\small, rotate=90] at (0,0.5) {Count};
            \end{scope}
        \end{tikzpicture}
        \caption*{(c) Logistic regression (Parametric)}
        %\label{subfig3}
    \end{minipage}
    \hfill
    \begin{minipage}[t]{0.43\textwidth}
        \centering
        \begin{tikzpicture}
            \node[anchor=south west,inner sep=0] (image) at (0,0) {\includegraphics[width=\textwidth]{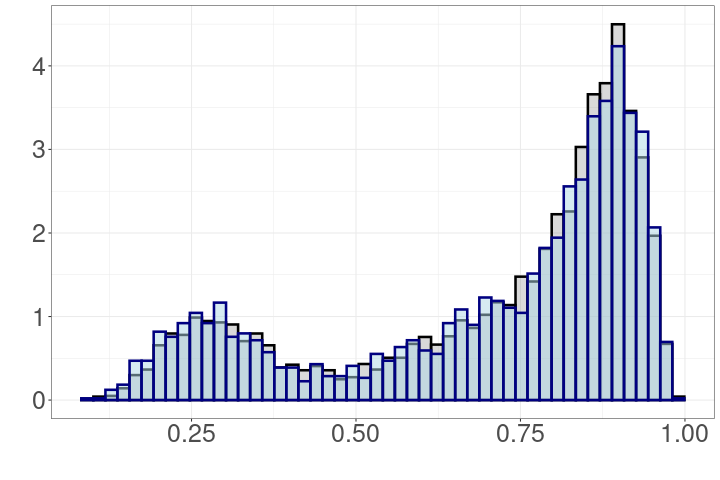}};
            \begin{scope}[x={(image.south east)},y={(image.north west)}]
                % Add annotation
                \node[black, font=\small] at (0.575,0) {$\hat{p}(X, Y_0)$};
                % Add "Count" annotation
                \node[black, font=\small, rotate=90] at (0,0.5) {Count};
            \end{scope}
        \end{tikzpicture}
        \caption*{(d) Random forest}
        %\label{subfig3}
    \end{minipage}
    \hfill
    \begin{minipage}[t]{0.43\textwidth}
        \centering
        \begin{tikzpicture}
            \node[anchor=south west,inner sep=0] (image) at (0,0) {\includegraphics[width=\textwidth]{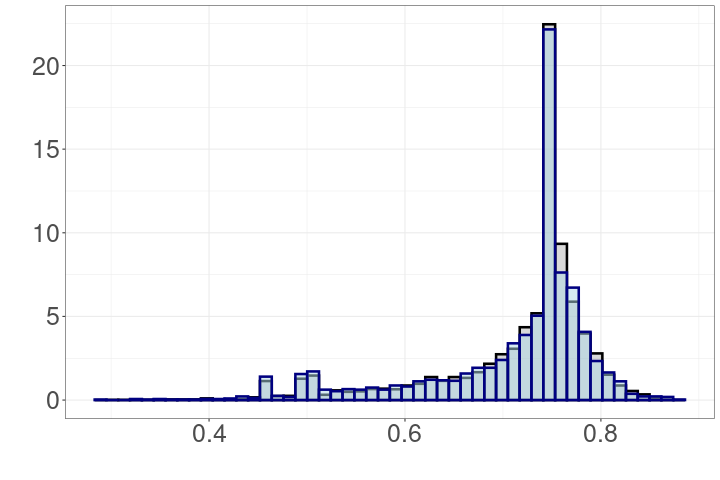}};
            \begin{scope}[x={(image.south east)},y={(image.north west)}]
                % Add annotation
                \node[black, font=\small] at (0.575,0) {$\hat{p}(X, Y_0)$};
                % Add "Count" annotation
                \node[black, font=\small, rotate=90] at (0,0.5) {Count};
            \end{scope}
        \end{tikzpicture}
        \caption*{(e) Support vector machine}
        %\label{subfig3}
    \end{minipage}
    \label{fig:psJCunconf}
    
    \begin{notes}
    The figure displays the common support for the propensity scores $\hat{p}(X, Y_0)$. Treated observations are color-coded in gray, and control observations are in blue. The sample size is $n = 9240$, with $6574$ treated units.
    \end{notes}

\end{figure}

% --------------------------------- DiD ---------------------------------------------- %

\begin{figure}[H]
    \caption{Job Corps data (proportion of weeks employed): Common support of $\hat{\pi}(X)$}
    
    \begin{minipage}[t]{0.43\textwidth}
        \centering
        \begin{tikzpicture}
            \node[anchor=south west,inner sep=0] (image) at (0,0) {\includegraphics[width=\textwidth]{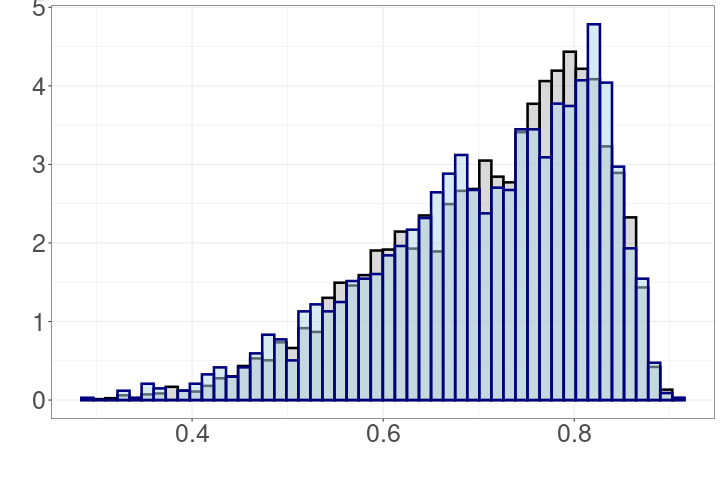}};
            \begin{scope}[x={(image.south east)},y={(image.north west)}]
                % Add annotation
                \node[black, font=\small] at (0.575,0) {$\hat{\pi}(X)$};
                % Add "Count" annotation
                \node[black, font=\small, rotate=90] at (0,0.5) {Count};
            \end{scope}
        \end{tikzpicture}
        \caption*{(a) Ensemble learner}
        %\label{subfig1}
    \end{minipage}
    \hfill
    \begin{minipage}[t]{0.43\textwidth}
        \centering
        \begin{tikzpicture}
            \node[anchor=south west,inner sep=0] (image) at (0,0) {\includegraphics[width=\textwidth]{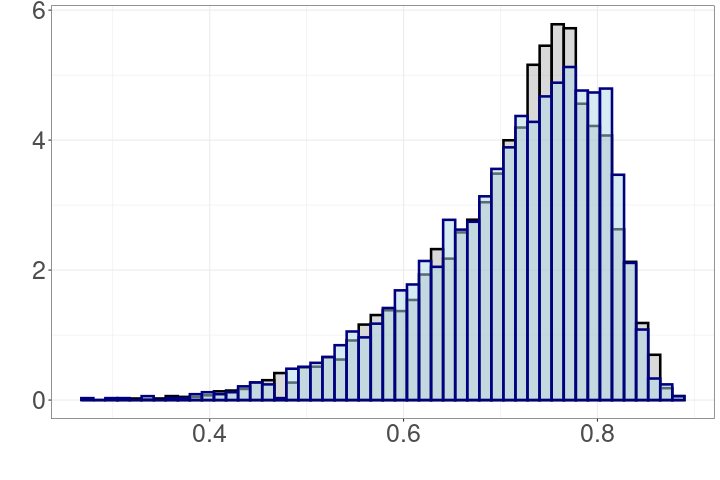}};
            \begin{scope}[x={(image.south east)},y={(image.north west)}]
                % Add annotation
                \node[black, font=\small] at (0.575,0) {$\hat{\pi}(X)$};
                % Add "Count" annotation
                \node[black, font=\small, rotate=90] at (0,0.5) {Count};
            \end{scope}
        \end{tikzpicture}
        \caption*{(b) Lasso}
        %\label{subfig2}
    \end{minipage}
    \hfill
    \begin{minipage}[t]{0.43\textwidth}
        \centering
        \begin{tikzpicture}
            \node[anchor=south west,inner sep=0] (image) at (0,0) {\includegraphics[width=\textwidth]{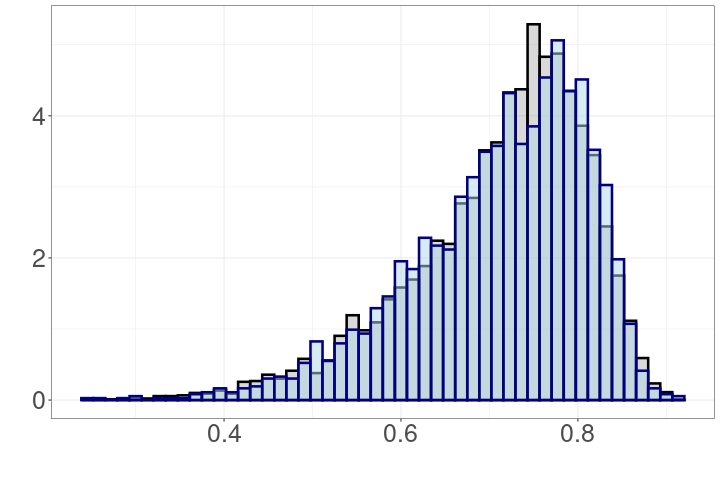}};
            \begin{scope}[x={(image.south east)},y={(image.north west)}]
                % Add annotation
                \node[black, font=\small] at (0.575,0) {$\hat{\pi}(X)$};
                % Add "Count" annotation
                \node[black, font=\small, rotate=90] at (0,0.5) {Count};
            \end{scope}
        \end{tikzpicture}
        \caption*{(c) Logistic regression (Parametric)}
        %\label{subfig3}
    \end{minipage}
    \hfill
    \begin{minipage}[t]{0.43\textwidth}
        \centering
        \begin{tikzpicture}
            \node[anchor=south west,inner sep=0] (image) at (0,0) {\includegraphics[width=\textwidth]{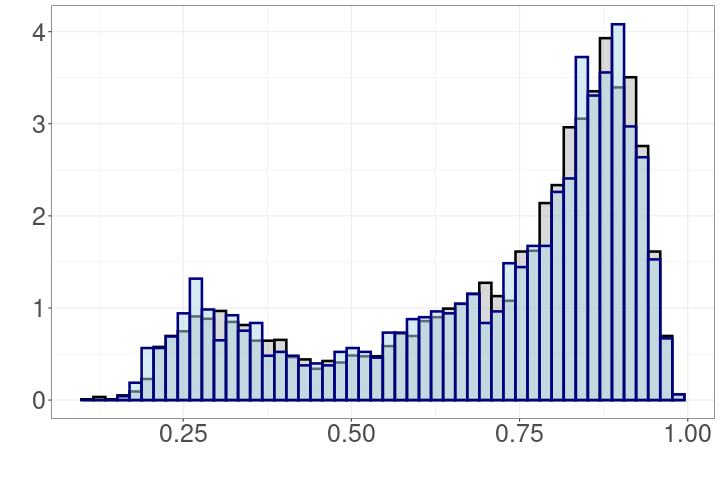}};
            \begin{scope}[x={(image.south east)},y={(image.north west)}]
                % Add annotation
                \node[black, font=\small] at (0.575,0) {$\hat{\pi}(X)$};
                % Add "Count" annotation
                \node[black, font=\small, rotate=90] at (0,0.5) {Count};
            \end{scope}
        \end{tikzpicture}
        \caption*{(d) Random forest}
        %\label{subfig3}
    \end{minipage}
    \hfill
    \begin{minipage}[t]{0.43\textwidth}
        \centering
        \begin{tikzpicture}
            \node[anchor=south west,inner sep=0] (image) at (0,0) {\includegraphics[width=\textwidth]{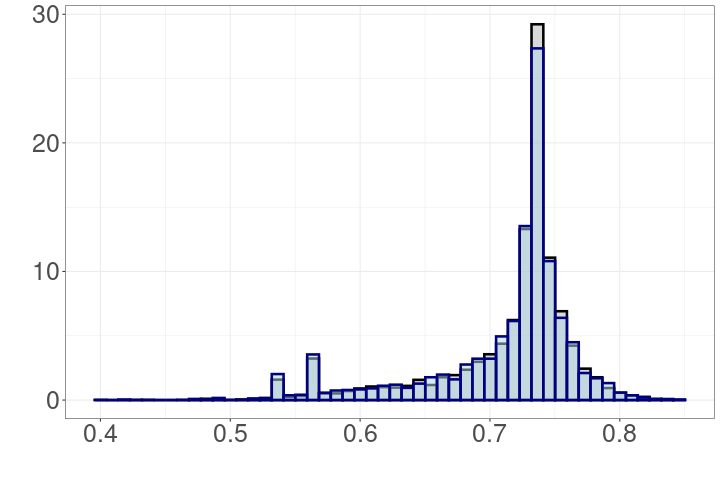}};
            \begin{scope}[x={(image.south east)},y={(image.north west)}]
                % Add annotation
                \node[black, font=\small] at (0.575,0) {$\hat{p}(X, Y_0)$};
                % Add "Count" annotation
                \node[black, font=\small, rotate=90] at (0,0.5) {Count};
            \end{scope}
        \end{tikzpicture}
        \caption*{(e) Support vector machine}
        %\label{subfig3}
    \end{minipage}

    \label{fig:psJCDiD}
    \begin{notes}
    The figure displays the common support for the propensity scores $\hat{\pi}(X)$. Treated observations are color-coded in gray, and control observations are in blue. The sample size is $n = 9240$, with $6574$ treated units.
    \end{notes}

\end{figure}

% ------------------------------------------------------------------------------ %
% Health status

\begin{figure}[H]
    \caption{Job Corps data (general health status): Common support of $\hat{p}(X, Y_0)$}
    
    \begin{minipage}[t]{0.43\textwidth}
        \centering
        \begin{tikzpicture}
            \node[anchor=south west,inner sep=0] (image) at (0,0) {\includegraphics[width=\textwidth]{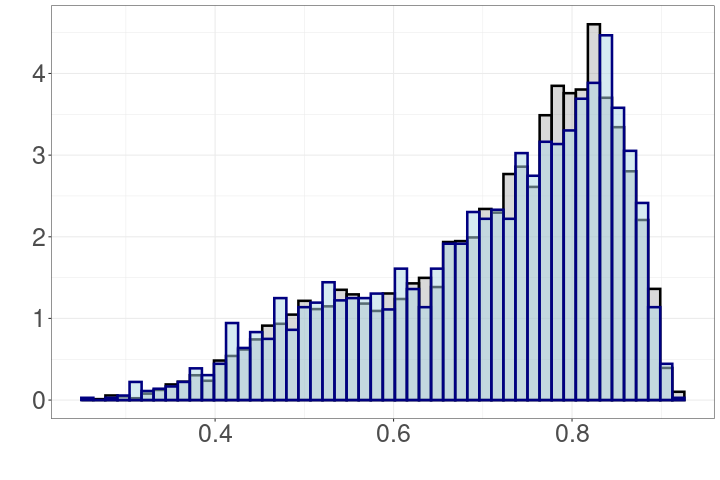}};
            \begin{scope}[x={(image.south east)},y={(image.north west)}]
                % Add annotation
                \node[black, font=\small] at (0.575,0) {$\hat{p}(X, Y_0)$};
                % Add "Count" annotation
                \node[black, font=\small, rotate=90] at (0,0.5) {Count};
            \end{scope}
        \end{tikzpicture}
        \caption*{(a) Ensemble learner}
        %\label{subfig1}
    \end{minipage}
    \hfill
    \begin{minipage}[t]{0.43\textwidth}
        \centering
        \begin{tikzpicture}
            \node[anchor=south west,inner sep=0] (image) at (0,0) {\includegraphics[width=\textwidth]{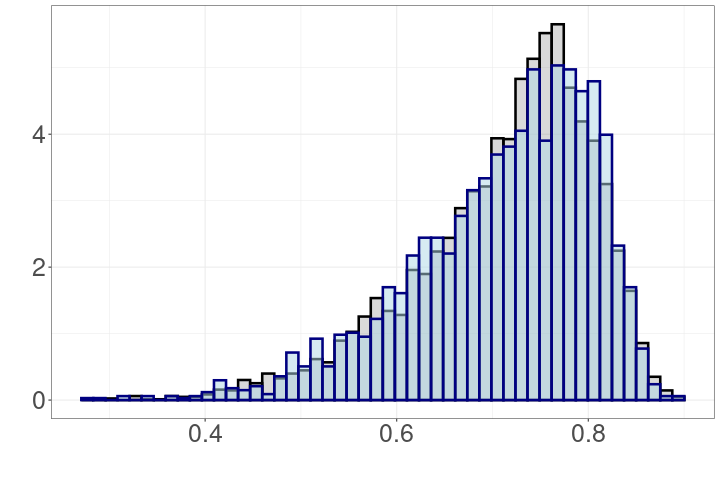}};
            \begin{scope}[x={(image.south east)},y={(image.north west)}]
                % Add annotation
                \node[black, font=\small] at (0.575,0) {$\hat{p}(X, Y_0)$};
                % Add "Count" annotation
                \node[black, font=\small, rotate=90] at (0,0.5) {Count};
            \end{scope}
        \end{tikzpicture}
        \caption*{(b) Lasso}
        %\label{subfig2}
    \end{minipage}
    \hfill
    \begin{minipage}[t]{0.43\textwidth}
        \centering
        \begin{tikzpicture}
            \node[anchor=south west,inner sep=0] (image) at (0,0) {\includegraphics[width=\textwidth]{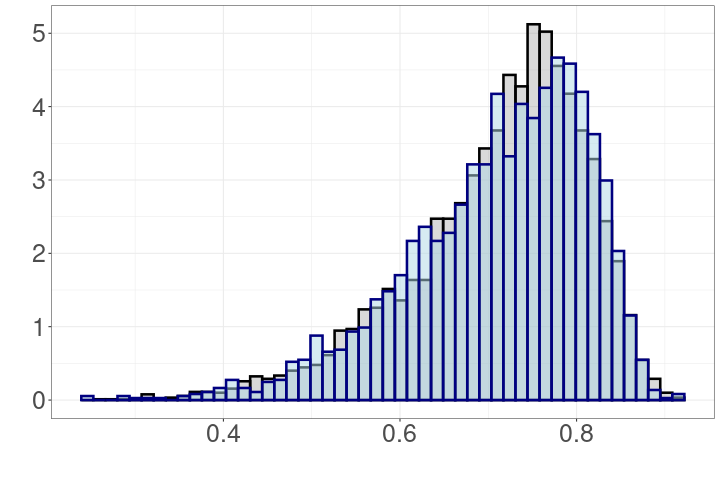}};
            \begin{scope}[x={(image.south east)},y={(image.north west)}]
                % Add annotation
                \node[black, font=\small] at (0.575,0) {$\hat{p}(X, Y_0)$};
                % Add "Count" annotation
                \node[black, font=\small, rotate=90] at (0,0.5) {Count};
            \end{scope}
        \end{tikzpicture}
        \caption*{(c) Logistic regression (Parametric)}
        %\label{subfig3}
    \end{minipage}
    \hfill
    \begin{minipage}[t]{0.43\textwidth}
        \centering
        \begin{tikzpicture}
            \node[anchor=south west,inner sep=0] (image) at (0,0) {\includegraphics[width=\textwidth]{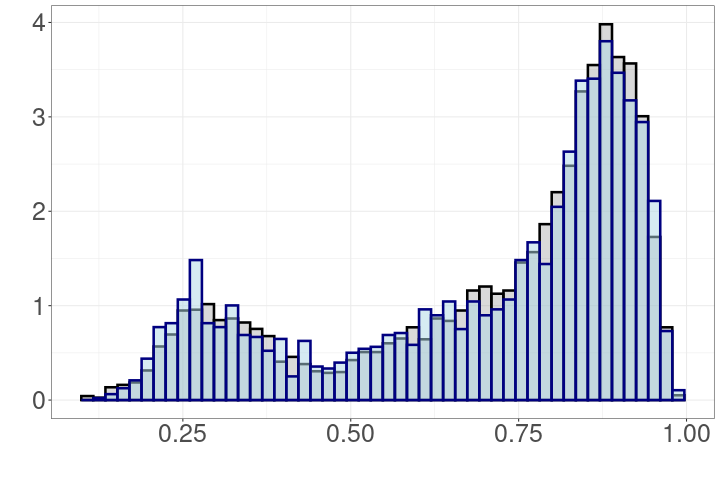}};
            \begin{scope}[x={(image.south east)},y={(image.north west)}]
                % Add annotation
                \node[black, font=\small] at (0.575,0) {$\hat{p}(X, Y_0)$};
                % Add "Count" annotation
                \node[black, font=\small, rotate=90] at (0,0.5) {Count};
            \end{scope}
        \end{tikzpicture}
        \caption*{(d) Random forest}
        %\label{subfig3}
    \end{minipage}
    \hfill
    \begin{minipage}[t]{0.43\textwidth}
        \centering
        \begin{tikzpicture}
            \node[anchor=south west,inner sep=0] (image) at (0,0) {\includegraphics[width=\textwidth]{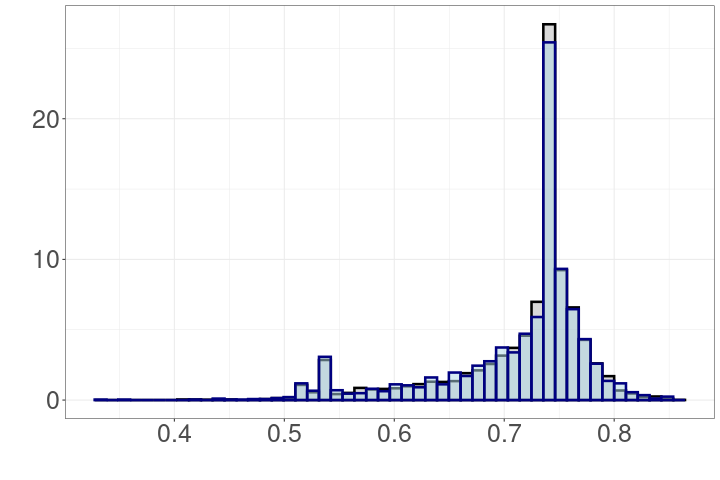}};
            \begin{scope}[x={(image.south east)},y={(image.north west)}]
                % Add annotation
                \node[black, font=\small] at (0.575,0) {$\hat{p}(X, Y_0)$};
                % Add "Count" annotation
                \node[black, font=\small, rotate=90] at (0,0.5) {Count};
            \end{scope}
        \end{tikzpicture}
        \caption*{(e) Support vector machine}
        %\label{subfig3}
    \end{minipage}
    \label{fig:psJCunconf}
    
    \begin{notes}
    The figure displays the common support for the propensity scores $\hat{p}(X, Y_0)$. Treated observations are color-coded in gray, and control observations are in blue. The sample size is $n = 9240$, with $6574$ treated units.
    \end{notes}

\end{figure}

% --------------------------------- DiD ---------------------------------------------- %

\begin{figure}[H]
    \caption{Job Corps data (general health status): Common support of $\hat{\pi}(X)$}
    
    \begin{minipage}[t]{0.43\textwidth}
        \centering
        \begin{tikzpicture}
            \node[anchor=south west,inner sep=0] (image) at (0,0) {\includegraphics[width=\textwidth]{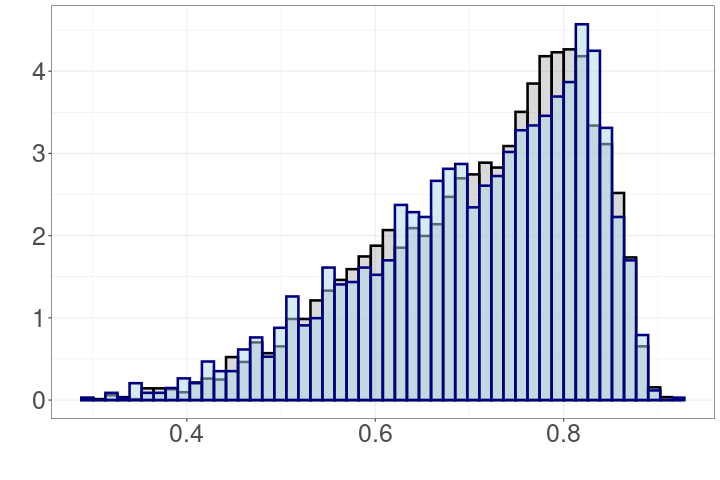}};
            \begin{scope}[x={(image.south east)},y={(image.north west)}]
                % Add annotation
                \node[black, font=\small] at (0.575,0) {$\hat{\pi}(X)$};
                % Add "Count" annotation
                \node[black, font=\small, rotate=90] at (0,0.5) {Count};
            \end{scope}
        \end{tikzpicture}
        \caption*{(a) Ensemble learner}
        %\label{subfig1}
    \end{minipage}
    \hfill
    \begin{minipage}[t]{0.43\textwidth}
        \centering
        \begin{tikzpicture}
            \node[anchor=south west,inner sep=0] (image) at (0,0) {\includegraphics[width=\textwidth]{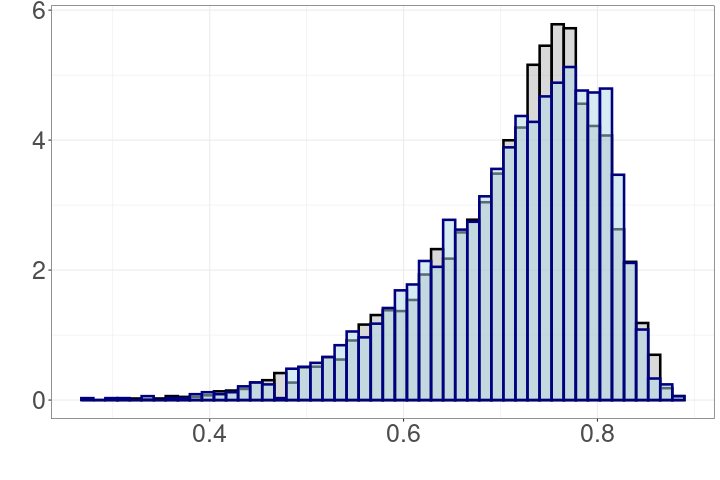}};
            \begin{scope}[x={(image.south east)},y={(image.north west)}]
                % Add annotation
                \node[black, font=\small] at (0.575,0) {$\hat{\pi}(X)$};
                % Add "Count" annotation
                \node[black, font=\small, rotate=90] at (0,0.5) {Count};
            \end{scope}
        \end{tikzpicture}
        \caption*{(b) Lasso}
        %\label{subfig2}
    \end{minipage}
    \hfill
    \begin{minipage}[t]{0.43\textwidth}
        \centering
        \begin{tikzpicture}
            \node[anchor=south west,inner sep=0] (image) at (0,0) {\includegraphics[width=\textwidth]{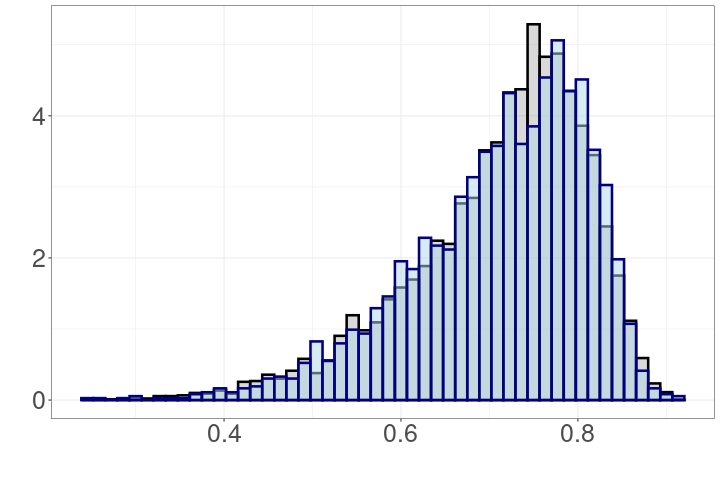}};
            \begin{scope}[x={(image.south east)},y={(image.north west)}]
                % Add annotation
                \node[black, font=\small] at (0.575,0) {$\hat{\pi}(X)$};
                % Add "Count" annotation
                \node[black, font=\small, rotate=90] at (0,0.5) {Count};
            \end{scope}
        \end{tikzpicture}
        \caption*{(c) Logistic regression (Parametric)}
        %\label{subfig3}
    \end{minipage}
    \hfill
    \begin{minipage}[t]{0.43\textwidth}
        \centering
        \begin{tikzpicture}
            \node[anchor=south west,inner sep=0] (image) at (0,0) {\includegraphics[width=\textwidth]{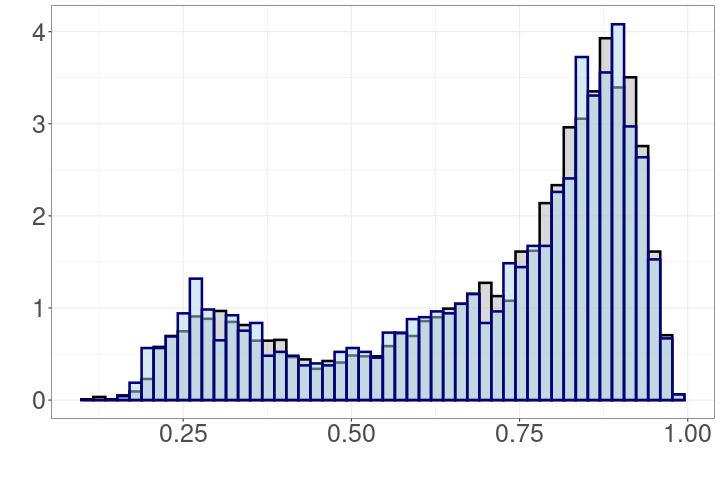}};
            \begin{scope}[x={(image.south east)},y={(image.north west)}]
                % Add annotation
                \node[black, font=\small] at (0.575,0) {$\hat{\pi}(X)$};
                % Add "Count" annotation
                \node[black, font=\small, rotate=90] at (0,0.5) {Count};
            \end{scope}
        \end{tikzpicture}
        \caption*{(d) Random forest}
        %\label{subfig3}
    \end{minipage}
    \hfill
    \begin{minipage}[t]{0.43\textwidth}
        \centering
        \begin{tikzpicture}
            \node[anchor=south west,inner sep=0] (image) at (0,0) {\includegraphics[width=\textwidth]{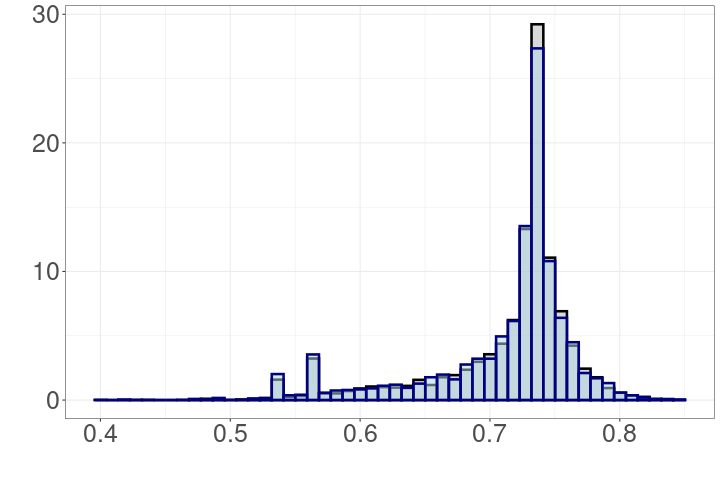}};
            \begin{scope}[x={(image.south east)},y={(image.north west)}]
                % Add annotation
                \node[black, font=\small] at (0.575,0) {$\hat{p}(X, Y_0)$};
                % Add "Count" annotation
                \node[black, font=\small, rotate=90] at (0,0.5) {Count};
            \end{scope}
        \end{tikzpicture}
        \caption*{(e) Support vector machine}
        %\label{subfig3}
    \end{minipage}

    \label{fig:psJCDiD}
    \begin{notes}
    The figure displays the common support for the propensity scores $\hat{\pi}(X)$. Treated observations are color-coded in gray, and control observations are in blue. The sample size is $n = 9240$, with $6574$ treated units.
    \end{notes}

\end{figure}

% ------------------------------------------------------------------------------ %
% NJmin
% ------------------------------------------------------------------------------ %
\subsection{Card and Krueger (1994) data}
\begin{figure}[H]
    \caption{Card and Krueger (1994) data: Common support of $\hat{p}(X, Y_0)$}
    
    \begin{minipage}[t]{0.43\textwidth}
        \centering
        \begin{tikzpicture}
            \node[anchor=south west,inner sep=0] (image) at (0,0) {\includegraphics[width=\textwidth]{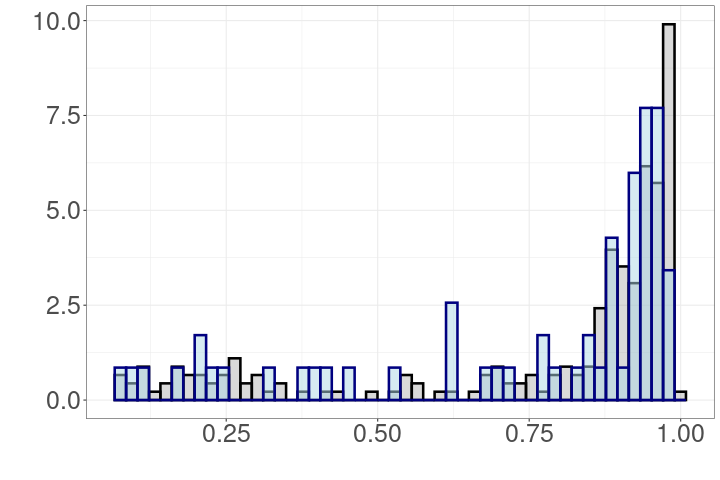}};
            \begin{scope}[x={(image.south east)},y={(image.north west)}]
                % Add annotation
                \node[black, font=\small] at (0.575,0) {$\hat{p}(X, Y_0)$};
                % Add "Count" annotation
                \node[black, font=\small, rotate=90] at (0,0.5) {Count};
            \end{scope}
        \end{tikzpicture}
        \caption*{(a) Ensemble learner}
        %\label{subfig1}
    \end{minipage}
    \hfill
    \begin{minipage}[t]{0.43\textwidth}
        \centering
        \begin{tikzpicture}
            \node[anchor=south west,inner sep=0] (image) at (0,0) {\includegraphics[width=\textwidth]{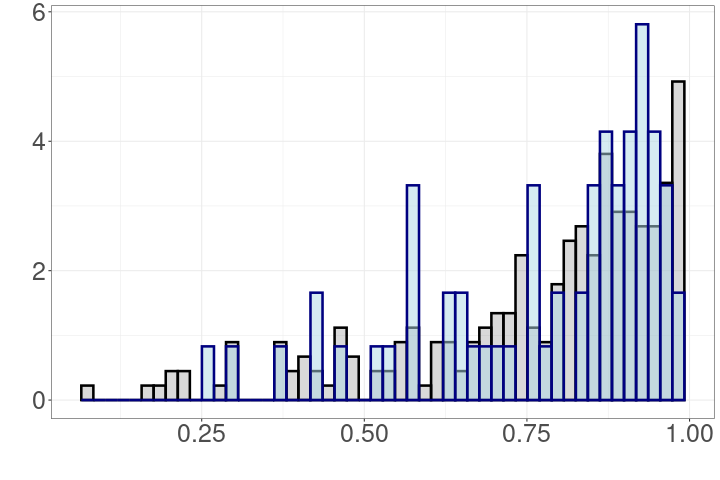}};
            \begin{scope}[x={(image.south east)},y={(image.north west)}]
                % Add annotation
                \node[black, font=\small] at (0.575,0) {$\hat{p}(X, Y_0)$};
                % Add "Count" annotation
                \node[black, font=\small, rotate=90] at (0,0.5) {Count};
            \end{scope}
        \end{tikzpicture}
        \caption*{(b) Lasso}
        %\label{subfig2}
    \end{minipage}
    \hfill
    \begin{minipage}[t]{0.43\textwidth}
        \centering
        \begin{tikzpicture}
            \node[anchor=south west,inner sep=0] (image) at (0,0) {\includegraphics[width=\textwidth]{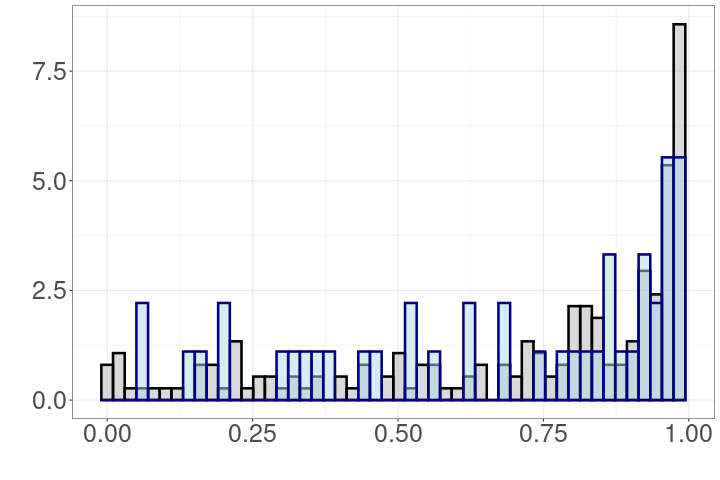}};
            \begin{scope}[x={(image.south east)},y={(image.north west)}]
                % Add annotation
                \node[black, font=\small] at (0.575,0) {$\hat{p}(X, Y_0)$};
                % Add "Count" annotation
                \node[black, font=\small, rotate=90] at (0,0.5) {Count};
            \end{scope}
        \end{tikzpicture}
        \caption*{(c) Logistic regression (Parametric)}
        %\label{subfig3}
    \end{minipage}
    \hfill
    \begin{minipage}[t]{0.43\textwidth}
        \centering
        \begin{tikzpicture}
            \node[anchor=south west,inner sep=0] (image) at (0,0) {\includegraphics[width=\textwidth]{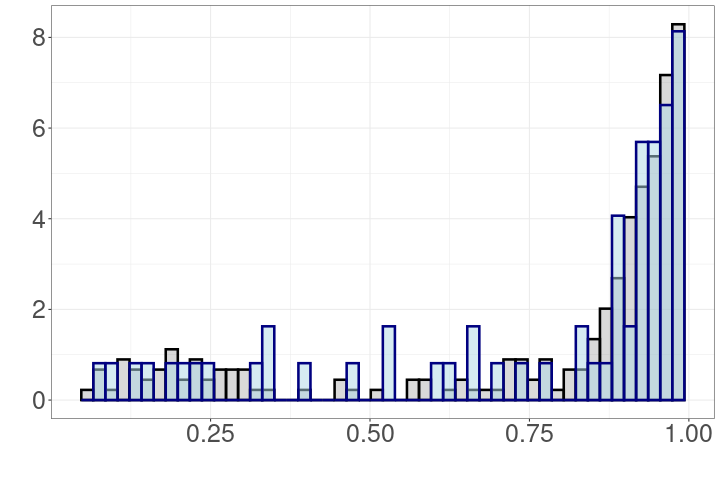}};
            \begin{scope}[x={(image.south east)},y={(image.north west)}]
                % Add annotation
                \node[black, font=\small] at (0.575,0) {$\hat{p}(X, Y_0)$};
                % Add "Count" annotation
                \node[black, font=\small, rotate=90] at (0,0.5) {Count};
            \end{scope}
        \end{tikzpicture}
        \caption*{(d) Random forest}
        %\label{subfig3}
    \end{minipage}

    \label{fig:psNjminUnconf}
    \begin{notes}
    The figure displays the common support for the propensity scores $\hat{p}(X, Y_0)$. Treated observations are color-coded in gray, and control observations are in blue. The sample size is $n = 334$, with $265$ treated units.
    \end{notes}

\end{figure}

% --------------------------------- DiD ---------------------------------------------- %

\begin{figure}[H]
    \caption{Card and Krueger (1994) data: Common support of $\hat{\pi}(X)$}
    
    \begin{minipage}[t]{0.43\textwidth}
        \centering
        \begin{tikzpicture}
            \node[anchor=south west,inner sep=0] (image) at (0,0) {\includegraphics[width=\textwidth]{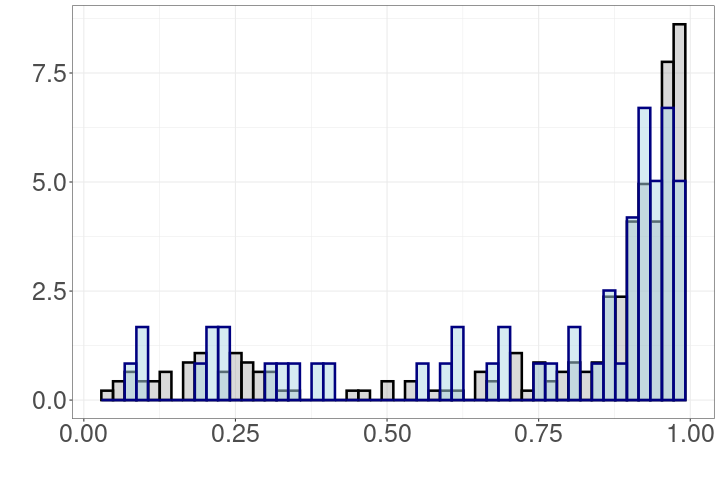}};
            \begin{scope}[x={(image.south east)},y={(image.north west)}]
                % Add annotation
                \node[black, font=\small] at (0.575,0) {$\hat{\pi}(X)$};
                % Add "Count" annotation
                \node[black, font=\small, rotate=90] at (0,0.5) {Count};
            \end{scope}
        \end{tikzpicture}
        \caption*{(a) Ensemble learner}
        %\label{subfig1}
    \end{minipage}
    \hfill
    \begin{minipage}[t]{0.43\textwidth}
        \centering
        \begin{tikzpicture}
            \node[anchor=south west,inner sep=0] (image) at (0,0) {\includegraphics[width=\textwidth]{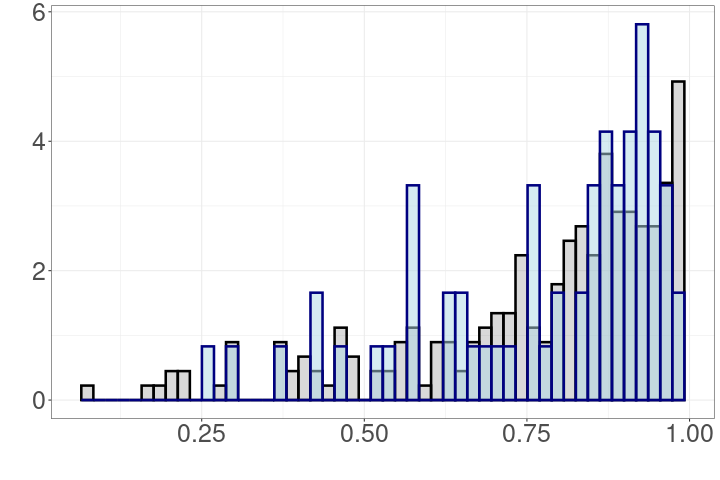}};
            \begin{scope}[x={(image.south east)},y={(image.north west)}]
                % Add annotation
                \node[black, font=\small] at (0.575,0) {$\hat{\pi}(X)$};
                % Add "Count" annotation
                \node[black, font=\small, rotate=90] at (0,0.5) {Count};
            \end{scope}
        \end{tikzpicture}
        \caption*{(b) Lasso}
        %\label{subfig2}
    \end{minipage}
    \hfill
    \begin{minipage}[t]{0.43\textwidth}
        \centering
        \begin{tikzpicture}
            \node[anchor=south west,inner sep=0] (image) at (0,0) {\includegraphics[width=\textwidth]{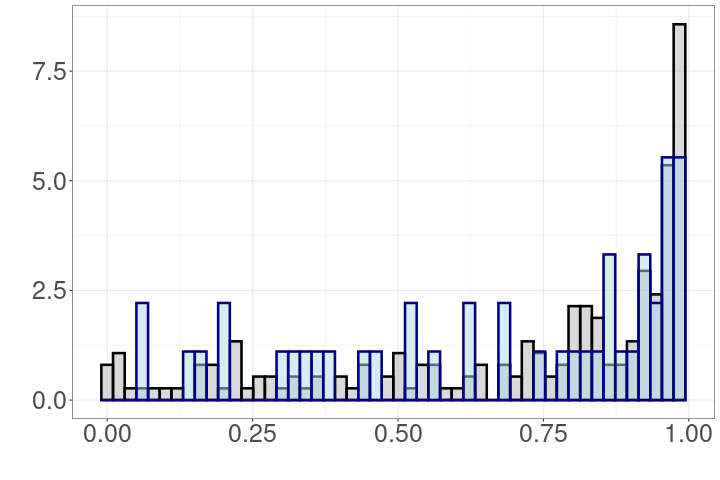}};
            \begin{scope}[x={(image.south east)},y={(image.north west)}]
                % Add annotation
                \node[black, font=\small] at (0.575,0) {$\hat{\pi}(X)$};
                % Add "Count" annotation
                \node[black, font=\small, rotate=90] at (0,0.5) {Count};
            \end{scope}
        \end{tikzpicture}
        \caption*{(c) Logistic regression (Parametric)}
        %\label{subfig3}
    \end{minipage}
    \hfill
    \begin{minipage}[t]{0.43\textwidth}
        \centering
        \begin{tikzpicture}
            \node[anchor=south west,inner sep=0] (image) at (0,0) {\includegraphics[width=\textwidth]{psnjminrandomforestunconf.png}};
            \begin{scope}[x={(image.south east)},y={(image.north west)}]
                % Add annotation
                \node[black, font=\small] at (0.575,0) {$\hat{\pi}(X)$};
                % Add "Count" annotation
                \node[black, font=\small, rotate=90] at (0,0.5) {Count};
            \end{scope}
        \end{tikzpicture}
        \caption*{(d) Random forest}
        %\label{subfig3}
    \end{minipage}

    \label{fig:psNjminDiD}
    \begin{notes}
    The figure displays the common support for the propensity scores $\hat{\pi}(X)$. Treated observations are color-coded in gray, and control observations are in blue. The sample size is $n = 334$, with $265$ treated units.
    \end{notes}

\end{figure}

% ------------------------------------------------------------------------------ %
% NHEFS
% ------------------------------------------------------------------------------ %
\subsection{NHEFS data}\begin{figure}[H]
    \caption{NHEFS data: Common support of $\hat{p}(X, Y_0)$}
    
    \begin{minipage}[t]{0.43\textwidth}
        \centering
        \begin{tikzpicture}
            \node[anchor=south west,inner sep=0] (image) at (0,0) {\includegraphics[width=\textwidth]{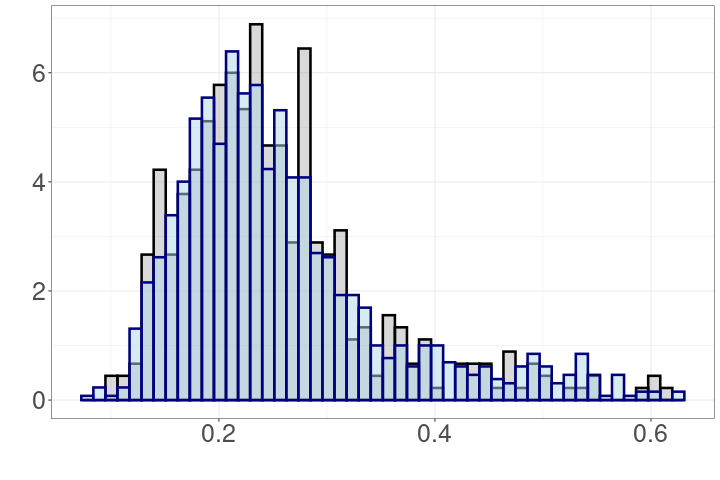}};
            \begin{scope}[x={(image.south east)},y={(image.north west)}]
                % Add annotation
                \node[black, font=\small] at (0.575,0) {$\hat{p}(X, Y_0)$};
                % Add "Count" annotation
                \node[black, font=\small, rotate=90] at (0,0.5) {Count};
            \end{scope}
        \end{tikzpicture}
        \caption*{(a) Ensemble learner}
        %\label{subfig1}
    \end{minipage}
    \hfill
    \begin{minipage}[t]{0.43\textwidth}
        \centering
        \begin{tikzpicture}
            \node[anchor=south west,inner sep=0] (image) at (0,0) {\includegraphics[width=\textwidth]{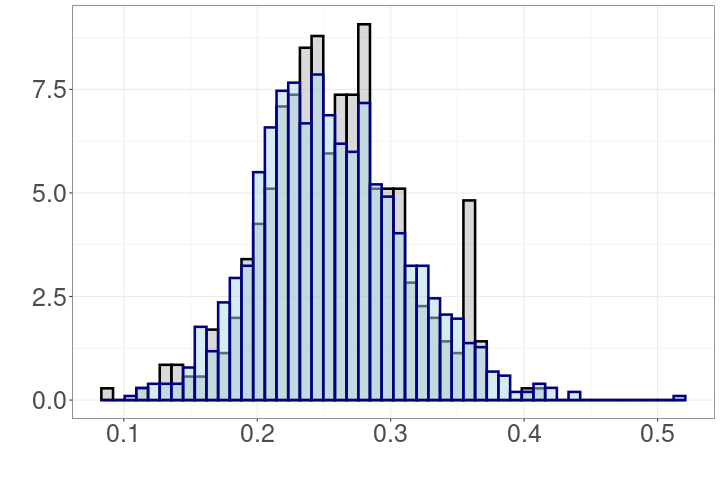}};
            \begin{scope}[x={(image.south east)},y={(image.north west)}]
                % Add annotation
                \node[black, font=\small] at (0.575,0) {$\hat{p}(X, Y_0)$};
                % Add "Count" annotation
                \node[black, font=\small, rotate=90] at (0,0.5) {Count};
            \end{scope}
        \end{tikzpicture}
        \caption*{(b) Lasso}
        %\label{subfig2}
    \end{minipage}
    \hfill
    \begin{minipage}[t]{0.43\textwidth}
        \centering
        \begin{tikzpicture}
            \node[anchor=south west,inner sep=0] (image) at (0,0) {\includegraphics[width=\textwidth]{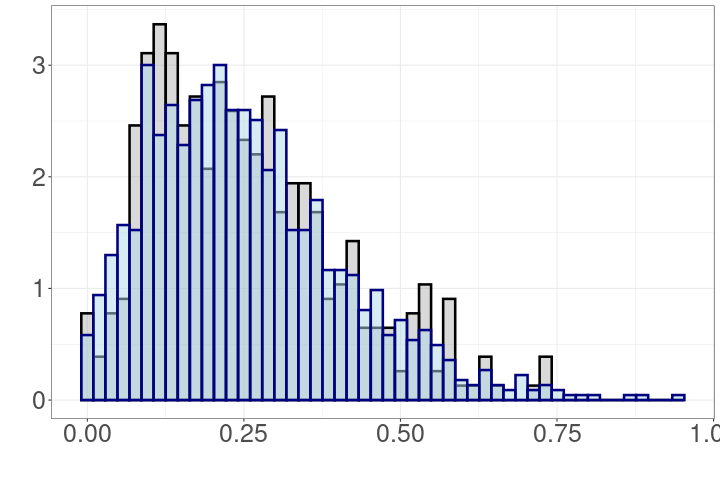}};
            \begin{scope}[x={(image.south east)},y={(image.north west)}]
                % Add annotation
                \node[black, font=\small] at (0.575,0) {$\hat{p}(X, Y_0)$};
                % Add "Count" annotation
                \node[black, font=\small, rotate=90] at (0,0.5) {Count};
            \end{scope}
        \end{tikzpicture}
        \caption*{(c) Logistic regression (Parametric)}
        %\label{subfig3}
    \end{minipage}
    \hfill
    \begin{minipage}[t]{0.43\textwidth}
        \centering
        \begin{tikzpicture}
            \node[anchor=south west,inner sep=0] (image) at (0,0) {\includegraphics[width=\textwidth]{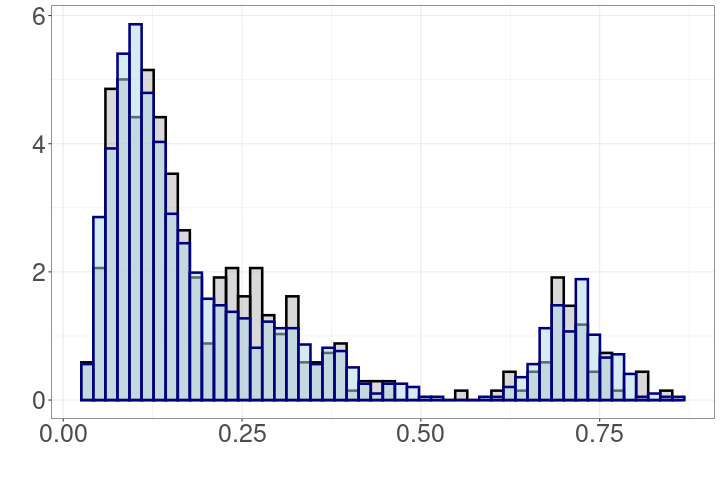}};
            \begin{scope}[x={(image.south east)},y={(image.north west)}]
                % Add annotation
                \node[black, font=\small] at (0.575,0) {$\hat{p}(X, Y_0)$};
                % Add "Count" annotation
                \node[black, font=\small, rotate=90] at (0,0.5) {Count};
            \end{scope}
        \end{tikzpicture}
        \caption*{(d) Random forest}
        %\label{subfig3}
    \end{minipage}

    \label{fig:psNHEFSUnconf}
    \begin{notes}
    The figure displays the common support for the propensity scores $\hat{p}(X, Y_0)$. Treated observations are color-coded in gray, and control observations are in blue. The sample size is $n = 1629$, with $403$ treated units.
    \end{notes}

\end{figure}

% --------------------------------- DiD ---------------------------------------------- %

\begin{figure}[H]
    \caption{NHEFS data: Common support of $\hat{\pi}(X)$}
    
    \begin{minipage}[t]{0.43\textwidth}
        \centering
        \begin{tikzpicture}
            \node[anchor=south west,inner sep=0] (image) at (0,0) {\includegraphics[width=\textwidth]{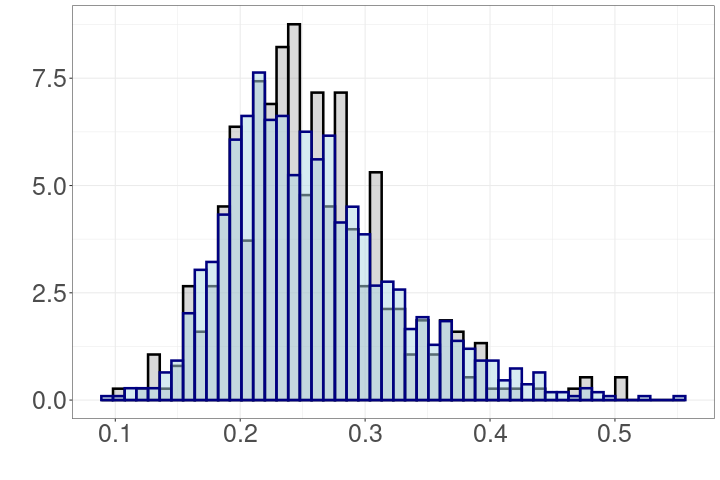}};
            \begin{scope}[x={(image.south east)},y={(image.north west)}]
                % Add annotation
                \node[black, font=\small] at (0.575,0) {$\hat{\pi}(X)$};
                % Add "Count" annotation
                \node[black, font=\small, rotate=90] at (0,0.5) {Count};
            \end{scope}
        \end{tikzpicture}
        \caption*{(a) Ensemble learner}
        %\label{subfig1}
    \end{minipage}
    \hfill
    \begin{minipage}[t]{0.43\textwidth}
        \centering
        \begin{tikzpicture}
            \node[anchor=south west,inner sep=0] (image) at (0,0) {\includegraphics[width=\textwidth]{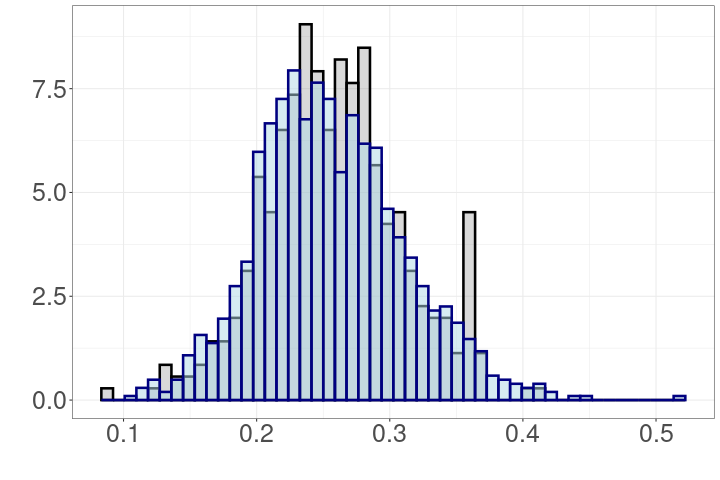}};
            \begin{scope}[x={(image.south east)},y={(image.north west)}]
                % Add annotation
                \node[black, font=\small] at (0.575,0) {$\hat{\pi}(X)$};
                % Add "Count" annotation
                \node[black, font=\small, rotate=90] at (0,0.5) {Count};
            \end{scope}
        \end{tikzpicture}
        \caption*{(b) Lasso}
        %\label{subfig2}
    \end{minipage}
    \hfill
    \begin{minipage}[t]{0.43\textwidth}
        \centering
        \begin{tikzpicture}
            \node[anchor=south west,inner sep=0] (image) at (0,0) {\includegraphics[width=\textwidth]{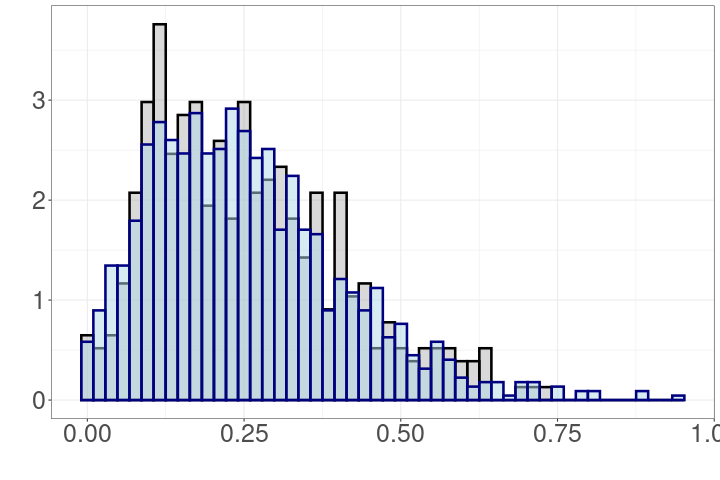}};
            \begin{scope}[x={(image.south east)},y={(image.north west)}]
                % Add annotation
                \node[black, font=\small] at (0.575,0) {$\hat{\pi}(X)$};
                % Add "Count" annotation
                \node[black, font=\small, rotate=90] at (0,0.5) {Count};
            \end{scope}
        \end{tikzpicture}
        \caption*{(c) Logistic regression (Parametric)}
        %\label{subfig3}
    \end{minipage}
    \hfill
    \begin{minipage}[t]{0.43\textwidth}
        \centering
        \begin{tikzpicture}
            \node[anchor=south west,inner sep=0] (image) at (0,0) {\includegraphics[width=\textwidth]{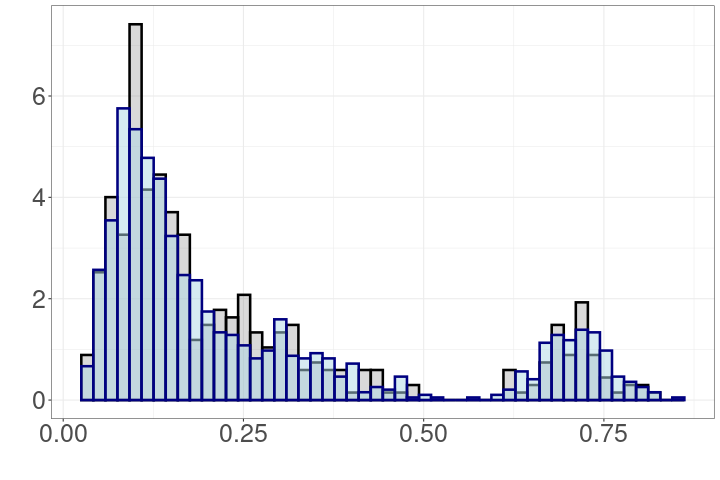}};
            \begin{scope}[x={(image.south east)},y={(image.north west)}]
                % Add annotation
                \node[black, font=\small] at (0.575,0) {$\hat{\pi}(X)$};
                % Add "Count" annotation
                \node[black, font=\small, rotate=90] at (0,0.5) {Count};
            \end{scope}
        \end{tikzpicture}
        \caption*{(d) Random forest}
        %\label{subfig3}
    \end{minipage}

    \label{fig:psNHEFSDiD}
    \begin{notes}
    The figure displays the common support for the propensity scores $\hat{\pi}(X)$. Treated observations are color-coded in gray, and control observations are in blue. The sample size is $n = 1629$, with $403$ treated units.
    \end{notes}

\end{figure}

\end{appendix}

\end{document}